\documentstyle[twocolumn,epsf]{article}

\newcommand{\ket}[1]{\left| {#1} \right>}
\newcommand{\bra}[1]{\left< {#1} \right|}
\newcommand{\beq}{\begin{eqnarray}}
\newcommand{\eeq}{\end{eqnarray}}
\newcommand{\eq}[1]{(\ref{#1})}
\newcommand{\lab}[1]{\label{#1}}

\begin{document}

\title{Quantum computing}
\author{Andrew Steane\\
Department of Atomic and Laser Physics, University of Oxford\\
Clarendon Laboratory, Parks Road, Oxford, OX1 3PU, England.}
\date{July 1997}
\maketitle

\onecolumn


\begin{center} {\bf Abstract} \end{center}

The subject of quantum computing brings together
ideas from classical information theory, computer science,
and quantum physics. This review aims to summarise not just quantum
computing, but the whole subject of quantum information theory.
Information can be identified as the most general thing which
must propagate from a cause to an effect. It therefore has
a fundamentally important role in the science of physics. 
However, the mathematical treatment of information,
especially information processing, is quite recent,
dating from the mid-twentieth century. This has meant that the full 
significance of information as a basic concept in physics is only now being 
discovered. This is especially true in quantum mechanics. The theory of 
quantum information and computing puts this significance on a firm footing, 
and has lead to some profound and exciting new insights into the natural
world. Among these are the use of quantum states to permit the
secure transmission of classical information (quantum cryptography), the 
use of quantum entanglement to permit reliable transmission of quantum 
states (teleportation), the
possibility of preserving quantum coherence in 
the presence of irreversible noise processes (quantum error correction), 
and the use of controlled quantum evolution for 
efficient computation (quantum computation). 
The common theme of all these insights is the use of quantum
entanglement as a computational resource.

It turns out that information theory and quantum mechanics fit together very 
well. In order to explain their relationship, this review begins with
an introduction to classical information theory and computer science,
including Shannon's theorem, error correcting codes, Turing machines
and computational complexity. The principles of quantum mechanics
are then outlined, and the EPR experiment described. The EPR-Bell
correlations, and quantum entanglement in general, form the essential
new ingredient which distinguishes quantum from classical information
theory, and, arguably, quantum from classical physics.

Basic quantum information ideas are next outlined, including qubits and data 
compression, quantum gates, the `no cloning' property, and teleportation. 
Quantum cryptography is briefly sketched. The universal quantum computer is 
described, based on the Church-Turing Principle and a network model of 
computation. Algorithms for such a computer are discussed, especially those 
for finding the period of a function, and searching a random list. Such 
algorithms prove that a quantum computer of sufficiently precise 
construction is not only fundamentally different from any computer which can 
only manipulate classical information, but can compute a small class of 
functions with greater efficiency. This implies that some important 
computational tasks are impossible for any device apart from a quantum 
computer. 

To build a universal quantum computer is well beyond the abilities of current 
technology. However, the principles of quantum information physics can be 
tested on smaller devices. The current experimental situation is reviewed, 
with emphasis on the linear ion trap, high-$Q$ optical cavities, and nuclear 
magnetic resonance methods. These allow coherent control in a Hilbert space 
of eight dimensions (3 qubits), and should be extendable up to a thousand or 
more dimensions (10 qubits). Among other things, these systems will allow 
the feasibility of quantum computing to be assessed. In fact such 
experiments are so difficult that it seemed likely until recently that a 
practically useful quantum computer (requiring, say, 1000 qubits) was 
actually ruled out by considerations of experimental imprecision and the 
unavoidable coupling between any system and its environment. However, a 
further fundamental part of quantum information physics provides a solution 
to this impasse. This is quantum error correction (QEC).

An introduction to quantum error correction is provided. The evolution of 
the quantum computer is restricted to a carefully chosen sub-space of its 
Hilbert space. Errors are almost certain to cause a departure from this 
sub-space. QEC provides a means to detect and undo such departures without 
upsetting the quantum computation. This achieves the apparently impossible, 
since the computation preserves quantum coherence even though during its 
course all the qubits in the computer will have relaxed spontaneously many 
times. 

The review concludes with an
outline of the main features of quantum information
physics, and avenues for future research.



\newpage
\twocolumn

\tableofcontents
\newpage

\section{Introduction}

The science of physics seeks to ask, and find precise answers to,
basic questions about why nature is as it is. Historically, 
the fundamental principles of physics have been 
concerned with questions such as ``what are things made of?'' and
``why do things move as they do?'' In his {\em Principia}, Newton
gave very wide-ranging answers to some of these questions. By showing
that the same mathamatical equations could describe the motions 
of everyday objects and of planets, he showed that an everyday
object such as a tea pot is made of essentially the {\em same sort of
stuff} as a planet: the motions of both can be described in terms
of their mass and the forces acting on them. Nowadays we would say that
both move in such a way as to conserve energy and momentum. In this
way, physics allows us to abstract from nature concepts such as
energy or momentum which
always obey fixed equations, although the same energy might be
expressed in many different ways: for example, an electron
in the large electron-positron collider at CERN, Geneva, can
have the same kinetic energy as a slug on a lettuce leaf.

Another thing which can be expressed in many different ways is
{\em information}. For example, the two statements
``the quantum computer is very interesting'' and ``l'ordinateur quantique
est tr\`es int\'eressant'' have something in common, although they
share no words. The thing they have in common is their {\em information}
content.
Essentially the same information could be expressed in many other
ways, for example by substituting numbers for letters in a 
scheme such as $a \rightarrow 97,\;b \rightarrow 98,\;c \rightarrow 99$
and so on, in which case the english version of the above statement
becomes 116 104 101 32 113 117 97 110 116 117 109 \ldots. 
It is very significant that information can be expressed
in different ways without losing its essential nature, since this leads
to the possibility of the automatic manipulation of information: a
machine need only be able to manipulate quite simple things like
integers in order to do surprisingly powerful information processing,
from document preparation to differential calculus, even
to translating between human languages. We are familiar with this
now, because of the ubiquitous computer, but even fifty years ago
such a widespread significance of automated information processing
was not forseen.

However, there is one thing that all ways of expressing information
must have in common: they all use real physical things to do the
job. Spoken words are conveyed by air pressure fluctuations, written
ones by arrangements of ink molecules on paper, even thoughts
depend on neurons (Landauer 1991). The rallying cry of the
information physicist
is ``no information without physical representation!'' 
Conversely, the fact that information is insensitive to exactly
how it is expressed, and can be freely translated from one form
to another, makes it an obvious candidate for a
fundamentally important role in physics, like energy
and momentum and other such abstractions. However, until the
second half of this century, the precise mathematical
treatment of information, especially information processing,
was undiscovered, so the significance of information in
physics was only hinted at in concepts such as entropy
in thermodynamics. It now appears that information may have a much deeper 
significance. Historically, much of fundamental physics has been concerned 
with discovering the fundamental particles of nature and the equations which 
describe their motions and interactions. It now appears that a different 
programme may be equally important: to discover the ways that nature allows, 
and prevents, {\em information} to be expressed and manipulated, rather than 
particles to move. For example, the best way to state exactly what can and
cannot travel faster than light is to identify information as the
speed-limited entity. In quantum
mechanics, it is highly significant that the state vector
must not contain, whether explicitly or implicitly, more
information than can meaningfully be associated with a given system.
Among other things this produces the wavefunction symmetry requirements
which lead to Bose Einstein and Fermi Dirac
statistics, the periodic structure of atoms, and so on.

The programme to re-investigate the fundamental principles of
physics from the standpoint of information theory is still
in its infancy. However, it already appears to be highly fruitful,
and it is this ambitious programme that I aim to summarise.

Historically, the concept of information in physics does not have
a clear-cut origin. An important thread can be traced if we
consider the paradox of Maxwell's demon of 1871 (fig. 1) (see also
Brillouin 1956). Recall that Maxwell's 
demon is a creature that opens and closes a trap door between two 
compartments of a chamber containing gas, and pursues the subversive policy 
of only opening the door when fast molecules approach it from the right, or 
slow ones from the left. In this way the demon establishes a temperature 
difference between the two compartments without doing any work, in 
violation of the second law of thermodynamics, and consequently permitting 
a host of contradictions. 

A number of attempts were made to exorcise Maxwell's demon (see
Bennett 1987), such as 
arguments that the demon cannot gather information without doing work, or 
without disturbing (and thus heating) the gas, both of which
are untrue. Some were tempted to
propose that the 2nd law of thermodynamics could indeed be violated
by the actions of an ``intelligent being.'' It was not until 1929
that Leo Szilard made progress by reducing the
problem to its essential components, in which the demon need
merely identify whether a single molecule is to the right or left
of a sliding partition, and its action
allows a simple heat engine, called Szilard's engine, to be run. Szilard
still had not solved the problem, since his analysis 
was unclear about whether or not the
act of measurement, whereby the demon
learns whether the molecule is to the left or the right, must involve
an increase in entropy.

A definitive and clear answer was not forthcoming, surprisingly, 
until a further fifty years had passed. In the intermediate years digital 
computers were developed, and the physical implications
of information gathering and processing were carefully considered.
The thermodynamic costs of elementary information manipulations
were analysed by Landauer and others during the 1960s (Landauer 1961,
Keyes and Landauer 1970),
and those of general computations by Bennett, Fredkin, Toffoli and others 
during the 1970s (Bennett 1973, Toffoli 1980, Fredkin and Toffoli 1982). It was 
found that almost anything can in principle be done in a reversible
manner, i.e. with no entropy cost at all (Bennett and Landauer 1985).
Bennett (1982) made explicit the relation between 
this work and Maxwell's paradox by proposing that the demon can indeed learn 
where the molecule is in Szilard's engine without doing any work or increasing 
any entropy in the environment, and so obtain useful work during one stroke of 
the engine. However, the information about the molecule's location must then be 
present in the demon's memory (fig. 1). As more and more strokes are performed, 
more and more information gathers in the demon's memory. To complete a 
thermodynamic cycle, the demon must {\em erase} its memory, and it is during 
this erasure operation that we identify an increase in entropy in the 
environment, as required by the 2nd law. This completes the essential physics 
of Maxwell's demon; further subtleties are discussed by 
Zurek (1989), Caves (1990), and Caves, Unruh and Zurek (1990). 

The thread we just followed was instructive, but to provide
a complete history of ideas relevent to quantum computing
is a formidable task. Our subject brings together what are arguably
two of the greatest revolutions in twentieth-century science,
namely quantum mechanics and information science (including computer
science). The relationship between these two giants is illustrated
in fig. 2. 

Classical information theory is founded on the
definition of information. A warning is in order here.
Whereas the theory tries to capture much of the normal
meaning of the term `information', it can no more do justice to the
full richness of that term in everyday language than particle
physics can encapsulate the everyday meaning of `charm'. 
`Information' for us will be an abstract term, defined in detail
in section \ref{s:mi}.
Much of information theory dates back to seminal work of Shannon in 
the 1940's (Slepian 1974). The observation that information
can be translated from one form to another is encapsulated and
quantified in Shannon's noiseless coding theorem (1948), which quantifies
the resources needed to store or transmit a given body
of information. Shannon also considered the fundamentally important
problem of communication in the presence of noise, and established
Shannon's main theorem (section \ref{s:ecc}) which is the
central result of classical information theory. Error-free
communication even in the presence of noise is achieved by means
of `error-correcting codes', and their study is a branch of mathematics
in its own right. Indeed, the journal {\em IEEE Transactions on Information
Theory} is almost totally taken up with the discovery and analysis
of error-correction by coding.
Pioneering work in this area was done by Golay (1949) and
Hamming (1950).

The foundations of computer science were formulated at roughly the same 
time as Shannon's information theory, and this is no coincidence. The 
father of computer science is arguably Alan Turing (1912-1954), and its prophet 
is Charles Babbage (1791-1871). Babbage conceived of most of the essential 
elements of a modern computer, though in his day there was not the technology 
available to implement his ideas. A century passed before Babbage's 
Analytical Engine was improved upon when Turing described the Universal Turing 
Machine in the mid 1930s. Turing's genius (see Hodges 1983)
was to clarify exactly what a 
calculating machine might be capable of, and to emphasise the role of 
programming, i.e. software, even more than Babbage had done. The giants
on whose 
shoulders Turing stood in order to get a better view were chiefly the 
mathematicians David Hilbert and Kurt G\"odel. Hilbert had emphasised 
between the 1890s and 1930s the importance of asking fundamental questions 
about the nature of mathematics. Instead of asking ``is this mathematical 
proposition true?'' Hilbert wanted to ask ``is it the case that every 
mathematical proposition can in principle be proved or disproved?''
This was unknown, but Hilbert's feeling, and that of most mathematicians, was 
that mathematics was indeed complete, so that conjectures such as Goldbach's 
(that every even number can be written as the sum of two primes) could be 
proved or disproved somehow, although the logical steps might be as yet 
undiscovered. 

G\"odel destroyed this hope by establishing
the existence of mathematical propositions which were 
undecidable, meaning that they could be neither proved nor disproved. The 
next interesting question was whether it would be easy to identify such 
propositions.
Progress in mathematics had always relied on the use of creative 
imagination, yet with hindsight mathematical proofs appear to be automatic, 
each step following inevitably from the one before. Hilbert asked whether 
this `inevitable' quality could be captured by a `mechanical' process. In 
other words, was there a universal mathematical method, which would 
establish the truth or otherwise of every mathematical assertion? After 
G\"odel, Hilbert's problem was re-phrased into that of establishing 
decidability rather than truth, and this is what Turing sought to address.

In the words of Newman, Turing's bold innovation was to introduce `paper
tape' into symbolic logic. In the search for an automatic process
by which mathematical questions could be decided, Turing envisaged
a thoroughly mechanical device, in fact a kind of glorified typewriter
(fig. 7).
The importance of the {\em Turing machine} (Turing 1936)
arises from the fact that it is
sufficiently complicated to address highly sophisticated mathematical
questions, but sufficiently simple to be subject to detailed analysis.
Turing used his machine as a theoretical construct to show that the assumed 
existence of a mechanical means to establish decidability leads to a 
contradiction (see section \ref{s:halting}). In other words, he was
initially concerned with quite 
abstract mathematics rather than practical computation. However, by 
seriously establishing the idea of automating abstract mathematical proofs 
rather than merely arithmatic, Turing greatly stimulated the development of 
general purpose information processing. This was in the days when a 
``computer'' was a person doing mathematics. 

Modern computers are neither Turing machines nor Babbage engines, though they 
are based on broadly similar principles, and their computational power
is equivalent (in a technical sense) to that of a Turing machine.
I will not trace their development 
here, since although this is a wonderful story, it would take too long to do 
justice to the many people involved. Let us just remark that all of this 
development represents a great improvement in speed and size, but does not 
involve any change in the essential idea of what a computer is, or how it 
operates. Quantum mechanics raises the possibility of such a change, however. 

Quantum mechanics is the mathematical structure which embraces, in 
principle, the whole of physics. We will not be directly concerned 
with gravity, high velocities, or exotic elementary particles, so the 
standard non-relativistic quantum mechanics will suffice. The
significant feature of quantum theory for our purpose is not the precise
details of the equations of motion, but the fact that they treat quantum
amplitudes, or state vectors in a Hilbert space, rather than classical
variables. It is this that allows new types of information
and computing.

There is a parallel between Hilbert's questions about mathematics and the 
questions we seek to pose in quantum information theory. Before Hilbert, 
almost all mathematical work had been concerned with establishing or 
refuting particular hypotheses, but Hilbert wanted to ask what general
type of hypothesis was even amenable to mathematical proof. Similarly,
most research in quantum physics has been concerned
with studying the evolution of specific physical systems, but
we want to ask what general type of evolution is even conceivable
under quantum mechanical rules.

The first deep insight into quantum information theory came with Bell's 
1964 analysis of the paradoxical thought-experiment proposed by Einstein, 
Podolsky and Rosen (EPR) in 1935. 
Bell's inequality draws attention to 
the importance of {\em correlations} between separated quantum systems 
which have interacted (directly or indirectly) in the past, but which no 
longer influence one another. In essence his argument shows that the degree 
of correlation which can be present in such systems exceeds that which 
could be predicted on the basis of {\em any} law of physics which describes 
particles in terms of classical variables rather than quantum states. 
Bell's argument was clarified by Bohm (1951, also Bohm and Aharonov 1957) 
and by Clauser, Holt, Horne and Shimony (1969), and experimental tests were 
carried out in the 1970s (see Clauser and Shimony (1978) and references
therein). Improvements in such 
experiments are largely concerned with preventing the possibility of any 
interaction between the separated quantum systems, and a significant step 
forward was made in the experiment of Aspect, Dalibard and Roger (1982), 
(see also Aspect 1991) since in their work any purported interaction would 
have either to travel faster than light, or possess other almost equally 
implausible qualities. 

The next link between quantum mechanics and information theory came
about when it was realised that simple properties of quantum systems,
such as the unavoidable disturbance involved in measurement, 
could be put to practical use, in {\em quantum cryptography}
(Wiesner 1983, Bennett {\em et. al.} 1982, Bennett and Brassard 1984;
for a recent review see Brassard and Crepeau 1996). 
Quantum cryptography covers several ideas, of which the most firmly
established is quantum key distribution.
This is an ingenious method in which transmitted quantum states are used to
perform a very particular communication task: to establish
at two separated locations a pair of identical, but
otherwise random, sequences of binary digits, without allowing
any third party to learn the sequence. This is very useful
because such a random sequence can be used as a cryptographic
key to permit secure communication. The significant feature
is that the principles of quantum mechanics guarantee
a type of conservation of quantum information, so that
if the necessary quantum information arrives at the parties
wishing to establish a random key, they can be sure it has
not gone elsewhere, such as to a spy. Thus the whole problem
of compromised keys, which fills the annals of espionage, is
avoided by taking advantage of the structure of the natural world.

While quantum cryptography was being analysed and demonstrated, the quantum 
computer was undergoing a quiet birth. Since quantum mechanics underlies
the behaviour of all systems, including those we call classical
(``even a screwdriver is quantum mechanical'', Landauer (1995)), it
was not obvious how to conceive of a distinctively quantum mechanical
computer, i.e. one which did not merely reproduce the action of a
classical Turing machine. Obviously it is not sufficient merely to
identify a quantum mechanical system whose evolution could be interpreted
as a computation; one must prove a much stronger result than this.
Conversely, we know that classical computers can simulate, by their 
computations, the evolution of any quantum system \ldots with one reservation: 
no classical process will allow one to prepare separated systems whose 
correlations break the Bell inequality. It appears from this that the
EPR-Bell correlations are the quintessential quantum-mechanical property
(Feynman 1982).

In order to think about computation from a quantum-mechanical point of view, 
the first ideas involved converting the action of a Turing machine into an 
equivalent reversible process, and then inventing a Hamiltonian which would 
cause a quantum system to evolve in a way which mimicked a reversible Turing 
machine. This depended on the work of Bennett (1973; see
also Lecerf 1963) who had shown that a 
universal classical computing machine (such as Turing's) could be made 
reversible while retaining its simplicity. Benioff (1980, 1982)
and others proposed such Turing-like Hamiltonians in the early 1980s. Although
Benioff's ideas did not allow the full analysis of quantum computation,
they showed that unitary quantum evolution is at least as powerful 
computationally as a classical computer.

A different approach was taken by Feynman (1982, 1986) who considered the 
possibility not of universal computation, but of universal {\em 
simulation}---i.e. a purpose-built quantum system which could simulate the {\em 
physical behaviour} of any other. Clearly, such a simulator would be a 
universal computer too, since any computer must be a physical system. Feynman 
gave arguments which suggested that quantum evolution could be used to 
compute certain problems more efficiently than any classical computer, but his 
device was not sufficiently specified to be called a computer, since he assumed 
that any interaction between adjacent two-state systems could be `ordered', 
without saying how. 

In 1985 an important step forward was taken by Deutsch. Deutsch's proposal 
is widely considered to represent the first blueprint for a quantum 
computer, in that it is sufficiently specific and simple to allow real 
machines to be contemplated, but sufficiently versatile to be a universal 
quantum simulator, though both points are debatable. Deutsch's
system is essentially a line of two-state systems, 
and looks more like a register machine than a Turing machine 
(both are universal classical computing machines). Deutsch proved that if 
the two-state systems could be made to evolve by means of a specific small 
set of simple operations, then {\em any} unitary evolution could be 
produced, and therefore the evolution could be made to simulate that of any 
physical system. He also discussed how to produce Turing-like behaviour 
using the same ideas. 

Deutsch's simple operations are now called quantum `gates', since they
play a role analogous to that of binary logic gates in classical
computers. Various authors have investigated the minimal class of gates
which are sufficient for quantum computation.

The two questionable aspects of Deutsch's proposal are its
efficiency and realisability. The question of efficiency is absolutely
fundamental in computer science, and on it the concept of `universality'
turns. A {\em universal} computer is one that not only
can reproduce (i.e. simulate) the action of any other, but can do so
without running too slowly. The `too slowly' here is defined in
terms of the number of computational steps required: this number must
not increase exponentially with the size of the input (the precise
meaning will be explained in section \ref{s:UTM}). Deutsch's
simulator is not universal in this strict sense, though
it was shown to be efficient for simulating a wide class of
quantum systems by Lloyd (1996). However, Deutsch's
work has established the concepts of quantum networks 
(Deutsch 1989) and quantum
logic gates, which are extremely important in that they allow us
to think clearly about quantum computation.

In the early 1990's several authors (Deutsch and Jozsa 1992, Berthiaume
and Brassard 1992, Bernstein and Vazirani 1993) sought computational 
tasks which could be solved by a quantum computer more efficiently than 
{\em any} classical computer. Such a quantum algorithm would play a 
conceptual role similar to that of Bell's inequality, in defining something 
of the essential nature of quantum mechanics. 
Initially only very small differences in performance were found, in which 
quantum mechanics permitted an answer to be found with certainty, as long 
as the quantum system was noise-free, where a probabilistic classical 
computer could achieve an answer `only' with high probability. 
An important advance was made by Simon (1994), who described an efficient 
quantum algorithm for a (somewhat abstract) problem for which no efficient 
solution was possible classically, even by probabilistic methods. This 
inspired Shor (1994) who astonished the community by describing an 
algorithm which was not only efficient on a quantum computer, but also 
addressed a central problem in computer science: that of factorising large 
integers.

Shor discussed both factorisation and discrete logarithms,
making use of a quantum Fourier transform method discovered by Coppersmith 
(1994) and Deutsch. Further important quantum algorithms were
discovered by Grover (1997) and Kitaev (1995).

Just as with classical computation and information theory, once
theoretical ideas about computation had got under way, an effort
was made to establish the essential nature of quantum information---the
task analogous to Shannon's work. The difficulty here
can be seen by considering the simplest quantum system, a two-state
system such as a spin half in a magnetic field. The quantum
state of a spin is a continuous quantity defined by two real numbers, 
so in principle it can store an infinite amount of classical information.
However, a measurement of a spin will only provide a single
two-valued answer (spin up$/$spin down)---there is no way to
gain access to the infinite information
which appears to be there, therefore it is incorrect to consider
the information content in those terms. This is reminiscent of the 
renormalisation problem in quantum electrodynamics. How much information
can a two-state quantum system store, then? The answer, provided by
Jozsa and Schumacher (1994) and 
Schumacher (1995), is {\em one two-state system's worth}!
Of course Schumacher and Jozsa did more than propose this simple answer,
rather they showed that the two-state system plays the role in
quantum information theory analogous to that of the bit in classical
information theory, in that the quantum information content of
{\em any} quantum system can be meaningfully measured
as the minimum number of two-state systems, now called quantum bits or
qubits, which would be needed to store or transmit the system's state
with high accuracy.

Let us return to the question of realisability of quantum computation. It is 
an elementary, but fundamentally important, observation that the quantum 
interference effects which permit algorithms such as Shor's are extremely 
fragile: the quantum computer is ultra-sensitive to experimental noise and 
impression. It is not true that early workers were unaware of this 
difficulty, rather their first aim was to establish whether a quantum 
computer had any fundamental significance at all. Armed with Shor's 
algorithm, it now appears that such a fundamental significance is 
established, by the following argument: either nature does allow a device to 
be run with sufficient precision to perform Shor's algorithm for large 
integers (greater than, say, a googol, $10^{100}$), or there are fundamental 
natural limits to precision in real systems. Both eventualities represent an 
important insight into the laws of nature. 

At this point, ideas of quantum information and quantum computing come 
together. For, a quantum computer can be made much less sensitive
to noise by means of a new idea which comes directly
from the marriage of quantum mechanics with classical information
theory, namely {\em quantum error correction}. Although the phrase
`error correction' is a natural one and was used with reference
to quantum computers prior to 1996, it was only in that year that
two important papers, of Calderbank and Shor, and independently
Steane, established a general framework whereby quantum information
processing can be used to combat a very wide class of noise
processes in a properly designed quantum system. Much progress
has since been made in generalising these ideas (Knill and Laflamme 1997, 
Ekert and Macchiavello 1996, Bennett {\em et. al.} 1996b,
Gottesman 1996, Calderbank {\em et. al.} 1997). An important
development was the demonstration by Shor (1996) and Kitaev (1996)
that correction can be achieved even when the corrective operations
are themselves imperfect. Such methods lead to a general concept
of `fault tolerant' computing, of which a helpful review is provided by 
Preskill (1997). 

If, as seems almost certain, quantum computation will only work in conjunction
with quantum error correction, it appears that the relationship between quantum 
information theory and quantum computers is even more intimate than that 
between Shannon's information theory and classical computers. Error correction 
does not in itself guarantee accurate quantum computation, since it cannot 
combat all types of noise, but the fact that it is possible at all is a 
significant development. 

A computer which only exists on paper will not actually perform
any computations, and in the end the only way to resolve the issue of
feasibility in quantum computer science is to build a quantum computer.
To this end, a number of authors proposed computer designs based on
Deutsch's idea, but with the physical details more fully worked out
(Teich {\em et. al.} 1988, Lloyd 1993, Berman {\em et. al.} 1994,
DiVincenco 1995b).
The great challenge is to find a sufficiently complex system whose
evolution is nevertheless both coherent (i.e. unitary) and controlable. 
It is not sufficient that only some aspects of a system should be
quantum mechanical, as in solid-state `quantum dots', or that there is an 
implicit assumption of unfeasible precision or cooling, which is often the
case for proposals using solid-state devices. Cirac and Zoller (1995)
proposed the use of a linear ion trap, which was a significant improvement in 
feasibility, since heroic efforts in the ion trapping community had already 
achieved the necessary precision and low temperature in experimental work, 
especially the group of Wineland who demonstrated cooling to the ground state 
of an ion trap in the same year (Diedrich {\em et. al.} 1989,
Monroe {\em et. al.} 1995). More recently, Gershenfeld
and Chuang (1997) and Cory {\em et. al.} (1996,1997) have
shown that nuclear magnetic resonance (NMR) techniques can be adapted
to fulfill the requirements of quantum computation, making this approach
also very promising. Other recent proposals of
Privman {\em et. al.} (1997) and Loss and DiVincenzo (1997) may also
be feasible.

As things stand, no quantum computer has been built, nor looks likely to be 
built in the author's lifetime, if we measure it in terms of Shor's algorithm, 
and ask for factoring of large numbers. However, if we ask instead for
a device in which quantum information ideas can be explored, then
only a few quantum bits are required, and this will certainly be achieved
in the near future. Simple two-bit operations have been carried out
in many physics experiments, notably magnetic resonance, and work
with three to ten qubits now seems feasible. Notable recent experiments
in this regard are those of Brune {\em et. al.} (1994),
Monroe {\em et. al.} (1995b), Turchette {\em et. al.} (1995)
and Mattle {\em et. al.} (1996).

\newpage

\section{Classical information theory}  \lab{s:cit}

This and the next section will summarise the classical theory of information 
and computing. This is textbook material (Minsky 1967, Hamming 1986) but is 
included here since it forms a background to quantum information and 
computing, and the article is aimed at physicists to whom the ideas may be 
new. 

\subsection{Measures of information}  \lab{s:mi}

The most basic problem in classical information theory is to obtain
a measure of information, that is, of amount of information. Suppose
I tell you the value of a number $X$. How much information have you
gained? That will depend on what you already knew about $X$. For example,
if you already knew $X$ was equal to $2$, you would learn nothing,
no information, from my revelation. On the other hand, if previously
your only knowledge was that $X$ was given by the throw of a die,
then to learn its value is to gain information. We have met here
a basic paradoxical property, which is that {\em information}
is often a measure of {\em ignorance}: the information content 
(or `self-information') of $X$
is defined to be the information you would gain if you learned the
value of $X$.

If $X$ is a random variable which has value $x$ with probability
$p(x)$, then the information content of $X$ is defined to be
  \beq
  S( \{p(x) \} ) = - \sum_x p(x) \log_2 p(x).    \lab{S}
  \eeq
Note that the logarithm is taken to base 2, and that $S$ is always positive 
since probabilities are bounded by $p(x) \le 1$. $S$ is a function of the 
{\em probability distribition} of values of $X$. It is important to remember 
this, since in what follows we will adopt the standard practice of using the 
notation $S(X)$ for $S( \{ p(x) \})$. It is understood that $S(X)$ does 
not mean a function of $X$, but rather the information content of the 
variable $X$. The quantity $S(X)$ is also referred to as an entropy, for 
obvious reasons. 

If we already know that $X=2$, then $p(2)=1$ and
there are no other terms in the sum, leading to $S=0$, so $X$
has no information content. If, on the other hand, $X$ is given
by the throw of a die, then $p(x)=1/6$ for $x \in \{1,2,3,4,5,6\}$
so $S = -\log_2(1/6) \simeq 2.58$. If $X$ can take $N$
different values, then the information content (or entropy) of $X$
is maximised when the probability distribution $p$ is flat,
with every $p(x) = 1/N$ (for example a fair die yields $S \simeq 2.58$,
but a loaded die with $p(6)=1/2, p(1 \cdots 5)=1/10$ yields $S \simeq
2.16$). This is consistent with the requirement that the information
(what we would gain if we learned $X$) is maximum when our prior
knowledge of $X$ is minimum.

Thus the maximum information which could in principle be
stored by a variable which can take on $N$ different values is
$\log_2(N)$. The logarithms are taken
to base 2 rather than some other base by convention. The choice
dictates the unit of information: $S(X) = 1$ when $X$ can take two values
with equal probability. A two-valued or binary variable thus can
contain one unit of information. This unit is called a {\em bit}.
The two values of a bit are typically written as the binary
digits 0 and 1.

In the case of a binary variable, we can define $p$ to be the
probability that $X=1$, then the probability that $X=0$ is $1-p$
and the information can be written as a function of $p$ alone:
\beq
H(p) = -p \log_2 p - (1-p) \log_2 (1-p)   \lab{H}
\eeq
This function is called the {\em entropy function}, $0 \le H(p) \le 1$.

In what follows, the subscript 2 will be dropped
on logarithms, it is assumed that all logarithms are
to base 2 unless otherwise indicated.

The probability that $Y=y$ given that $X=x$ is written $p(y | x)$.
The {\em conditional entropy} $S(Y | X)$ is defined by
  \begin{eqnarray}
S(Y | X) &=& -\sum_x p(x) \sum_y p(y | x) \log p(y | x) \lab{SYgX} \\
	 &=& -\sum_x \sum_y p(x,y) \log p(y | x)
  \end{eqnarray}
where the second line is deduced using $p(x,y) = p(x) p(y | x)$
(this is the probability that $X=x$ {\em and} $Y=y$). 
By inspection of the definition, we see that $S(Y | X)$
is a measure of how much information on average would remain
in $Y$ if we were to learn $X$. Note that $S(Y | X) \le S(Y)$
always and $S(Y | X) \ne S(X | Y)$ usually.

The conditional entropy is important mainly as a stepping-stone
to the next quantity, the {\em mutual information}, defined by
  \begin{eqnarray}
I(X: Y) &=& \sum_x \sum_y p(x,y) \log \frac{p(x,y)}{p(x) p(y)} \\
	&=& S(X) - S(X | Y)                    \lab{I}
  \end{eqnarray}
From the definition, $I(X:Y)$ is a measure of how much $X$ and
$Y$ contain information about each other\footnote{Many authors
write $I(X;Y)$ rather than $I(X:Y)$. I prefer the latter since
the symmetry of the colon reflects the fact that $I(X:Y) = I(Y:X)$.}.
If $X$ and $Y$ are independent then $p(x,y) = p(x) p(y)$ so $I(X:Y) = 0$. 
The relationships between the basic measures of information are
indicated in fig. 3. The reader may like to prove as an exercise 
that $S(X,Y)$, the information content of $X$ and $Y$ (the information we 
would gain if, initially knowing neither, we learned the value of both $X$ 
and $Y$) satisfies
$S(X,Y) = S(X) + S(Y) - I(X:Y).$

Information can disappear, but it cannot spring spontaneously from nowhere.
This important fact finds mathematical expression in the {\em data
processing inequality}:
  \beq
\mbox{if} \;\; X \rightarrow Y \rightarrow Z\;\;\; \mbox{then}
\;\;\; I(X:Z) \le I(X:Y).                         \lab{data}
  \eeq
The symbol $X \rightarrow Y \rightarrow Z$ means that $X, Y$ and $Z$
form a process (a Markov chain) in which $Z$ depends on $Y$ but not directly
on $X$: $p(x,y,z) = p(x) p(y | x) p (z | y)$. 
The content of the data processing inequality is that
the `data processor' $Y$ can pass on to $Z$ no more information 
about $X$ than it received.

\subsection{Data compression}   \lab{s:dc}

Having pulled the definition of information content, equation \eq{S},
out of a hat, our aim is now to prove that this is a good measure
of information. It is not obvious at first sight even how to think
about such a task. One of the main contributions of classical
information theory is to provide useful ways to think about
information. We will describe a simple situation in order to illustrate
the methods. Let us suppose one person,
traditionally called Alice, knows the value of $X$, and she
wishes to communicate it to Bob. We
restrict ourselves to the simple case that $X$ has
only two possible values: either `yes' or `no'. We say
that Alice is a `source' with an `alphabet' of two symbols. 
Alice communicates by sending binary digits (noughts and ones) to
Bob. We will measure the information content of $X$ by counting
how many bits Alice must send, {\em on average}, to allow Bob
to learn $X$. Obviously, she could just send 0 for `no' and 1 for
`yes', giving a `bit rate' of one bit per $X$ value communicated. However, 
what if $X$ were an essentially random variable, except that it is more 
likely to be `no' than `yes'? (think of the output of decisions from a grant 
funding body, for example). In this case, Alice can communicate more 
efficiently by adopting the following procedure. 

Let $p$ be the probability that $X=1$ and $1-p$ be the probability that $X=0$. 
Alice waits until $n$ values of $X$ are available to be sent, where $n$ will be 
large. The mean number of ones in such a sequence of $n$ values is $np$, and
it is likely that the number of ones in any given sequence is close to this
mean. Suppose $np$ is an integer, then the probability of obtaining any
given sequence containing $np$ ones is
  \beq
p^{np} (1-p)^{n-np} = 2^{-n H(p)}.
  \eeq
The reader should satisfy him or herself that the two sides of this equation
are indeed equal: the right hand side hints at how the argument can be 
generalised. Such a sequence is called a {\em typical sequence}. To be 
specific, we define the set of typical sequences to be all sequences such that 
  \beq
2^{-n( H(p) + \epsilon)} \le p(\mbox{sequence}) \le
2^{-n( H(p) - \epsilon)}
  \eeq
Now, it can be shown that the probability that Alice's $n$ values
actually form a typical sequence is greater than $1-\epsilon$,
for sufficiently large $n$, no matter how small $\epsilon$ is. 
This implies that Alice
need not communicate $n$ bits to Bob in order for him to learn $n$
decisions. She need only tell Bob {\em which typical sequence} she
has. They must agree together beforehand how the typical sequences
are to be labelled: for example, they may agree to number them in order
of increasing binary value. Alice just sends the label,
not the sequence itself. To deduce how well this works, it can
be shown that the typical sequences all have equal probability, and there
are $2^{n H(p)}$ of them. To communicate one of $2^{n H(p)}$ 
possibilities, clealy Alice must send $n H(p)$ bits. Also, Alice
cannot do better than this (i.e. send fewer bits) since the
typical sequences are equiprobable: there is nothing to be
gained by further manipulating the information. Therefore,
the information content of each value of $X$ in the original
sequence must be $H(p)$, which proves \eq{S}.

The mathematical details skipped over in the above argument
all stem from the law of large numbers, which states that,
given arbitrarily small $\epsilon$, $\delta$
  \beq
P\left( \left| m - n p \right| < n\epsilon \right) 
> 1 - \delta
  \eeq
for sufficiently large $n$, where $m$ is the number of ones obtained in a 
sequence of $n$ values. For large enough $n$, the number of ones $m$ will 
differ from the mean $np$ by an amount arbitrarily small compared to $n$. 
For example, in our case the noughts and ones will be distributed according to 
the binomial distribution 
  \begin{eqnarray}
P(n,m) &=& C(n,m) p^m (1-p)^{n-m}        \lab{binom}   \\
  &\simeq& \frac{1}{\sigma \sqrt{2 \pi}} e^{-(m-np)^2 / 2 \sigma^2} 
  \end{eqnarray}
where the Gaussian form is obtained in the limit $n, np \rightarrow \infty$,
with the standard deviation $\sigma = \sqrt{np(1-p)}$, and
$C(n,m) = n!/m!(n-m)!$. 

The above argument has already yielded a significant practical result
associated with \eq{S}. This is that to communicate $n$ values of
$X$, we need only send $n S(X) \le n$ bits down a communication channel. This 
idea is referred to as {\em data compression}, and is also called {\em 
Shannon's noiseless coding theorem}.

The typical sequences idea has given a means to calculate information
content, but it is not the best way to compress information in practice,
because Alice must wait for a large number of decisions to accumulate
before she communicates anything to Bob. A better method
is for Alice to accumulate a few decisions, say 4, and communicate this
as a single `message' as best she can. Huffman derived an optimal
method whereby Alice sends short strings to communicate the most likely
messages, and longer ones to communicate the least likely messages,
see table 1 for an example. The translation process is referred
to as `encoding' and `decoding' (fig. 4); this terminology
does not imply any wish to keep information secret.

For the case $p=1/4$ Shannon's noiseless coding theorem tells us
that the best possible data compression technique would
communicate each message of four $X$ values by sending on average
$4 H(1/4) \simeq 3.245$ bits. The Huffman code in table 1 gives
on average 3.273 bits per message. This is quite close
to the minimum, showing that practical methods like Huffman's
are powerful.

Data compression is a concept of great practical importance. It is used in 
telecommunications, for example to compress the information required to 
convey television pictures, and data storage in computers. From the point of 
view of an engineer designing a communication channel, data compression can 
appear miraculous. Suppose we have set up a telephone link to a mountainous 
area, but the communication rate is not high enough to send, say, the pixels 
of a live video image. The old-style engineering option would be to replace 
the telephone link with a faster one, but information theory suggests 
instead the possibility of using the same link, but adding data processing 
at either end (data compression and decompression). It comes as a great 
surprise that the usefulness of a cable can thus be improved by tinkering 
with the information instead of the cable. 

\subsection{The binary symmetric channel}  \lab{s:bin}

So far we have considered the case of communication down a perfect, i.e. 
noise-free channel. We have gained two main results of practical value: a 
measure of the best possible data compression (Shannon's noiseless coding 
theorem), and a practical method to compress data (Huffman coding). We now 
turn to the important question of communication in the presence of noise. 
As in the last section, we will analyse the simplest case in order
to illustrate principles which are in fact more general.

Suppose we have a binary channel, i.e. one which allows Alice to
send noughts and ones to Bob. The noise-free channel conveys
$0 \rightarrow 0$ and $1 \rightarrow 1$, but a noisy channel
might sometimes cause 0 to become 1 and vice versa. There is
an infinite variety of different types of noise. For example,
the erroneous `bit flip' $0 \rightarrow 1$ might be
just as likely as $1 \rightarrow 0$, or the channel might
have a tendency to `relax' towards 0, in which case
$1 \rightarrow 0$ happens but $0 \rightarrow 1$ does not.
Also, such errors might occur independently from bit to
bit, or occur in bursts. 

A very important type of noise is one which affects 
different bits independently, and causes
both $0 \rightarrow 1$ and $1 \rightarrow 
0$ errors. This is important because it captures the essential features
of many processes encountered in realistic situations. If the
two errors $0 \rightarrow 1$ and $1 \rightarrow 0$ are equally likely,
then the noisy channel is called a `binary symmetric channel'.
The binary symmetric channel has a single parameter, $p$,
which is the error probability per bit sent.
Suppose the message sent into the channel by Alice is $X$,
and the noisy message which Bob receives is $Y$. Bob is then faced
with the task of deducing $X$ as best he can from $Y$. If $X$
consists of a single bit, then Bob will
make use of the conditional probabilities
  \begin{eqnarray*}
p(x=0 | y=0) = p(x=1 | y=1) = 1-p \\
p(x=0 | y=1) = p(x=1 | y=0) = p 
  \end{eqnarray*}
giving $S(X | Y) = H(p)$ using equations (\ref{SYgX}) and (\ref{H}). 
Therefore, from the definition \eq{I} of mutual information, we have
  \beq
I(X:Y) = S(X) - H(p)   \lab{Ip}
  \eeq
Clearly, the presence of noise in the channel limits the information
about Alice's $X$ contained in Bob's received $Y$. Also, because
of the data processing inequality, equation (\ref{data}), Bob cannot increase 
his information about $X$ by manipulating $Y$. However, \eq{Ip} shows that 
Alice and Bob can communicate better if $S(X)$ is large. The general
insight is that the information communicated depends both on the
source and the properties of the channel. It would be useful
to have a measure of the channel alone, to tell us how well it
conveys information. This quantity is called the {\em capacity}
of the channel and it is defined to be the maximum possible mutual
information $I(X:Y)$ between the input and output of the channel,
maximised over all possible sources:
  \beq
\mbox{Channel capacity}\;\; C \equiv \max_{\{p(x)\}} I(X:Y)
\lab{C}
  \eeq
Channel capacity is measured in units of `bits out per symbol in'
and for binary channels must lie between zero and one. 

It is all very well to have a definition, but \eq{C} does not allow us to 
compare channels very easily, since we have to perform the maximisation over 
input strategies, which is non-trivial. To establish
the capacity $C(p)$ of the binary symmetric channel
is a basic problem in information theory, but fortunately this
case is quite simple. From 
equations (\ref{Ip}) and (\ref{C}) one may see that the answer is
  \beq
C(p) = 1 - H(p), 
  \eeq
obtained when $S(X) = 1$ (i.e. $P(x=0) = P(x=1) = 1/2$).

\subsection{Error-correcting codes}   \lab{s:ecc}

So far we have investigated how much information gets through a noisy channel, 
and how much is lost. Alice cannot convey to Bob more information than $C(p)$ 
per symbol communicated. However, suppose Bob is busy defusing a bomb and Alice 
is shouting from a distance which wire to cut : she will not say
``the blue wire'' just once, and hope that Bob heard correctly. She will repeat
the message many times, and Bob will wait until he
is sure to have got it right. Thus error-free communication can be
achieved even over a noisy channel. In this example one obtains the benefit of
reduced error rate at the sacrifice of reduced information rate. The next
stage of our information theoretic programme is to identify more powerful
techniques to circumvent noise (Hamming 1986, Hill 1986, Jones 1979,
MacWilliams and Sloane 1977). 

We will need the following concepts. The set $\{0, 1\}$ is considered
as a group (a Galois field GF(2)) where the operations $+,-,\times,\div$
are carried out modulo 2 (thus, $1+1=0$). An $n$-bit binary word
is a vector of $n$ components, for example 011 is the vector $(0,1,1)$.
A set of such vectors forms a vector space under addition,
since for example $011 + 101$ means $(0,1,1)+(1,0,1) = (0+1,1+0,1+1)
= (1,1,0) = 110$ by the standard rules of vector addition. This 
is equivalent to the exclusive-or operation carried out 
bitwise between the two binary words. 

The effect of noise on a word $u$ can be expressed $u \rightarrow
u' = u + e$, where the error vector $e$ indicates which bits in $u$
were flipped by the noise. For example, $u = 1001101 \rightarrow
u' = 1101110$ can be expressed $u' = u + 0100011$. 
An error correcting code $C$ is a set of words such that
  \beq
u + e \! \ne \! v + f  \;\; \forall u,v \in C\; (u\ne v),
\;\; \forall e,f \in E \lab{code}
  \eeq
where $E$ is the set of errors correctable by $C$, which includes the
case of no error, $e=0$. To use such a code, Alice and Bob agree
on which codeword $u$ corresponds to which message, and Alice
only ever sends codewords down the channel. Since the channel
is noisy, Bob receives not $u$ but $u + e$. However, Bob can deduce
$u$ unambiguously from $u+e$ since by condition
(\ref{code}), no other codeword $v$ sent by Alice
could have caused Bob to receive $u+e$.

An example error-correcting code is shown in the right-hand column of table 1. 
This is a $[7,4,3]$ Hamming code, named after its discoverer. The notation 
$[n,k,d]$ means that the codewords are $n$ bits long, there are $2^k$ of them, 
and they all differ from each other in at least $d$ places. Because of the 
latter feature, the condition (\ref{code}) is satisfied for any error which 
affects at most one bit. In other words the set $E$ of correctable errors is 
$\{0000000$,$1000000$,$0100000$,$0010000$, $0001000$,$0000100$,$0000010$,
$0000001\}$. Note 
that $E$ can have at most $2^{n-k}$ members. The 
ratio $k/n$ is called the {\em rate} of the code, since each block of $n$ 
transmitted bits conveys $k$ bits of information, thus $k/n$ bits
per bit. 

The parameter $d$ is called the `minimum distance' of the code, and is 
important when encoding for noise which affects successive bits independently, 
as in the binary symmetric channel. For, a code of minumum distance $d$ can 
correct all errors affecting less than $d/2$ bits of the transmitted codeword,
and for independent noise this is the {\em most likely} set of errors.
In fact, the probability that an $n$-bit word receives $m$ errors
is given by the binomial distribution \eq{binom}, so if the code can
correct more than the mean number of errors $np$, the correction is
highly likely to succeed.

The central result of classical information theory is that powerful
error correcting codes exist:

\begin{quote}
Shannon's theorem: If the rate $k/n < C(p)$
and $n$ is sufficiently large, there exists a binary
code allowing transmission with an arbitrarily small error
probability.
\end{quote}

The error probability here is the probability that an uncorrectable
error occurs, causing Bob to misinterpret the received word.
Shannon's theorem is highly surprising, since it implies that
it is not necessary to engineer very low-noise communication channels,
an expensive and difficult task. Instead, we can compensate noise
by error correction coding and decoding, that is, by information processing!
The meaning of Shannon's theorem is illustrated by fig. 5.

The main problem of coding theory is to identify codes with
large rate $k/n$ and large distance $d$. These two conditions are
mutually incompatible, so a compromise is needed. The problem
is notoriously difficult and has no
general solution. To make connection with
quantum error correction, we will need to mention one important concept,
that of the {\em parity check matrix}. An error correcting code
is called linear if it is closed under addition, i.e.  $u + v
\in C \; \forall u,v \in C$. Such a code is completely specified by
its parity check matrix $H$, which is a set of $(n-k)$ 
linearly independent $n$-bit words
satisfying $H \cdot u = 0 \; \forall u \in C$. The important property
is encapsulated by the following equation:
  \beq
H \cdot (u + e) = (H \cdot u) + (H \cdot e) = H \cdot e.  \lab{syn}
  \eeq
This states that if Bob evaluates $H \cdot u'$ for his noisy received word $u' 
= u+e$, he will obtain the same answer $H \cdot e$, no matter what word $u$ 
Alice sent him! If this evaluation were done automatically, Bob could learn $H 
\cdot e$, called the {\em error syndrome}, without learning $u$. If Bob can 
deduce the error $e$ from $H \cdot e$, which one can show is possible for all 
correctable errors, then he can correct the message (by subtracting $e$ from 
it) without ever learning what it was! In quantum error correction, this is the 
origin of the reason one can correct a quantum state without
disturbing it.

\section{Classical theory of computation}

We now turn to the theory of computation. This is mostly concerned with
the questions ``what is computable?'' and ``what resources are necessary?''

The fundamental resources required for computing are a means to store
and to manipulate symbols. The important questions are such things as
how complicated must the symbols be, how many will we need, how
complicated must the manipulations be, and how many of them will we need?

The general insight is that computation is deemed {\em hard} or inefficient
if the amount of resources required rises exponentially with a measure
of the size of the problem to be addressed. The size of the problem
is given by the amount of {\em information} required to specify the
problem. Applying this idea at the most basic level, we find that
a computer must be able to manipulate binary symbols, not just
unary symbols\footnote{Unary notation has a single symbol, 1.
The positive integers are written 1,11,111,1111,\ldots}, otherwise the number 
of memory locations needed would grow exponentially with the amount of 
information to be manipulated.
On the other hand, it is not necessary to work in decimal notation
(10 symbols) or any other notation with an `alphabet' of more
than two symbols. This greatly simplifies computer design and analysis.

To manipulate $n$ binary symbols, it is not necessary to manipulate
them all at once, since it can be shown that any transformation can be brought 
about by manipulating the binary symbols one at a time or in pairs. A binary 
`logic gate' takes two bits $x,y$ as inputs, and calculates a function 
$f(x,y)$. Since $f$ can be 0 or 1, and there are four possible inputs, there 
are 16 possible functions $f$. This set of 16 different logic gates is called a 
`universal set', since by combining such gates in series, any transformation of 
$n$ bits can be carried out. Futhermore, the action of some of the 16 gates 
can be reproduced by combining others, so we do not need all 16, and in fact 
only one, the {\sc nand} gate, is necessary ({\sc nand} is {\sc not and}, for 
which the output is 0 if and only if both inputs are 1). 

By concatenating logic gates, we can manipulate $n$-bit
symbols (see fig. 6). This general approach is called the network 
model of computation, and is useful for our purposes because it suggests 
the model of quantum computation which is currently most feasible 
experimentally. In this model, the essential components of a computer are a 
set of bits, many copies of the universal logic gate, and connecting wires. 

\subsection{Universal computer; Turing machine}  \lab{s:UTM}

The word `universal' has a further significance in relation to
computers. Turing showed that
it is possible to construct a {\em universal} computer,
which can simulate the action of any other, in the following sense.
Let us write $T(x)$ for the output
of a Turing machine $T$ (fig. 7) acting on input tape $x$.
Now, a Turing machine can be completely specified by
writing down how it responds to 0 and 1 on the input tape,
for every possible internal configuration of the machine
(of which there are a finite number). This specification
can itself be written as a binary number $d[T]$. Turing showed
that there exists a machine $U$, called a universal Turing
machine, with the properties
  \beq
U(d[T],x) = T(x)
  \eeq
and the number of steps taken by $U$ to simulate each step
of $T$ is only a polynomial (not exponential) function of
the length of $d[T]$. In other words,
if we provide $U$ with an input tape containing both a description
of $T$ and the input $x$, then $U$ will compute the same function
as $T$ would have done, for {\em any} machine $T$, without an
exponential slow-down. 

To complete the argument, it can be
shown that other models of computation, such as the network
model, are {\em computationally equivalent} to the Turing model:
they permit the same functions to be computed, with the
same computational efficiency (see next section).
Thus the concept of the univeral machine establishes that a certain
finite degree of complexity of construction is sufficient to allow
very general information processing. This is the fundamental
result of computer science. Indeed, the power of the Turing machine
and its cousins is so great that Church (1936) and Turing (1936) framed 
the ``Church-Turing thesis,'' to the effect that

{\em Every function `which would naturally be regarded as computable'
can be computed by the universal Turing machine}.

This thesis is unproven, but has survived many attempts to find
a counterexample, making it a very powerful result.
To it we owe the versatility of the modern general-purpose computer,
since `computable functions' include tasks such as word processing,
process control, and so on. The quantum computer, to be described
in section \ref{s:uqc} will
throw new light on this central thesis.

\subsection{Computational complexity}   \lab{s:cc}

Once we have established the idea of a universal computer, computational 
tasks can be classified in terms of their difficulty
in the following manner.  
A given algorithm is deemed to address not
just one instance of a problem, such as ``find the square of 237,''
but one class of problem, such as ``given $x$, find its square.''
The amount of information given to the computer in order to 
specify the problem is $L = \log x$, i.e. the number of bits needed
to store the value of $x$. The {\em computational complexity}
of the problem is determined by the number of steps $s$ a Turing
machine must make in order to complete any algorithmic method to
solve the problem. In the network model, the complexity is determined by
the number of logic gates required. If an algorithm exists with
$s$ given by any polynomial function of $L$ (eg $s \propto L^3 + L$)
then the problem is deemed tractable
and is placed in the complexity class ``{\sc p}''. If $s$ rises
exponentially with $l$ (eg $s \propto 2^L = x$) then the
problem is hard and is in another complexity class. It is often easier
to verify a solution, that is, to test whether or not it is correct,
than to find one. The class ``{\sc np}'' is the set of problems for
which solutions can be verified in polynomial time. Obviously
{\sc p} $\in$ {\sc np}, and one would guess that
there are problems in {\sc np} which are not in {\sc p},
(i.e. {\sc np} $\ne$ {\sc p}) though
surprisingly the latter has never been proved, since it is very
hard to rule out the possible existence of as yet undiscovered algorithms.
However, the important point is that the membership of these classes
does not depend on the model of computation, i.e. the physical
realisation of the computer, since the Turing machine can
simulate any other computer with only a polynomial, rather than
exponential slow-down.

An important example of an intractable problem is that of factorisation: 
given a composite (i.e. non-prime) number $x$, the task is to find one of 
its factors. If $x$ is even, or a multiple of any small number, then it is 
easy to find a factor. The interesting case is when the prime factors of 
$x$ are all themselves large. In this case there is no known simple
method. The best known method, the {\em number field sieve}
(Menezes {\em et. al.} 1997) requires a number of computational steps
of order $s \sim \exp( 2 L^{1/3} (\log L)^{2/3} )$ where $L = \ln x$.
By devoting a substantial machine network to this task, one 
can today factor a number of 130 decimal digits (Crandall 1997), i.e. $L \simeq 
300$, giving $s \sim 10^{18}$. This is time-consuming but possible (for example 
42 days at $10^{12}$ operations per second). However, if we double $L$, $s$ 
increases to $\sim 10^{25}$, so now the problem is intractable: it would take a 
million years with current technology, or would require computers running a 
million times faster than current ones. 
The lesson is an important one: a computationally `hard' problem is one which 
in practice is not merely difficult but impossible to solve. 

The factorisation problem has acquired great practical importance
because it is at the heart of widely used cyptographic systems such as
that of Rivest, Shamir and Adleman (1979) (see Hellman 1979).
For, given a message $M$ (in the form of a long binary number), it is
easy to calculate an encrypted version $E = M^s \;{\rm mod}\;c$ where
$s$ and $c$ are well-chosen large integers which can be made public. To 
decrypt the message, the receiver calculates $E^t \;{\rm mod}\; c$ which is 
equal to $M$ for a value of $t$ which can be quickly deduced from $s$ and the 
factors of $c$ (Schroeder 1984). In practice $c=pq$ is chosen to be the product 
of two large primes $p,q$ known only to the user who published $c$, so only 
that user can read the messages---unless someone manages to factorise $c$. 
It is a very useful feature that no secret keys need be distributed in
such a system: the `key' $c,s$ allowing encryption is public knowledge.

\subsection{Uncomputable functions}  \label{s:halting}

There is an even stronger way in which a task may be impossible for a 
computer. In the quest to solve some problem, we could `live 
with' a slow algorithm, but what if one does not exist at all? Such 
problems are termed {\em uncomputable}. The most important example is
the ``halting problem'', a rather beautiful result.
A feature of computers familiar to programmers is that they
may sometimes be thrown into a never-ending loop. Consider,
for example, the instruction ``while $x > 2$,
divide $x$ by 1'' for $x$ initially greater than 2.
We can see that this algorithm will never halt,
without actually running it.
More interesting from a mathematical point of view is an
algorithm such as ``while $x$ is equal to the sum of
two primes, add 2 to $x$, otherwise print $x$ and halt'',
beginning at $x=8$. The algorithm is certainly feasible since all pairs of 
primes less than $x$ can be found and added systematically. Will such an 
algorithm ever halt? If so, then a counterexample to the Goldbach 
conjecture exists. Using such techniques, a vast section of 
mathematical and physical theory could be reduced to the question
``would such and such an algorithm halt if we were to run it?''
If we could find a general way to
establish whether or not algorithms will halt, we would have an 
extremely powerful mathematical tool. In a certain sense, it
would solve all of mathematics!

Let us suppose that it is possible to find a general algorithm which will 
work out whether any Turing machine will halt on any input. Such an 
algorithm solves the problem ``given $x$ and $d[T]$, would Turing machine 
$T$ halt if it were fed $x$ as input?''. Here $d[T]$ is the description of 
$T$. If such an algorithm exists, then it is possible
to make a Turing machine $T_H$ which halts
if and only if $T( d[T] )$ does not halt, where $d[T]$
is the description of $T$. Here $T_H$ takes as input
$d[T]$, which is sufficient to tell $T_H$ about
both the Turing machine $T$ and the input to $T$. Hence we have
  \beq
T_H( d[T] ) \;\;\mbox{halts} \leftrightarrow T( d[T] ) \;\;
\mbox{does not halt}
  \eeq
So far everything is ok. However, what if we feed $T_H$
the description of itself, $d[T_H]$? Then
  \beq
T_H\left( d[T_H] \right) \;\;\mbox{halts} \leftrightarrow
T_H\left( d[T_H] \right) \;\; \mbox{does not halt}
  \eeq
which is a contradiction. By this argument Turing showed that 
there is no automatic means to establish whether Turing machines
will halt in general: the ``halting problem'' is uncomputable.
This implies that mathematics, and information
processing in general, is a rich body of different ideas
which cannot all be summarised in one grand
algorithm. This liberating observation is closely related to
G\"odel's theorem.

\section{Quantum verses classical physics} \label{s:qvc}

In order to think about quantum information theory, let us first
state the principles of non-relativisitic quantum mechanics,
as follows (Shankar 1980).

\begin{enumerate}
\item The state of an isolated system $\cal Q$ is represented by a vector
$\ket{\psi(t)}$ in a Hilbert space.
\item Variables such as position and momentum are termed
observables and are represented by Hermitian operators.
The position and momentum operators $X,P$ have the following
matrix elements in the eigenbasis of $X$:
  \begin{eqnarray*}
    \bra{x} X \ket{x'} &=& x \delta (x-x') \\
    \bra{x} P \ket{x'} &=& -i \hbar \delta' (x-x')
  \end{eqnarray*}
\item The state vector obeys the Schr\"odinger equation
\beq
i \hbar \frac{d}{dt} \ket{\psi(t)} = {\cal H} \ket{\psi(t)}  \label{Sch}
\eeq
where ${\cal H}$ is the quantum Hamiltonian operator.
\item Measurement postulate.
  \end{enumerate}

The fourth postulate, which has not been made explicit, is a subject of some 
debate, since quite different interpretive approaches lead to the same 
predictions, and the concept of `measurement' is fraught with ambiguities in 
quantum mechanics (Wheeler and Zurek 1983, Bell 1987, Peres 1993).
A statement which is valid for most practical purposes is 
that certain physical interactions are recognisably `measurements',
and their effect on the state vector $\ket{\psi}$ is to change
it to an eigenstate $\ket{k}$ of the variable being measured, 
the value of $k$ being randomly chosen with 
probability $P \propto |\left< k \right. \ket{\psi}|^2$. The
change $\ket{\psi} \rightarrow \ket{k}$ can be expressed by the
projection operator $(\ket{k}\bra{k})/\left< k \right. \ket{\psi}$.

Note that according to the above equations, the evolution of an isolated 
quantum system is always {\em unitary}, in other words $\ket{\psi(t)} = U(t) 
\ket{\psi(0)}$ where $U(t) = \exp(-i \int {\cal H} dt / \hbar)$ is a unitary 
operator, $U U^{\dagger} = I$. This is true, but there is a difficulty that 
there is no such thing as a truly isolated system (i.e. one which experiences 
no interactions with any other systems), except possibly the whole universe. 
Therefore there is always some approximation involved in using the 
Schr\"odinger equation to describe real systems. 

One way to handle this approximation is to speak of the system $\cal Q$ and 
its environment $\cal T$. The evolution of $\cal Q$ is primarily that given 
by its Schr\"odinger equation, but the interaction between $\cal Q$ and 
$\cal T$ has, in part, the character of a measurement of $\cal Q$. 
This produces a non-unitary contribution to the evolution of $\cal Q$
(since projections are not unitary), and this ubiquitous phenomenon
is called {\em decoherence}. I have underlined these elementary ideas
because they are central in what follows.

We can now begin to bring together ideas of physics and of
information processing. For, it is clear that much of the wonderful
behaviour we see around us in Nature could be understood as a form
of information processing, and conversely our computers are able
to simulate, by their processing, many of the patterns of Nature.
The obvious, if somewhat imprecise, questions are 

\begin{enumerate}
\item ``can Nature
usefully be regarded as essentially an information processor?''
\item
``could a computer simulate the whole of Nature?''
\end{enumerate}

The principles of quantum mechanics suggest that the answer to the first 
quesion is {\em yes}\footnote{This does not necessarily imply that such 
language captures everthing that can be said about Nature, merely that this 
is a useful abstraction at the descriptive level of physics. I
do not believe any physical `laws' could be adequate to 
completely describe human behaviour, for example, since they are 
sufficiently approximate or non-prescriptive to leave us room for 
manoeuvre (Polkinghorne 1994).}. For, the state vector $\ket{\psi}$ so 
central to quantum mechanics is a concept very much like those of 
information science: it is an abstract entity which contains exactly all 
the information about the system $\cal Q$. The word `exactly' here is a 
reminder that not only is $\ket{\psi}$ a complete description of $Q$, it 
is also one that does not contain any extraneous information which can not 
meaningfully be associated with $\cal Q$. The importance of this in 
quantum statistics of Fermi and Bose gases was mentioned in the 
introduction. 

The second question can be made more precise by converting
the Church-Turing thesis into a principle of physics,

{\em Every finitely realizible physical system can be 
simulated arbitrarily closely
by a universal model computing machine operating by
finite means.}

This statement is based on that of Deutsch (1985). The idea is to propose 
that a principle like this is not derived from quantum mechanics, but rather 
underpins it, like other principles such as that of conservation of energy. 
The qualifications introduced by `finitely realizible' and `finite means' 
are important in order to state something useful.

The new version of the Church-Turing thesis (now called the 
`Church-Turing Principle') does not refer to Turing machines. This is 
important because there are fundamental differences between the very nature 
of the Turing machine and the principles of quantum mechanics. One is 
described in terms of operations on classical bits, the other in terms of 
evolution of quantum states. Hence there is the possibility that the 
universal Turing machine, and hence all classical computers, might not be 
able to simulate some of the behaviour to be found in Nature. Conversely, 
it may be physically possible (i.e. not ruled out by the laws of Nature) to 
realise a new type of computation essentially different from that of 
classical computer science. This is the central aim of quantum computing.

\subsection{EPR paradox, Bell's inequality}  \lab{s:EPR}

In 1935 Einstein, Podolski and Rosen (EPR) drew attention to an important 
feature of non-relativistic quantum mechanics. Their argument, and Bell's 
analysis, can now be recognised as one of the seeds from which quantum 
information theory has grown. The EPR paradox should be familiar to any 
physics graduate, and I will not repeat the argument in detail. However, the 
main points will provide a useful way in to quantum information concepts. 

The EPR thought-experiment can be reduced in essence to an experiment 
involving pairs of two-state quantum systems (Bohm 1951, Bohm and Aharonov
1957). Let us consider a 
pair of spin-half particles $A$ and $B$, writing the ($m_z = +1/2$)
spin `up' state  $\ket{\uparrow}$ and the ($m_z = -1/2$) spin `down' state 
$\ket{\downarrow}$. The particles are prepared initially in the singlet state 
$(\ket{\uparrow}\ket{\downarrow} - \ket{\downarrow}\ket{\uparrow})/ \sqrt{2}$, 
and they subsequently fly apart, propagating in opposite directions along the 
$y$-axis. Alice and Bob are widely separated, and they receive particle $A$ and 
$B$ respectively. EPR were concerned with whether quantum mechanics provides a 
complete description of the particles, or whether something was left out, some 
property of the spin angular momenta ${\bf s}_A,{\bf s}_B$ which quantum theory 
failed to describe. Such a property has since become known as a `hidden 
variable'. They argued that something was left out, because this experiment 
allows one to predict with certainty the result of measuring any component of 
${\bf s}_B$, without causing any disturbance of $B$. Therefore all the
components of ${\bf s}_B$ have definite values, say EPR, and the
quantum theory only provides an incomplete description. To make the 
certain prediction without disturbing $B$, one chooses any axis $\eta$ along 
which one wishes to know $B$'s angular momentum, and then measures not $B$ but 
$A$, using a Stern-Gerlach apparatus aligned along $\eta$. Since the singlet 
state carries no net angular momentum, one can be sure that the corresponding 
measurement on $B$ would yield the opposite result to the one obtained for $A$. 

The EPR paper is important because it is carefully argued, and the
fallacy is hard to unearth. The fallacy can be exposed in one
of two ways: one can say either that Alice's measurement does influence
Bob's particle, or (which I prefer) that the quantum state vector
$\ket{\phi}$ is not an intrinsic property of a quantum system, but
an expression for the information content of a quantum variable.
In a singlet state
there is mutual information between $A$ and $B$, so the information
content of $B$ changes when we learn something about $A$. So far
there is no difference from the behaviour of classical information,
so nothing surprising has occurred.

A more thorough analysis of the EPR experiment yields a big surprise. This 
was discovered by Bell (1964,1966). Suppose Alice and Bob measure 
the spin component of $A$ and $B$ along different axes $\eta_A$ and 
$\eta_B$ in the $x$-$z$ plane. Each measurement yields an answer $+$
or $-$. Quantum theory and experiment agree that the 
probability for the two measurements to yield the same result is 
$\sin^2((\phi_A - \phi_B)/2)$, where $\phi_A$ ($\phi_B$) is the angle between 
$\eta_A$ ($\eta_B$) and the $z$ axis. However, there is no way to assign 
{\em local} properties, that is properties of $A$ and $B$ independently, 
which lead to this high a correlation, in which the results are
certain to be opposite when $\phi_A = \phi_B$, certain to be
equal when $\phi_A = \phi_B + 180^{\circ}$, and also, for example, have a 
$\sin^2(60^{\circ}) = 3/4$ chance of being equal when $\phi_A - \phi_B = 
120^{\circ}$. Feynman (1982) gives a particularly clear analysis. At $\phi_A - 
\phi_B = 120^{\circ}$ the highest correlation which local hidden variables
could produce is $2/3$. 

The Bell-EPR argument allows us to identify a task which is physically 
possible, but which no classical computer could perform: when repeatedly 
given inputs $\phi_A$, $\phi_B$ at completely separated locations, respond 
quickly (i.e. too quick to allow light-speed communication between the 
locations) with yes/no responses which are perfectly correlated when 
$\phi_A = \phi_B + 180^{\circ}$, anticorrelated when $\phi_A = \phi_B$,
and more than $\sim 70\%$ correlated when $\phi_A - \phi_B = 
120^{\circ}$. 

Experimental tests of Bell's argument were carried out in the 1970's and 80's 
and the quantum theory was verified (Clauser and Shimony 1978, Aspect {\em 
et. al.} 1982; for more recent work see Aspect (1991),
Kwiat {\em et. al.} 1995 and 
references therein). This was a significant new probe into the logical 
structure of quantum mechanics. The argument can be made even stronger by 
considering a more complicated system. In 
particular, for three spins prepared in a state such as 
$(\ket{\uparrow}\ket{\uparrow} \ket{\uparrow} + 
\ket{\downarrow}\ket{\downarrow}\ket{\downarrow}) / \sqrt{2}$, Greenberger, 
Horne and Zeilinger (1989) (GHZ) showed that a single measurement along 
a horizontal 
axis for two particles, and along a vertical axis for the third, will yield 
with certainty a result which is the exact opposite of what a local 
hidden-variable theory would predict. A wider discussion and references are 
provided by Greenberger {\em et. al.} (1990), Mermin (1990).

The Bell-EPR correlations show that quantum mechanics permits
at least one simple task which is beyond the capabilities of
classical computers, and they hint at a new type of mutual information
(Schumacher and Nielsen 1996).
In order to pursue these ideas, we will need to construct a 
complete theory of quantum information.

\section{Quantum Information}

Just as in the discussion of classical information theory, quantum
information ideas are best introduced by stating them, and
then showing afterwards how they link together. 
Quantum communication is treated in a special issue of {\em J. Mod. Opt.},
volume 41 (1994); reviews and references for quantum
cryptography are given by Bennett {\em et. al.} (1992);
Hughes {\em et. al.} (1995); Phoenix and
Townsend (1995); Brassard and Crepeau (1996); Ekert (1997). Spiller (1996) reviews both communication and computing. 
\subsection{Qubits}

The elementary unit of quantum information is the {\em qubit}
(Schumacher 1995).
A single qubit can be envisaged as a two-state system such as a spin-half
or a two-level atom (see fig. 12), but when we measure quantum 
information in qubits we are really doing something more abstract: a quantum 
system is said to have $n$ qubits if it has a Hilbert space of $2^n$ 
dimensions, and so has available $2^n$ {\em mutually orthogonal} quantum states 
(recall that $n$ classical bits can represent up to $2^n$ different things). 
This definition of the qubit will be elaborated in section \ref{s:qdc}. 

We will write two orthogonal states of a single qubit as $\{ \ket{0},
\ket{1} \}$. More generally, $2^n$ mutually orthogonal
states of $n$ qubits can be written $\{ \ket{i} \}$, where $i$ is
an $n$-bit binary number. For example, for three qubits we
have
$\{ \ket{000},\ket{001}, \ket{010}, \ket{011},$ 
$\ket{100},\ket{101}, \ket{110}, \ket{111} \}$.

\subsection{Quantum gates}

Simple unitary operations on qubits are called quantum `logic gates'
(Deutsch 1985, 1989).
For example, if a qubit evolves as $\ket{0} \rightarrow \ket{0}$,
$\ket{1} \rightarrow \exp(i\omega t)\ket{1}$, then after time
$t$ we may say that the operation, or `gate'
  \beq
P(\theta) = \left( \begin{array}{cc}
1 & 0 \\
0 & e^{i \theta} \end{array} \right)
  \eeq
has been applied to the qubit, where $\theta = \omega t$. 
This can also be written $P(\theta) = \ket{0}\bra{0}
+ \exp(i\theta) \ket{1}\bra{1}$. Here are some other
elementary quantum gates:
  \begin{eqnarray}
I &\equiv& \ket{0}\bra{0} + \ket{1}\bra{1} \;\; = \mbox{identity} \\
X &\equiv& \ket{0}\bra{1} + \ket{1}\bra{0} \;\; = \mbox{\sc not}\\
Z &\equiv& P(\pi) \\
Y &\equiv& X Z \\
H &\equiv& \frac{1}{\sqrt{2}}\left[ \rule{0em}{1.3em}
\left(\ket{0} + \ket{1}\right)\bra{0} + 
\left(\ket{0} - \ket{1}\right)\bra{1} \right]
  \end{eqnarray}
these all act on a single qubit, and can be achieved by the action
of some Hamiltonian in Schr\"odinger's equation, since they
are all unitary operators\footnote{The letter $H$ is adopted
for the final gate here because its effect is
a {\em Hadamard} transformation. This is not to be confused with the
Hamiltonian ${\cal H}$.}. There are an infinite number of single-qubit
quantum gates, in contrast to classical information theory,
where only two logic gates are possible for a single bit, namely
the identity and the logical {\sc not} operation. The quantum {\sc not}
gate carries $\ket{0}$ to $\ket{1}$ and vice versa, and so
is analagous to a classical {\sc not}. This gate is also called
$X$ since it is the Pauli $\sigma_x$ operator. Note that the set
$\{ I, X, Y, Z \}$ is a group under multiplication. 

Of all the possible unitary operators acting on a pair of qubits,
an interesting subset is those which can be written
$\ket{0}\bra{0}\otimes I + \ket{1}\bra{1}\otimes U$, where $I$ is the
single-qubit identity operation, and $U$ is some other single-qubit
gate. Such a two-qubit gate is called a ``controlled $U$''
gate, since the action $I$ or $U$ on the second qubit is controlled
by whether the first qubit is in the state $\ket{0}$ or $\ket{1}$.
For example, the effect of controlled-{\sc not} (``{\sc cnot}'') is
  \begin{eqnarray}
\ket{00} &\rightarrow& \ket{00} \nonumber \\
\ket{01} &\rightarrow& \ket{01} \nonumber \\
\ket{10} &\rightarrow& \ket{11} \nonumber \\
\ket{11} &\rightarrow& \ket{10}       \label{cnot}
  \end{eqnarray}
Here the second qubit undergoes a {\sc not} if and only if the first
qubit is in the state $\ket{1}$.
This list of state changes is the analogue of the truth table
for a classical binary logic gate. 
The effect of controlled-{\sc not} acting on a state
$\ket{a}\ket{b}$ can be written
$a \rightarrow a$, $b \rightarrow a \oplus b$, where
$\oplus$ signifies the exclusive or ({\sc xor}) operation.
For this reason, this gate is also called the {\sc xor} gate.

Other logical operations require further qubits. For 
example, the {\sc and} operation is achieved by use of the 3-qubit 
``controlled-controlled-{\sc not}'' gate, in which the 
third qubit experiences {\sc not} if and only if both the others are in the 
state $\ket{1}$. This gate is named a Toffoli gate, after Toffoli (1980)
who showed that the classical version is universal for classical
reversible computation.
The effect on a state $\ket{a}\ket{b}\ket{0}$
is $a \rightarrow a, b \rightarrow b, 0 \rightarrow a \cdot b$.
In other words if the third qubit is prepared in $\ket{0}$ then
this gate computes the {\sc and} of the first two qubits. The use
of three qubits is necessary in order to permit the whole operation
to be unitary, and thus allowed in quantum mechanical evolution.

It is an amusing excercise to find the combinations of gates which perform 
elementary arithmatical operations such as binary addition and 
multiplication. Many basic constructions are given by Barenco {\em et. al.} 
(1995b), further general design considerations are discussed
by Vedral {\em et. al.} (1996) and Beckman {\em et. al.} (1996).

The action of a sequence of quantum gates can be written in
operator notation, for example $X_1 H_2 \mbox{\sc xor}_{1,3} \ket{\phi}$
where $\ket{\phi}$ is some state of three qubits, and the subscripts
on the operators indicate to which qubits they apply. However, once
more than a few quantum gates are involved, this notation is
rather obscure, and can usefully be replaced by a diagram known
as a quantum network---see fig. 8. These diagrams will be used
hereafter. 

\subsection{No cloning}

{\em No cloning theorem:} An unknown quantum state cannot be cloned.

This states that it is impossible to generate copies of a quantum state 
reliably, unless the state is already known (i.e. unless there exists 
classical information which specifies it). Proof: to generate a copy of a 
quantum state $\ket{\alpha}$, we must cause a pair of quantum systems to 
undergo the evolution $U (\ket{\alpha} \ket{0}) = \ket{\alpha} 
\ket{\alpha}$ where $U$ is the unitary evolution operator. If this is to 
work for any state, then $U$ must not depend on $\alpha$, and therefore $U 
(\ket{\beta}  \ket{0}) = \ket{\beta}  \ket{\beta}$ for $\ket{\beta} \ne 
\ket{\alpha}$. However, if we consider the state $\ket{\gamma} = 
(\ket{\alpha} + \ket{\beta})/\sqrt{2}$, we have $U (\ket{\gamma}  \ket{0}) 
= (\ket{\alpha}\ket{\alpha} + \ket{\beta}\ket{\beta})/\sqrt{2} \ne 
\ket{\gamma}\ket{\gamma}$ so the cloning operation fails. This argument 
applies to any purported cloning method (Wooters and Zurek 1982, Dieks 
1982).

Note that any given `cloning' operation $U$ can work on some states
($\ket{\alpha}$ and $\ket{\beta}$ in the above example), though
since $U$ is trace-preserving, two different clonable states must
be orthogonal, $\left< \alpha \right| \left. \beta \right> = 0$. 
Unless we already know that the state to be copied is one of these states,
we cannot guarantee that the chosen $U$ will correctly clone it. This is
in contrast to classical information, where machines like
photocopiers can easily copy whatever classical information is
sent to them. The controlled-{\sc not} or {\sc xor} operation
of equation (\ref{cnot}) is a copying operation for the states
$\ket{0}$ and $\ket{1}$, but not for states such as $\ket{+}
\equiv (\ket{0} + \ket{1}) / \sqrt{2}$ and $\ket{-}
\equiv (\ket{0} - \ket{1}) / \sqrt{2}$.

The no-cloning theorem and the EPR paradox together reveal a rather subtle 
way in which non-relativistic quantum mechanics is a consistent theory. 
For, if cloning were possible, then EPR correlations could be used to 
communicate faster than light, which leads to a contradiction (an effect 
preceding a cause) once the principles of special relativity are taken into 
account. To see this, observe that by generating many clones, and then 
measuring them in different bases, Bob could deduce unambiguously whether 
his member of an EPR pair is in a state of the basis $\{\ket{0}, \ket{1}\}$ 
or of the basis $\{\ket{+},\ket{-}\}$. Alice would communicate 
instanteously by forcing the EPR pair into one basis or the other through 
her choice of measurement axis (Glauber 1986). 

\subsection{Dense coding}

We will discuss the following statement:

{\em Quantum entanglement is an information resource.}

Qubits can be used to store and transmit classical information.
To transmit a classical bit string 00101, for example, Alice can send
5 qubits prepared in the state $\ket{00101}$. The receiver Bob
can extract the information by measuring each qubit in the basis
$\{ \ket{0}, \ket{1} \}$ (i.e. these are the eigenstates of the measured
observable). The measurement results yield the classical bit string
with no ambiguity. No more than one classical
bit can be communicated for each qubit sent.

Suppose now that Alice and Bob are in possession of an entangled pair of 
qubits, in the state $\ket{00} + \ket{11}$ (we will usually drop
normalisation factors such as $\sqrt{2}$ from now on, to keep
the notation uncluttered). Alice and Bob need never have 
communicated: we imagine a mechanical central facility generating entangled 
pairs and sending one qubit to each of Alice and Bob, who store them
(see fig. 9a). In this 
situation, Alice can communicate {\em two} classical bits by sending Bob 
only {\em one} qubit (namely her half of the entangled pair). This 
idea due to Wiesner (Bennett and Wiesner 1992) is 
called ``dense coding'', since only  one quantum bit travels from
Alice to Bob in order to convey two classical bits.
Two quantum bits are involved, but Alice only ever sees one of them.
The method relies on the following fact:
the four mutually orthogonal states $\ket{00} + \ket{11},\;
\ket{00} - \ket{11}$, $\ket{01} + \ket{10},\;\ket{01} - \ket{10}$
can be generated from each other by operations on a single qubit. This set
of states is called the Bell basis, since they exhibit the strongest
possible Bell-EPR correlations (Braunstein {\em et. al.} 1992).
Starting from
$\ket{00} + \ket{11}$, Alice can generate any of the Bell basis
states by operating on her qubit with one of the operators $\{I,X,Y,Z\}$. 
Since there are four possibilities, her choice of operation
represents two bits of classical information. She then sends her qubit
to Bob, who must deduce which Bell basis state the qubits are in. This
he does by operating on the pair with the {\sc xor} gate, and measuring
the target bit, thus distinguishing $\ket{00} \pm \ket{11}$ from
$\ket{01} \pm \ket{10}$. To find the sign in the superposition, he
operates with $H$ on the remaining qubit, and measures it. Hence
Bob obtains two classical bits with no ambiguity.

Dense coding is difficult to implement, and so has no practical value
merely as a standard communication method. However, it can permit secure
communication: the qubit sent by Alice will only yield the
two classical information bits to someone in possession of the
entangled partner qubit. More generally, dense coding is an example
of the statement which began this section.
It reveals a relationship between 
classical information, qubits, and the information content of quantum 
entanglement (Barenco and Ekert 1995). A laboratory demonstration of the 
main features is described by Mattle {\em et. al.} (1996); Weinfurter (1994) 
and Braunstein and Mann (1995) discuss some of the methods employed, based 
on a source of EPR photon pairs from parametric down-conversion. 

\subsection{Quantum teleportation}

{\em It is possible to transmit qubits without sending qubits!}

Suppose Alice wishes to communicate to Bob a single qubit in the state
$\ket{\phi}$. If Alice already knows what state she has, for
example $\ket{\phi} = \ket{0}$, she can communicate it to Bob
by sending just classical information, eg ``Dear Bob, I have the
state $\ket{0}$. Regards, Alice.'' However, if $\ket{\phi}$ is
unknown there is no way for Alice to learn it with
certainty: any measurement she may perform may change the state,
and she cannot clone it and measure the copies. Hence it appears
that the only way to transmit $\ket{\phi}$ to Bob is to send
him the physical qubit (i.e. the electron or atom or whatever), or
possibly to swap the state into another quantum system and send
that. In either case a quantum system is transmitted.

Quantum teleportation (Bennett {\em et. al.} 1993, Bennett 1995)
permits a way around this limitation.
As in dense coding, we will use quantum entanglement as an information
resource. Suppose Alice and Bob possess an entangled pair in the
state $\ket{00} + \ket{11}$. Alice wishes to transmit to Bob
a qubit in an unknown state $\ket{\phi}$. Without loss of
generality, we can write $\ket{\phi} = a \ket{0} + b \ket{1}$
where $a$ and $b$ are unknown coefficients. Then the initial
state of all three qubits is 
  \beq
a\ket{000} + b\ket{100} + a\ket{011} + b\ket{111}
  \eeq
Alice now measures in the Bell basis
the first two qubits, i.e. the unknown
one and her member of the entangled pair. The network to do this
is shown in fig. 9b. After Alice has applied the
{\sc xor} and Hadamard gates, and just before she measures her
qubits, the state is
  \begin{eqnarray}
\lefteqn{}&& 
  \ket{00}\left( a\ket{0} + b \ket{1}\right)
+ \ket{01}\left( a\ket{1} + b \ket{0}\right)         \nonumber \\
&& \rule{-2ex}{0em} 
+ \ket{10}\left( a\ket{0} - b \ket{1}\right)
+ \ket{11}\left( a\ket{1} - b \ket{0}\right).
  \end{eqnarray}
Alice's measurements collapse the state onto one of four
different possibilities, and yield two classical bits. The
two bits are sent to Bob, who uses them to learn which
of the operators $\{I,X,Z,Y\}$ he must apply to his qubit in
order to place it in the state $a\ket{0} + b \ket{1} = \ket{\phi}$. Thus
Bob ends up with the qubit (i.e. the quantum information, not
the actual quantum system) which Alice wished to transmit.

Note that the quantum information can only arrive at Bob if it disappears 
from Alice (no cloning). Also, quantum information is complete information: 
$\ket{\phi}$ is the complete description of Alice's qubit. The use of the 
word `teleportation' draws attention to these two facts. Teleportation
becomes an especially important idea when we come to consider communication
in the presence of noise, section \ref{s:qec}. 

\subsection{Quantum data compression}  \lab{s:qdc}

Having introduced the qubit, we now wish to show that it is a useful 
measure of quantum information content. The proof of this is due to Jozsa 
and Schumacher (1994) and Schumacher (1995), building on work of Kholevo 
(1973) and Levitin (1987). To begin the argument, we first need a quantity 
which expresses how much information you would gain if you were to learn 
the quantum state of some system $\cal Q$. A suitable quantity is the Von 
Neumann entropy 
  \beq
    S(\rho) = - {\rm Tr} \rho \log \rho
  \eeq
where Tr is the trace operation, and $\rho$ is the density operator
describing an ensemble of states of the quantum system. This is
to be compared with the classical Shannon entropy, equation (\ref{S}).
Suppose a classical random variable $X$ has a
probability distribution $p(x)$. If a quantum system is
prepared in a state $\ket{x}$ dictated by the value of $X$,
then the density matrix is $\sum_x p(x)
\ket{x} \bra{x}$, where the states $\ket{x}$ need not be orthogonal. 
It can be shown (Kholevo 1973, Levitin 1987) that $S(\rho)$ is an upper 
limit on the classical mutual information $I(X:Y)$ between $X$ and the 
result $Y$ of a measurement on the system. 

To make connection with qubits, we consider the resources needed
to store or transmit the state of a quantum system $q$ of
density matrix $\rho$. The idea is to collect $n \gg 1$ such systems, and
transfer (`encode') the joint state into some smaller system.
The smaller system is transmitted down the channel, and at
the receiving end the joint state is `decoded' into $n$ systems
$q'$ of the same type as $q$ (see fig. 9c).
The final density matrix of each $q'$ is $\rho'$, and the
whole process is deemed successful if $\rho'$ is sufficiently close
to $\rho$. The measure of the similarity between two density
matrices is the {\em fidelity} defined by
  \beq
f(\rho, \rho') =
\left( {\rm Tr} \sqrt{\rho^{1/2} \rho' \rho^{1/2} } \right)^2
  \eeq
This can be interpreted as the probability that $q'$ passes
a test which ascertained if it was in the state $\rho$. When
$\rho$ and $\rho'$ are both pure states, $\ket{\phi}\bra{\phi}$
and $\ket{\phi'}\bra{\phi'}$, the fidelity is
none other than the familiar overlap: $f = | \left< \phi \right| \left. 
\phi' \right> |^2$.

Our aim is to find the smallest transmitted system which permits $f = 1 - 
\epsilon$ for $\epsilon \ll 1$. The argument is analogous to the `typical 
sequences' idea used in section \ref{s:dc}. Restricting ourselves for 
simplicity to two-state systems, the total state of $n$ systems is 
represented by a vector in a Hilbert space of $2^n$ dimensions. However, if 
the von Neumann entropy $S(\rho) < 1$ then it is highly likely (i.e. tends to 
certainty in the limit of large $n$) that, in any given realisation, the 
state vector actually falls in a {\em typical sub-space} of Hilbert space. 
Schumacher and Jozsa showed that the dimension of the typical sub-space is 
$2^{n S(\rho)}$. Hence only $n S(\rho)$ qubits are required to represent the 
quantum information faithfully, and the qubit (i.e. the logarithm of the 
dimensionality of Hilbert space) is a useful measure of quantum information. 
Furthermore, the encoding and decoding operation is `blind': it does not 
depend on knowledge of the exact states being transmitted. 

Schumacher and Josza's result is
powerful because it is general: no assumptions are made about
the exact nature of the quantum states involved. 
In particular, they need not be orthogonal. 
If the states to be transmitted were mutually orthogonal,
the whole problem would reduce to one of classical information.

The `encoding' and `decoding' required to achieve such quantum data 
compression and decompression is technologically very demanding. It cannot 
at present be done at all using photons. However, it is the ultimate 
compression allowed by the laws of physics. The details of the required 
quantum networks have been deduced by Cleve and DiVincenzo (1996). 

As well as the essential concept of information, other classical ideas such 
as Huffman coding have their quantum counterparts. Furthermore, Schumacher 
and Nielson (1996) derive a quantity which they call `coherent information' 
which is a measure of mutual information for quantum systems.
It includes that part of the mutual information between entangled systems 
which cannot be accounted for classically. This is a helpful way to 
understand the Bell-EPR correlations. 

\subsection{Quantum cryptography} 

No overview of quantum information is complete without a mention
of quantum cryptography. This area stems from an unpublished paper
of Wiesner written around 1970 (Wiesner 1983). It includes various
ideas whereby
the properties of quantum systems are used to achieve useful
cryptographic tasks, such as secure (i.e. secret) communication. The subject
may be divided into quantum {\em key distribution}, and a collection of
other ideas broadly related to {\em bit commitment}. Quantum key
distribution will be outlined below. Bit commitment refers to the 
scenario in which Alice must make some decision, such as a
vote, in such a way that Bob can be sure that Alice fixed her vote before
a given time, but where Bob
can only learn Alice's vote at some later time which she chooses. A classical, cumbersome method to achieve bit commitment
is for Alice to write down her vote
and place it in a safe which she gives to Bob. When she wishes Bob, later,
to learn the information, she gives him the key to the safe. A typical
quantum protocol is a carefully constructed variation on the idea
that Alice provides Bob with
a prepared qubit, and only later tells him in what basis it was prepared.

The early contributions
to the field of quantum cryptography
were listed in the introduction, further references may be
found in the reviews mentioned at the beginning of this section.
Cryptography has the unusual feature that it is not possible to prove by
experiment that a cryptographic procedure is secure: who knows whether a spy
or cheating person managed to beat the system? Instead, the users' confidence
in the methods must rely on mathematical proofs of security, and it is
here that much important work has been done. A concerted effort
has enabled proofs to be established for the security
of correctly implemented quantum key distribution. However, the
bit commitment idea, long thought to be secure through quantum
methods, was recently proved to be insecure
(Mayers 1997, Lo and Chau 1997) because the participants can
cheat by making use of quantum entanglement.

Quantum key distribution is a method in which quantum states
are used to establish a random secret key for cryptography. The essential
ideas are as follows: Alice and Bob are, as usual, widely seperated
and wish to communicate. Alice sends to Bob $2n$ qubits,
each prepared in
one of the states $\ket{0},\ket{1},\ket{+},\ket{-}$, randomly
chosen\footnote{Many other methods are possible, we adopt this
one merely to illustrate the concepts.}.
Bob measures his received bits, choosing the measurement 
basis randomly between $\{\ket{0},\ket{1}\}$ and $\{\ket{+},\ket{-}\}$.
Next, Alice and Bob inform each other publicly (i.e. anyone can listen in)
of the basis they used to prepare or measure each qubit. They
find out on which occasions they by chance used the same basis, which
happens on average half the time, and retain just those results.
In the absence of errors or interference, they now share the same
random string of $n$ classical bits (they agree for example
to associate $\ket{0}$ and $\ket{+}$ with 0; $\ket{1}$ and $\ket{-}$
with 1). This classical bit string is often called the {\em raw quantum
transmission}, RQT.

So far nothing has been gained by using qubits. The important
feature is, however, that it is impossible for anyone to learn
Bob's measurement results by observing the qubits {\em en route},
without leaving evidence of their presence. The crudest way
for an eavesdopper Eve to attempt to discover the key would be
for her to intercept the qubits and measure them, then pass them on to Bob.
On average half the time Eve guesses Alice's basis correctly and thus
does not disturb the qubit. However, Eve's correct guesses do not coincide 
with Bob's, so Eve learns the state of half of the $n$ qubits which
Alice and Bob later decide to 
trust, and disturbs the other half, for example sending to Bob
$\ket{+}$ for Alice's $\ket{0}$. Half of those disturbed will be
projected by Bob's measurement back onto the original state
sent by Alice, so overall Eve corrupts $n/4$ bits of the RQT.

Alice and Bob can now detect Eve's presence simply by randomly choosing 
$n/2$ bits of the RQT and announcing publicly the values they have. If they 
agree on all these bits, then they can trust that no eavesdropper was 
present, since the probability that Eve was present and they happened to 
choose $n/2$ uncorrupted bits is $(3/4)^{n/2} \simeq 10^{-125}$ for 
$n=1000$. The $n/2$ undisclosed bits form the secret key.

In practice the protocol is more complicated since Eve might adopt
other strategies (e.g. not intercept all the qubits), and 
noise will currupt some 
of the qubits even in the absence of an evesdropper. Instead of rejecting 
the key if many of the disclosed bits differ, Alice and Bob retain it as 
long as they find the error rate to be well below $25\%$. They then process 
the key in two steps. The first is to detect and remove errors, which is 
done by publicly comparing parity checks on publicly chosen random subsets 
of the bits, while discarding bits to prevent increasing Eve's information. 
The second step is to decrease Eve's knowledge of the key, by distilling 
from it a smaller key, composed of parity values calculated from the 
original key. In this way a key of around $n/4$ bits is obtained, of which 
Eve probably knows less than $10^{-6}$ of one bit (Bennett {\em et. al.} 
1992). 

The protocol just described is not the only one possible. Another approach 
(Ekert 1991) involves the use of EPR pairs, which Alice and Bob measure 
along one of three different axes. To rule out eavesdropping they check for 
Bell-EPR correlations in their results. 

The great thing about quantum key distribution is that it is feasible with 
current technology. A pioneering experiment (Bennett and Brassard 1989) 
demonstrated the principle, and much progress has been made since then. 
Hughes {\em et. al.} (1995) and Phoenix and Townsend (1995) summarised the 
state of affairs two years ago, and recently Zbinden {\em et. al.} (1997) 
have reported excellent key distribution through 23 km of standard telecom 
fibre under lake Geneva. The qubits are stored in the polarisation states of 
laser pulses, i.e. coherent states of light, with on average $0.1$ photons per 
pulse. This low light level is necessary so that pulses containing more than 
one photon are unlikely. Such pulses would provide duplicate qubits, and 
hence a means for an evesdropper to go undetected. The system achieves
a bit error rate of $1.35\%$, which is low enough to 
guarantee privacy in the full protocol. The data transmission rate is rather 
low: MHz as opposed to the GHz rates common in classical communications, but 
the system is very reliable. 

Such spectacular experimental mastery is
in contrast to the subject of the next section.

\section{The universal quantum computer}  \lab{s:uqc}

We now have sufficient concepts to understand the jewel at the
heart of quantum information theory, namely, the quantum computer (QC).
Ekert and Jozsa (1996) and Barenco (1996) give
introductory reviews concentrating on the quantum
computer and factorisation; a review with emphasis on practicalities
is provided by Spiller (1996). Introductory material is also provided
by DiVincenzo (1995b) and Shor (1996).

The QC is first and foremost a machine which is a theoretical construct, like 
a thought-experiment, whose purpose is to allow quantum information 
processing to be formally analysed. In particular it establishes
the Church-Turing Principle introduced in section \ref{s:qvc}.

Here is a prescription for a quantum computer, based on that of
Deutsch (1985, 1989):

A quantum computer is a set of $n$ qubits in which the following
operations are experimentally feasible:
  \begin{enumerate}
\item Each qubit can be prepared in some known state $\ket{0}$.
\item Each qubit can be measured in the basis $\{ \ket{0}, \ket{1} \}$.
\item A universal quantum gate (or set of gates) can be applied
at will to any fixed-size subset of the qubits.
\item The qubits do not evolve other than via the above transformations.
  \end{enumerate}

This prescription is incomplete in certain technical ways to be
discussed, but it encompasses the main ideas. The model of
computation we have in mind is a network model, in which logic
gates are applied sequentially to a set of bits (here, quantum
bits). In an electronic classical computer, logic gates are
spread out in space on a circuit board, but in the QC
we typically imagine the logic gates to be interactions turned
on and off in time, with the qubits at fixed positions,
as in a quantum network diagram (fig. 8, 12).
Other models of quantum computation can be conceived, such
as a cellular automaton model (Margolus 1990).

\subsection{Universal gate}

The universal quantum gate is the quantum equivalent of the
classical universal gate, namely a gate which 
by its repeated use on different combinations of bits
can generate the action of any other gate. What is the set
of all possible quantum gates, however? To answer
this, we appeal to the principles of quantum mechanics
(Schr\"odinger's equation), 
and answer that since all quantum evolution is unitary,
it is sufficient to
be able to generate {\em all unitary transformations}
of the $n$ qubits in the computer. This might seem a tall
order, since we have a continuous and therefore infinite
set. However, it turns out that quite simple quantum gates
can be universal, as Deutsch showed in 1985. 

The simplest way to think about universal gates is
to consider the pair of gates $V(\theta, \phi)$ and
controlled-not (or {\sc xor}), where $V(\theta, \phi)$
is a general rotation of a single qubit, ie
  \beq
V(\theta, \phi) = \left( \begin{array}{lr}
\cos (\theta/2) & -i e^{-i\phi} \sin (\theta/2) \\
 -i e^{i\phi} \sin (\theta/2) & \cos (\theta/2) 
\end{array} \right).    \lab{V}
  \eeq
It can be shown that any $n \times n$ unitary matrix can
be formed by composing 2-qubit {\sc xor} gates and single-qubit
rotations. Therefore, this pair of operations is universal for quantum 
computation. A purist may argue that $V(\theta, \phi)$ is an infinite set of 
gates since the parameters $\theta$ and $\phi$ are continuous, but it 
suffices to choose two particular irrational angles for $\theta$ and $\phi$, 
and the resulting single gate can generate all single-qubit rotations by 
repeated application; however, a practical system need not use
such laborious methods. 
The {\sc xor} and rotation operations can 
be combined to make a controlled rotation which is a single universal gate. 
Such universal quantum gates were discussed by Deutsch {\em et. al.} (1995),
Lloyd (1995), DiVincenzo (1995a) and Barenco (1995). 

It is remarkable that 2-qubit gates are sufficient for quantum
computation. This is why the quantum gate is a powerful
and important concept. 

\subsection{Church-Turing principle}

Having presented the QC, it is necessary to argue for its universality, i.e. 
that it fulfills the Church-Turing Principle as claimed. The two-step 
argument is very simple. First, the state of any finite quantum system is 
simply a vector in Hilbert space, and therefore can be represented to 
arbitrary precision by a finite number of qubits. Secondly, the evolution of 
any finite quantum system is a unitary transformation of the state, and 
therefore can be simulated on the QC, which can generate any unitary 
transformation with arbitrary precision. 

A point of principle is raised by Myers (1997), who points out that
there is a difficulty with computational tasks for which the number
of steps for completion cannot be predicted. We cannot in general
observe the QC to find out if it has halted, in contrast to a
classical computer. However, we will only be concerned with tasks
where either the number of steps is predictable, or the
QC can signal completion by setting a dedicated qubit which
is otherwise not involved in the computation (Deutsch 1985). This is a 
very broad class of problems. Nielsen and Chuang (1997) consider
the use of a {\em fixed} quantum gate array, showing that there is no
array which, operating on qubits representing
both data and program, can perform any unitary transformation
on the data. However, we consider a machine in which 
a classical computer controls the quantum gates applied to a quantum
register, so any gate array can be `ordered' by a classical program
to the classical computer.

The QC is certainly an interesting theoretical
tool. However, there hangs over it a large and important 
question-mark: what about imperfection? The prescription given above 
is written as if measurements and gates can be applied with
arbitrary precision, which is unphysical, as is the fourth
requirement (no extraneous evolution). The prescription can
be made realistic by attaching to each of the four requirements
a statement about the degree of allowable imprecision.
This is a subject of on-going research, and we will take it
up in section \ref{s:qec}. Meanwhile, let us investigate more
specifically what a sufficiently well-made quantum computer might do.

\section{Quantum algorithms}  \lab{s:qa}

It is well known that classical computers are able to calculate
the behaviour of quantum systems, so we have not yet demonstrated
that a quantum computer can do anything which a classical computer
can not. Indeed, since our theories of physics always involve
equations which we can write down and manipulate, it seems highly unlikely
that quantum mechanics, or any future physical theory, would permit
computational problems to be addressed which are not
in principle solvable on a large enough classical Turing machine.
However, as we saw in section \ref{s:cc}, those
words `large enough', and also `fast enough', are centrally important
in computer science. Problems which are computationally `hard' can be 
impossible in practice. In technical language, while quantum computing does not 
enlarge the set of computational problems which can be addressed (compared to 
classical computing), it does introduce the possibility of new complexity 
classes. Put more simply, tasks for which
classical computers are too slow may be solvable with quantum computers.

\subsection{Simulation of physical systems}

The first and most obvious application of a QC is
that of simulating some other quantum system. To simulate a
state vector in a $2^n$-dimensional Hilbert space, a classical
computer needs to manipulate vectors containing of order $2^n$
complex numbers, whereas a quantum computer requires just $n$
qubits, making it much more efficient in storage space. 
To simulate evolution, in general both the classical and
quantum computers will be inefficient. A classical computer
must manipulate 
matrices containing of order $2^{2n}$ elements, which requires
a number of operations (multiplication, addition) exponentially
large in $n$, while a quantum computer must build unitary
operations in $2^n$-dimensional Hilbert space, which 
usually requires an exponentially large number of elementary
quantum logic gates. 
Therefore the quantum computer is not guaranteed to simulate 
{\em every} physical system efficiently. However, it can be shown
that it can simulate a large
class of quantum systems efficiently, including many for
which there is no efficient classical algorithm, such
as many-body systems with local interactions (Lloyd 1996,
Zalka 1996, Wiesner 1996, Meyer 1996, Lidar and Biam 1996, Abrams
and Lloyd 1997, Boghosian and Taylor 1997). 

\subsection{Period finding and Shor's factorisation algorithm}

So far we have discussed simulation of Nature, which is a rather restricted
type of computation. We would like to let the QC loose
on more general problems, but it has so far proved hard to find ones
on which it performs better than classical computers. However, the
fact that there exist such problems at all is a profound insight into
physics, and has stimulated much of the recent interest in the field.

Currently one of the most important quantum algorithms is that for
finding the period of a function. Suppose a function $f(x)$ is periodic with 
period $r$, i.e. $f(x) = f(x + r)$. Suppose further that $f(x)$ can be 
efficiently computed from $x$, and all we know initially is that $N/2 < r < N$ 
for some $N$. Assuming there is no analytic technique to deduce the 
period of $f(x)$, the best we can do on a classical computer is to calculate 
$f(x)$ for of order $N/2$ values of $x$, and find out when the function
repeats itself (for well-behaved functions only $O(\sqrt{N})$ values may
be needed on average). This is inefficient since the 
number of operations is exponential in the input size $\log N$
(the information required to specify $N$).

The task can be solved efficiently on a QC by the elegant
method shown in fig. 10, due to Shor (1994), building on Simon (1994). The
QC requires $2n$ qubits, plus a further $0(n)$ for workspace,
where $n = \lceil 2 \log N \rceil$ 
(the notation $\lceil x \rceil$ means the nearest integer greater 
than $x$). These are divided into two `registers', each of $n$ qubits. They 
will be referred to as the $x$ and $y$ registers; both are initially prepared 
in the state $\ket{0}$ (i.e. all $n$ qubits in states $\ket{0}$). Next, the 
operation $H$ is applied to each qubit in the $x$ register, making the total 
state 
  \beq
\frac{1}{\sqrt{w}} \sum_{x=0}^{w-1} \ket{x} \ket{0}  \label{step1}
  \eeq
where $w = 2^n$. This operation is referred to as a Fourier transform
in fig. 10, for reasons that will shortly become apparant. 
The notation $\ket{x}$ means a state such as $\ket{0011010}$,
where $0011010$ is the integer $x$ in binary notation. In this
context the basis $\{ \ket{0}, \ket{1} \}$ is referred to as the
`computational basis.' It is convenient (though not of course necessary)
to use this basis when describing the computer.

Next,
a network of logic gates is applied to both $x$ and $y$ regisiters,
to perform the transformation $U_f \ket{x} \ket{0} = \ket{x} \ket{f(x)}$.
Note that this transformation can be unitary because the input
state $\ket{x} \ket{0}$ is in one to one correspondance with the
output state $\ket{x} \ket{f(x)}$, so the process is reversible. 
Now, applying $U_f$ to the state given in eq. (\ref{step1}), we
obtain
  \beq
\frac{1}{\sqrt{w}} \sum_{x=0}^{w-1} \ket{x} \ket{f(x)}  \label{step2}
  \eeq
This state is illustrated in fig. 11a.
At this point something rather wonderful has taken place: the value of
$f(x)$ has been calculated for $w=2^n$ values of $x$, all
in one go! This feature is referred to as {\em quantum parallelism}
and represents a huge parallelism because of the exponential
dependence on $n$ (imagine having $2^{100}$, i.e. a million
times Avagadro's number, of classical processors!)

Although the $2^n$ evaluations of $f(x)$ are in some sense `present'
in the quantum state in eq. (\ref{step2}), unfortunately we
cannot gain direct access to them. For, a measurement 
(in the computational basis) of the
$y$ register, which is the next step in the algorithm, will
only reveal one value of $f(x)$\footnote{It is not strictly
necessary to measure the $y$ register, but this simplifies
the description.}. Suppose the value obtained
is $f(x) = u$. The $y$ register state collapses onto $\ket{u}$,
and the total state becomes
  \beq
\frac{1}{\sqrt{M}} \sum_{j=0}^{M-1} \ket{d_u+jr} \ket{u} \label{step3}
  \eeq
where $d_u + j r$, for $j=0,1,2 \ldots M-1$, are all the values of
$x$ for which $f(x) = u$. In other words the periodicity of
$f(x)$ means that the $x$ register remains
in a superposition of $M \simeq w/r$ states, at values of
$x$ separated by the period $r$. Note that the offset $d_u$
of the set of $x$ values depends on the value $u$ obtained
in the measurement of the $y$ register.

It now remains to extract the periodicity of the state in the
$x$ register. This is done by applying a Fourier transform,
and then measuring the state. The discrete Fourier transform 
employed is the following unitary process: 
  \beq
U_{\cal FT} \ket{x} = \frac{1}{\sqrt{w}} \sum_{k=0}^{w-1}
e^{i 2\pi k x /w} \ket{k}
  \eeq
Note that eq. (\ref{step1}) is an example of this, operating
on the initial state $\ket{0}$. 
The quantum network to apply $U_{\cal FT}$ is based on the
fast Fourier transform algorithm (see, e.g., Knuth (1981)). The quantum
version was worked out by Coppersmith (1994) and Deutsch (1994)
independently, a clear presentation may also be found
in Ekert and Josza (1996), Barenco (1996)\footnote{An exact
quantum Fourier transform would require rotation operations
of precision exponential in $n$, which raises a problem with the
efficiency of Shor's algorithm. However, an approximate version
of the Fourier transform is sufficient (Barenco {\em et. al.} 1996)}.
Before applying $U_{\cal FT}$ to eq. (\ref{step3})
we will make the simplifying assumption that $r$ divides $w$
exactly, so $M = w/r$. The essential ideas are not
affected by this restriction; when it is relaxed some added
complications must be taken into account (Shor 1994, 1995a; Ekert and Josza 
1996). 

The $y$ register no longer concerns us, so we will just consider
the $x$ state from eq. (\ref{step3}):
  \beq
U_{\cal FT} \frac{1}{\sqrt{w/r}} \sum_{j=0}^{w/r-1} \ket{d_u + jr}
   = \frac{1}{\sqrt{r}}
\sum_k \tilde{f} (k) \ket{k}     \label{step4}
  \eeq
where
  \beq
|\tilde{f} (k)| = \left\{ \begin{array}{ll} 1 \;\;\; & \mbox{if $k$ is a
multiple of $w/r$} \\
0 & \mbox{otherwise}
  \end{array} \right.     
  \eeq
This state is illustrated in fig. 11b.
The final state of the $x$ register is now measured, and we see
that the value obtained must be a multiple of $w/r$. It remains
to deduce $r$ from this. We have $x = \lambda w/r$
where $\lambda$ is unknown. If $\lambda$ and $r$ have no
common factors, then we cancel $x/w$ down to an irreducible 
fraction and thus obtain $\lambda$ and $r$. If $\lambda$ and $r$ have a common 
factor, which is unlikely for large $r$, then the algorithm fails. 
In this case, the whole algorithm must be repeated from the start. After a 
number of repetitions no greater than $\sim \log r$, and usually much
less than this, the probability of success can be 
shown to be arbitrarily close to $1$ (Ekert and Josza 1996). 

The quantum period-finding algorithm we have described is efficient as long as 
$U_f$, the evaluation of $f(x)$, is efficient. The total number of 
elementary logic gates required is a polynomial rather than exponential 
function of $n$. As was emphasised in section \ref{s:cc}, this makes all the 
difference between tractable and intractable in practice, for sufficiently 
large $n$. 

To add the icing on the cake, it can be remarked that the important
factorisation problem mentioned in section \ref{s:cc} can be
reduced to one of finding the period of a simple function. This
and all the above ingredients were first brought together by Shor (1994),
who thus showed that the factorisation problem is tractable
on an ideal quantum computer. The function to be evaluated in this
case is $f(x) = a^x \;{\rm mod}\;N$ where $N$ is the number to
be factorised, and $a < N$ is chosen randomly. One can show
using elementary number theory (Ekert and Josza 1996) that for most choices of 
$a$, the period $r$ is even and $a^{r/2} \pm 1$ shares a common factor with 
$N$. The common factor (which is of course a factor $N$) 
can then be deduced rapidly using a classical algorithm due to Euclid 
({\em circa} 300 BC; see, e.g. Hardy and Wright 1965). 

To evaluate $f(x)$ efficiently, repeated squaring (modulo $N$) is used, 
giving powers $((a^2)^2)^2 \ldots$. Selected such powers of $a$, 
corresponding to the binary expansion of $a$, are then multiplied together. 
Complete networks for the whole of Shor's algorithm were described
by Miquel {\em et. al.} (1996), Vedral {\em et. al.} (1996) and Beckman
{\em et. al.} (1996). They require of order $300 (\log N)^3$ logic gates.
Therefore, to 
factorise numbers of order $10^{130}$, i.e. at the limit of current classical 
methods, would require $\sim 2 \times 10^{10}$ gates per run, or 7 hours if 
the `switching rate' is one megaHertz\footnote{The algorithm might need to be
run $\log r \sim 60$ times to ensure at least one successful run, but
the average number of runs required will be much less than this.}.
Considering how 
difficult it is to make a quantum computer, this offers no advantage over 
classical computation. However, if we double the number of digits to 260 then 
the problem is intractable classically (see section \ref{s:cc}), while the 
ideal quantum computer takes just 8 times longer than before. The existence 
of such a powerful method is an exciting and profound new insight into 
quantum theory. 

The period-finding algorithm appears at first sight like a conjuring trick: 
it is not quite clear how the quantum computer managed to produce the 
period like a rabbit out of a hat. Examining fig. 11 and 
equations (\ref{step1}) to (\ref{step4}), I would say that the most important 
features are contained in eq. (\ref{step2}). They are not only the {\em 
quantum parallelism} already mentioned, but also {\em quantum 
entanglement}, and, finally, quantum interference. Each value of $f(x)$ 
retains a link with the value of $x$ which produced it, through the 
entanglement of the $x$ and $y$ registers in eq. (\ref{step2}). The `magic' 
happens when a measurement of the $y$ register produces the special state 
$\ket{\psi}$ (eq. \ref{step3}) in the $x$ register, and it is quantum 
entanglement which permits this (see also Jozsa 1997a). The final Fourier 
transform can be regarded as an interference between the various superposed 
states in the $x$ register (compare with the action of a diffraction 
grating). 

Interference effects can be used for computational purposes with 
classical light fields, or water waves for that matter, so interference is not 
in itself the essentially quantum feature. Rather, the exponentially 
large number of interfering states, and the entanglement, are features
which do not arise in classical systems. 

\subsection{Grover's search algorithm}

Despite considerable efforts in the quantum computing community, the number
of useful quantum algorithms which have been discovered remains small.
They consist mainly of variants on the period-finding algorithm presented
above, and another quite different task: that of searching an unstructured
list. Grover (1997) presented a quantum algorithm for the following
problem: given an unstructured list of items $\{ x_i \}$, find
a particular item $x_j = t$. Think, for example, of looking for a particular
telephone number in the telephone directory (for someone whose name
you do not know). It is not hard to prove that classical
algorithms can do no better than searching through the list, requiring
on average $N/2$ steps, for a list of $N$ items. Grover's algorithm
requires of order $\sqrt{N}$ steps. The task remains computationally hard: it 
is not transferred to a new complexity class, but it is remarkable that such a 
seemingly hopeless task can be speeded up at all. 
The `quantum speed-up' $\sim \sqrt{N}/2$ is greater than that achieved
by Shor's factorisation algorithm ($\sim \exp( 2 (\ln N)^{1/3} )$), and
would be important for the
huge sets ($N \simeq 10^{16}$) which can arise, for example, in
code-breaking problems (Brassard 1997). 

An important further point was proved by Bennett {\em et. al.} (1997), namely
that Grover's algorithm is optimal: no quantum algorithm can do better
than $O(\sqrt{N})$. 

A brief sketch of Grover's algorithm is as follows. Each item
has a label $i$, and we must be able to test in a unitary
way whether any item is the one we are seeking. In other
words there must exist a unitary operator $S$ such that $S \ket{i}
= \ket{i}$ if $i \ne j$, and $S \ket{j} = - \ket{j}$, where $j$
is the label of the special item. For example, the test might establish
whether $i$ is the solution of some hard computational
problem\footnote{That is, an ``{\sc np}'' problem for which finding a
solution is hard, but testing a proposed solution is easy.}.
The method
begins by placing a single quantum register in a superposition
of all computational states, as in the period-finding algorithm
(eq. (\ref{step1})). Define
  \beq
\ket{ \Psi (\theta ) } \equiv  \sin \theta \ket{j} + 
\frac{ \cos \theta }{\sqrt{N-1}} \sum_{i \ne j} \ket{i}
  \eeq
where $j$ is the label of the element $t = x_j$ to be found.
The initially prepared state is an equally-weighted superposition, 
$\ket{ \Psi (\theta_0 ) }$ where $\sin \theta_0 = 1 / \sqrt{N}$. 
Now apply $S$, which reverses the sign of the
one special 
element of the superposition, then Fourier transform, change the
sign of all components except $\ket{0}$, and Fourier transform
back again. These operations represent a subtle interference
effect which achieves the following transformation:
  \beq
U_G \ket{ \theta } = \ket{ \Psi (\theta + \phi ) }  \label{UG}
  \eeq
where $\sin \phi = 2 \sqrt{N-1} / N$. 
The coefficient of the special element
is now slightly larger than that of all the other elements.
The method proceeds simply by applying $U_G$ $m$ times, where $m \simeq 
(\pi/4) \sqrt{N}$. The slow rotation brings $\theta$ very close to $\pi/2$,
so the quantum state becomes almost precisely equal to 
$\ket{j}$. After the $m$ iterations the state is measured and the value $j$ 
obtained (with error probability $O(1/N)$).
If $U_G$ is applied too many times, the success probability diminishes,
so it is important to know $m$, which was deduced by Boyer {\em et. al.} 
(1996). Kristen Fuchs compares the technique to cooking a souffl\'e. The 
state is placed in the `quantum oven' and the desired answer rises slowly. 
You must open the oven at the right time, neither too soon not too late, to 
guarantee success. Otherwise the souffl\'e will fall---the state collapses 
to the wrong answer.

The two algorithms I have presented are the easiest to describe, and 
illustrate many of the methods of quantum computation. However, just what 
further methods may exist is an open question. Kitaev (1996) has shown 
how to solve the factorisation and related problems using a technique 
fundamentally different from Shor's. His ideas have some similarities to 
Grover's. Kitaev's method is helpfully clarified by Jozsa (1997b) who 
also brings out the common features of several quantum algorithms based 
on Fourier transforms. The quantum programmer's toolbox is thus slowly 
growing. It seems safe to predict, however, that the class of problems 
for which quantum computers out-perform classical ones is a special and 
therefore small class. On the other hand, any problem for which finding 
solutions is hard, but testing a candidate solution is easy, can at last 
resort be solved by an exhaustive search, and here Grover's algorithm may 
prove very useful.

\section{Experimental quantum information processors}  \lab{s:qip}

The most elementary quantum logical operations have been demonstrated in 
many physics experiments during the past 50 years. For example, the {\sc 
not} operation ($X$) is no more than a stimulated transition between two 
energy levels $\ket{0}$ and $\ket{1}$. The important {\sc xor} operation 
can also be identified as a driven transition in a four-level system. 
However, if we wish to contemplate a quantum computer it is necessary to 
find a system which is sufficiently controllable to allow quantum logic 
gates to be applied at will, and yet is sufficiently complicated to store 
many qubits of quantum information. 

It is very hard to find such systems. One might hope to fabricate quantum 
devices on solid state microchips---this is the logical progression of the 
microfabrication techniques which have allowed classical computers to 
become so powerful. However, quantum computation relies on complicated 
interference effects and the great problem in realising it is the problem 
of noise. No quantum system is really isolated, and the coupling to the 
environment produces decoherence which destroys the quantum computation. In 
solid state devices the environment is the substrate, and the coupling to 
this environment is strong, producing typical decoherence times of the 
order of picoseconds. It is important to realise that it is not enough to 
have two different states $\ket{0}$ and $\ket{1}$ which are themselves 
stable (for example states of different current in a superconductor): we 
require also that superpositions such as $\ket{0} + \ket{1}$ preserve their 
phase, and this is typically where the decoherence timescale is so short. 

At present there are two candidate systems which should permit quantum 
computation on 10 to 40 qubits. These are the proposal of Cirac and Zoller 
(1995) using a line of singly charged atoms confined and cooled in vacuum 
in an ion trap, and the proposal of Gershenfeld and Chuang (1997), and 
simultaneously Cory {\em et. al.} (1996), using the methods of bulk nuclear 
magnetic resonance. In both cases the proposals rely on the impressive 
efforts of a large community of researchers which developed the 
experimental techniques. Previous proposals for experimental quantum 
computation (Lloyd 1993, Berman {\em et. al.} 1994, Barenco {\em et. al.}
1995a, DiVincenzo 1995b) touched on some of the important 
methods but were not experimentally feasible. Further recent proposals
(Privman {\em et. al.} 1997, Loss and DiVincenzo 1997) may become
feasible in the near future.

\subsection{Ion trap}

The ion trap method is illustrated in fig. 12,
and described in detail by Steane (1997b). A string
of ions is confined by a combination of oscillating and static electric
fields in a linear `Paul trap' in high vacuum ($10^{-8}$ Pa).
A single laser beam is split by beam splitters and acousto-optic
modulators into many beam pairs, one pair illuminating each ion.
Each ion has two long-lived states, for example
different levels of the ground state hyperfine structure (the
lifetime of such states against spontaneous decay can exceed thousands
of years). Let us refer to these two states as $\ket{g}$ and $\ket{e}$;
they are orthogonal and so together represent one qubit.
Each laser beam pair can drive coherent Raman transitions
between the internal states of the relevant ion. This allows
any single-qubit quantum gate to be applied to any ion, but not
two-qubit gates. The latter requires an interaction between ions,
and this is provided by their Coulomb repulsion. However, exactly
how to use this interaction is far from obvious; it required the
important insight of Cirac and Zoller.

Light carries not only energy but also momentum, so whenever a
laser beam pair interacts with an ion, it exchanges momentum with the ion. In 
fact, the mutual repulsion of the ions means that the whole string of ions 
moves {\em en masse} when the motion is quantised (M\"ossbauer effect).
The motion of
the ion string is quantised because the ion string is confined in the
potential provided by the Paul trap. The quantum states of motion
correspond to the different degrees of excitation (`phonons')
of the normal modes of vibration of the string.
In particular we focus on the ground state of the motion $\ket{n=0}$
and the lowest excited state $\ket{n=1}$ of the fundamental mode.
To achieve, for example, controlled-$Z$ between
ion $x$ and ion $y$, we start with the motion in the ground state
$\ket{n=0}$. A pulse of the laser beams on ion $x$ drives
the transition $\ket{n=0}\ket{g}_x \rightarrow \ket{n=0}\ket{g}_x$,
$\ket{n=0}\ket{e}_x \rightarrow \ket{n=1}\ket{g}_x$, so the ion
finishes in the ground state, and the motion finishes in the initial
state of the ion: this is a `swap' operation. Next a pulse
of the laser beams on ion $y$ drives the transition
  \begin{eqnarray*}
\ket{n=0}\ket{g}_y &\rightarrow& \ket{n=0}\ket{g}_y \\
\ket{n=0}\ket{e}_y &\rightarrow& \ket{n=0}\ket{e}_y \\
\ket{n=1}\ket{g}_y &\rightarrow& \ket{n=1}\ket{g}_y \\
\ket{n=1}\ket{e}_y &\rightarrow& -\ket{n=1}\ket{e}_y
  \end{eqnarray*}
Finally, we repeat the initial pulse on ion $x$. The overall effect
of the three pulses is
   \begin{eqnarray*}
\ket{n=0}\ket{g}_x\ket{g}_y &\rightarrow& \ket{n=0}\ket{g}_x\ket{g}_y \\
\ket{n=0}\ket{g}_x\ket{e}_y &\rightarrow& \ket{n=0}\ket{g}_x\ket{e}_y \\
\ket{n=0}\ket{e}_x\ket{g}_y &\rightarrow& \ket{n=0}\ket{e}_x\ket{g}_y \\
\ket{n=0}\ket{e}_x\ket{e}_y &\rightarrow& -\ket{n=0}\ket{e}_x\ket{e}_y \\
  \end{eqnarray*}
which is exactly a controlled-$Z$ between $x$ and $y$. 
Each laser pulse must have a precisely controlled frequency and
duration. The controlled-$Z$ gate and the single-qubit
gates together provide a universal set, so we can perform arbitrary
transformations of the joint state of all the ions!

To complete the prescription for a quantum computer (section \ref{s:uqc}),
we must be able to prepare the initial state and measure the final
state. The first is possible through the methods of optical pumping
and laser cooling, the second through the `quantum jump' 
or `electron shelving' measurement
technique. All these are powerful techniques developed in the
atomic physics community over the past twenty years. However, the
combination of all the techniques at once has only been achieved
in a single experiment, which demonstrated preparation, quantum
gates, and measurement for just a single trapped ion (Monroe {\em et. al}
1995b). 

The chief experimental difficulty in the ion trap method is to cool the 
string of ions to the ground state of the trap (a sub-microKelvin 
temperature), and the chief source of decoherence is the heating of this 
motion owing to the coupling between the charged ion string and noise 
voltages in the electrodes (Steane 1997, Wineland {\em et. al.} 1997). It is 
unknown just how much the heating can be reduced. A conservative statement 
is that in the next few years 100 quantum gates could be applied to a few 
ions without losing coherence. In the longer term one may hope for an order 
of magnitude increase in both figures. It seems clear that an ion trap 
processor will never achieve sufficient storage capacity and coherence to 
permit factorisation of hundred-digit numbers. However, it would be 
fascinating to try a quantum algorithm on just a few qubits (4 to 10) and 
thus to observe the principles of quantum information processing at work. We 
will discuss in section \ref{s:qec} methods which should allow the number of 
coherent gate operations to be greatly increased. 

\subsection{Nuclear magnetic resonance}

The proposal using nuclear magnetic resonance (NMR) is illustrated in fig. 
13. The quantum processor in this case is a molecule containing a `backbone' 
of about ten atoms, with other atoms such as hydrogen attached so as to use 
up all the chemical bonds. It is the nuclei which interest us. Each has a 
magnetic moment associated with the nuclear spin, and the spin states 
provide the qubits. The molecule is placed in a large magnetic field, and 
the spin states of the nuclei are manipulated by applying oscillating 
magnetic fields in pulses of controlled duration. 

So far, so good. The problem is that the spin state of the nuclei of a 
single molecule can be neither prepared nor measured. To circumvent this 
problem, we use not a single molecule, but a cup of liquid containing some 
$10^{20}$ molecules! We then measure the average spin state, which can be 
achieved since the average oscillating magnetic moment of all the nuclei is 
large enough to produce a detectable magnetic field. 
Some subtleties enter at this point. Each of the molecules in
the liquid has a very slightly different local magnetic field,
influenced by other molecules in the vicinity, so each `quantum
processor' evolves slightly differently. This problem 
is circumvented by the spin-echo technique, a standard tool in NMR
which allows the effects of free evolution of the spins to be reversed, 
without reversing the effect of the quantum gates. However, this 
increases the difficulty of applying long sequences of quantum gates. 

The remaining problem is to prepare the initial state. The cup of liquid is 
in thermal equilibrium to begin with, so the different spin states have 
occupation probabilities given by the Boltzman distribution. One makes use 
of the fact that spin states are close in energy, and so have nearly equal 
occupations initially. Thus the density matrix $\rho$ of the $O(10^{20})$ 
nuclear spins is very close to the identity matrix $I$. It is the small {\em 
difference} $\Delta = \rho - I$ which can be used to store quantum 
information. Although $\Delta$ is not the density matrix of any quantum 
system, it nevertheless transforms under well-chosen field pulses in the same 
way as a density matrix would, and hence can be considered to represent an 
effective quantum computer. The reader is referred to Gershenfeld and Chuang 
(1997) for a detailed description, including the further subtlety that an 
effective pure state must be distilled out of $\Delta$ by means of a pulse 
sequence which performs quantum data compression. 

NMR experiments have for some years routinely achieved
spin state manipulations and measurements equivalent in
complexity to those required for quantum information processing
on a few qubits, therefore the first few-qubit quantum processors 
will be NMR systems. The method does not scale very well as the
number of qubits is increased, however. For example, with
$n$ qubits the measured signal scales as $2^{-n}$. Also
the possibility to measure the state is limited, since only
the average state of many processors is detectable. This restricts
the ability to apply quantum error correction (section \ref{s:qec}),
and complicates the design of quantum algorithms.

\subsection{High-$Q$ optical cavities}

Both systems we have described permit simple quantum information 
processing, but not quantum communication. However, in a very high-quality 
optical cavity, a strong coupling can be achieved between a single atom or 
ion and a single mode of the electromagnetic field. This coupling can be 
used to apply quantum gates between the field mode and the ion, thus 
opening the way to transferring quantum information between separated ion 
traps, via high-$Q$ optical cavities and optical fibres (Cirac {\em et. 
al.} 1997). Such experiments are now being contemplated. The required 
strong coupling between a cavity field and an atom has been demonstrated by 
Brune {\em et. al.} (1994), and Turchette {\em et. al.} (1995). An 
electromagnetic field mode can also be used to couple ions within a single 
trap, providing a faster alternative to the phonon method (Pellizzari {\em 
et. al.} 1995). 

\section{Quantum error correction} \lab{s:qec}

In section \ref{s:qa} we discussed some beautiful quantum algorithms. Their 
power only rivals classical computers, however, on quite large problems, 
requiring thousands of qubits and billions of quantum gates (with the 
possible exception of algorithms for simulation of physical systems). In 
section \ref{s:qip} we examined some experimental systems, and found that 
we can only contemplate `computers' of a few tens of qubits and perhaps 
some thousands of gates. Such systems are not `computers' at all because 
they are not sufficiently versatile: they should at best be called modest 
quantum information processors. Whence came this huge disparity between the 
hope and the reality? 

The problem is that the prescription for the universal quantum computer, 
section \ref{s:uqc}, is unphysical in its fourth requirement. There is no 
such thing as a perfect quantum gate, nor is there such a thing as an 
isolated system. One may hope that it is possible in principle to achieve 
any degree of perfection in a real device, but in practice this is an 
impossible dream. Gates such as {\sc xor} rely on a coupling between 
separated qubits, but if qubits are coupled to each other, they will 
unavoidably be coupled to something else as well (Plenio and Knight 1996). 
A rough guide is that it is very hard to find a system in which the loss of 
coherence is smaller than one part in a million each time a {\sc xor} gate 
is applied. This means the decoherence is roughly $10^7$ times too fast to 
allow factorisation of a 130 digit number! It is an open question whether 
the laws of physics offer any intrinsic lower limit to the decoherence 
rate, but it is safe to say that it would be simpler to speed up classical 
computation by a factor of $10^6$ than to achieve such low decoherence in a 
large quantum computer. Such arguments were eloquently put forward by 
Haroche and Raimond (1996). Their work, and that of others
such as Landauer (1995,1996) sounds a helpful note of caution.
More detailed treatments of decoherence in quantum computers are given
by Unruh (1995), Palma {\em et. al.} (1996) and Chuang {\em et. al.}
(1995). Large numerical studies are described by
Miquel {\em et. al.} (1996) and Barenco {\em et. al.} (1997).

Classical computers are reliable not because they are perfectly
engineered, but because they are insensitive to noise. One way
to understand this is to examine in detail a device such as
a flip-flop, or even a humble mechanical switch. Their stability
is based on a combination of amplification and dissipation: a small
departure of a mechanical switch from `on' or `off' results in a large
restoring force from the spring. Amplifiers do the corresponding job in
a flip-flop. The restoring force is not sufficient
alone, however: with a conservative force, the switch would oscillate
between `on' and `off'. It is important also to have damping, supplied
by an inelastic collision which generates heat in the case of
a mechanical switch, and by resistors in the electronic flip-flop.
However, these methods are ruled out for a quantum computer
by the fundamental principles of quantum mechanics. The no-cloning
theorem means amplification of unknown quantum states is impossible,
and dissipation is incompatible with unitary evolution. 

Such fundamental considerations lead to the widely accepted belief that 
quantum mechanics rules out the possibility to stabilize a quantum computer 
against the effects of random noise. A repeated projection of the computer's
state by well-chosen measurements is not in itself sufficient 
(Berthiaume {\em et. al.} 1994, Miquel {\em et. al} 1997).
However, by careful application of 
information theory one can find a way around this impasse. The idea is to 
adapt the error correction methods of classical information theory to the 
quantum situation. 

Quantum error correction (QEC) was established as an important and general 
method by Steane (1996b) and independently Calderbank and Shor (1996). Some 
of the ideas had been introduced previously by Shor (1995b) and Steane 
(1996a). They are related to the `entanglement purification' 
introduced by Bennett {\em et. al.} (1996a) and independently Deutsch {\em 
et. al.} (1996). The theory of QEC was further advanced by Knill and 
Laflamme (1997), Ekert and Macchiavello (1996),
Bennett {\em et. al.} (1996b). The latter paper describes 
the optimal 5-qubit code also
independently discovered by Laflamme {\em et. al.} (1996).
Gottesman (1996) and Calderbank {\em et. al.} (1997) 
discovered a general group-theoretic framework, introducing the important 
concept of the stabilizer, which also enabled many more codes to be found
(Calderbank {\em et. al.} 1996, Steane 1996cd). Quantum coding theory reached
a further level of maturity with the discovery by Shor and Laflamme (1997)
of a quantum analogue to the MacWilliams identities of classical coding
theory.

QEC uses networks of quantum gates and measurements, and at first is was 
not clear whether these networks had themselves to be perfect in order for 
the method to work. An important step forward was taken by Shor (1996) and 
Kitaev (1996) who showed how to make error correcting networks tolerant of 
errors within the network. In other words, such `fault tolerant' networks 
remove more noise than they introduce. Shor's methods were generalised by 
DiVincenzo and Shor (1996) and made more efficient by Steane (1997a,c). Knill 
and Laflamme (1996) introduced the idea of `concatenated' coding, which is 
a recursive coding method. It has the advantage of allowing arbitrarily 
long quantum computations as long as the noise per elementary operation is 
below a finite threshold, at the cost of inefficient use of quantum memory 
(so requiring a large computer). This threshold result was derived by 
several authors (Knill {\em et al} 1996, Aharonov and Ben-Or 1996, 
Gottesman {\em et. al.} 1996). Further fault tolerant methods are described
by Knill {\em et. al.} (1997), Gottesman (1997), Kitaev (1997).

The discovery of QEC was roughly simultaneous
with that of a related idea which also permits 
noise-free transmission of quantum states over a noisy quantum channel. 
This is the `entanglement purification' (Bennett {\em et. al.} 1996a, Deutsch 
{\em et. al.} 1996). The central idea here is for Alice to generate many 
entangled pairs of qubits, sending one of each pair down the noisy channel 
to Bob. Bob and Alice store their qubits, and perform simple parity
checking measurements: for example, Bob's performs {\sc xor} between
a given qubit and the next he receives, then measures just the target qubit.
Alice does the same on her qubits, and they compare results. If they agree, the 
unmeasured qubits are (by chance) closer than average to the desired state 
$\ket{00} + \ket{11}$. If they disagree, the qubits are rejected. 
By recursive use of such checks,
a few `good' entangled pairs are distilled out of the many noisy ones. Once in 
possession of a good entangled state, Alice and Bob can communicate by 
teleportation. A thorough discussion is given by Bennett {\em et. al.} (1996b). 

Using similar ideas, with important improvements, van Enk {\em et. al.} 
(1997) have recently shown how quantum information might be reliably 
transmitted between atoms in separated high-$Q$ optical cavities via imperfect 
optical fibres, using imperfect gate operations. 

I will now outline the main principles of QEC.

Let us write down the 
worst possible thing which could happen to a single qubit: a completely 
general interaction between a qubit and its environment is
  \begin{eqnarray}
\ket{e_i}\left(a\ket{0} + b \ket{1} \right) \! &\rightarrow& \!
a\left(c_{00}\ket{e_{00}}\ket{0} + c_{01}\ket{e_{01}} \ket{1}\right) 
\nonumber \\
&& \rule{-8ex}{0em}
 + b\left(c_{10}\ket{e_{10}}\ket{1} + c_{11}\ket{e_{11}} \ket{0} \right)
  \end{eqnarray}
where $\ket{e_{\ldots}}$ 
denotes states of the environment and $c_{\ldots}$ are coefficients
depending on the noise.
The first significant point is to notice 
that this general interaction can be written
  \beq
 \ket{e_i} \ket{\phi}
\rightarrow \left(\ket{e_I} I + \ket{e_X} X + \ket{e_Y} Y
+ \ket{e_Z} Z \right) \ket{\phi}      \lab{IXYZ}
  \eeq
where $\ket{\phi} = a\ket{0} + b\ket{1}$ is the initial state of the
qubit, and $\ket{e_I} = c_{00}\ket{e_{00}} + c_{10}\ket{e_{10}}$,
$\ket{e_X} = c_{01}\ket{e_{01}} + c_{11}\ket{e_{11}}$, and so on.
Note that these environment states are not necessarily normalised.
Eq. (\ref{IXYZ}) tells us that we have essentially three types
of error to correct on each qubit: $X$, $Y$ and $Z$ errors. These
are `bit flip' ($X$) errors, phase errors ($Z$) or both ($Y=XZ$).

Suppose our computer $q$ is to manipulate $k$ qubits of quantum
information. Let a general state of the $k$ qubits be $\ket{\phi}$.
We first make the computer larger, introducing a further $n-k$
qubits, initially in the state $\ket{0}$. Call the enlarged
system $qc$. An `encoding' operation
is performed: $E (\ket{\phi} \ket{0}) = \ket{\phi_E}$. Now, let
noise affect the $n$ qubits of $qc$. Without loss of generality, the
noise can be written as a sum of `error operators' $M$, where each
error operator is a tensor product of $n$ operators (one for
each qubit), taken from the set $\{I,X,Y,Z\}$. For example
$M = I_1 X_2 I_3 Y_4 Z_5 X_6 I_7$ for the case $n=7$. A general noisy
state is
  \beq
\sum_s \ket{e_s} M_s \ket{\phi_E}   \lab{noise}
  \eeq
Now we introduce even more qubits: a further $n-k$, prepared in
the state $\ket{0}_a$. This additional set is called an `ancilla'. 
For any given encoding $E$, there exists a {\em syndrome extraction}
operation $A$, operating on the joint system of $qc$ and $a$.
whose effect is $A (M_s \ket{\phi_E} \ket{0}_a)
= (M_s \ket{\phi_E}) \ket{s}_a$ $\forall \; M_s \in {\cal S}$.
The set $\cal S$ is the set of correctable errors, which depends
on the encoding. In the notation $\ket{s}_a$, $s$ is just a binary
number which indicates which error operator $M_s$ we are dealing
with, so the states $\ket{s}_a$ are mutually orthogonal.
Suppose for simplicity that the general noisy state (\ref{noise})
only contains $M_s \in \cal S$, then the joint state of
environment, $qc$ and $a$ after syndrome extraction is
  \beq
\sum_s \ket{e_s} \left( M_s \ket{\phi_E} \right) \ket{s}_a 
  \eeq
We now measure the ancilla state, and something rather wonderful
happens: the whole state collapses onto
$\ket{e_s} \left( M_s \ket{\phi_E} \right) \ket{s}_a$,
for some particular value of $s$. Now, instead of general noise,
we have just one particular error operator $M_s$ to worry about.
Furthermore, the measurement tells
us the value $s$ (the `error syndrome') from which we can deduce which 
$M_s$ we have! Armed with this knowledge, we apply $M_s^{-1}$
to $qc$ by means of a few quantum gates ($X$, $Z$ or $Y$), thus
producing the final state $\ket{e_s} \ket{\phi_E} \ket{s}_a$. In other
words, we have recovered the noise-free state of $qc$! The
final environment state is immaterial, and we can re-prepare the
ancilla in $\ket{0}_a$ for further use. 

The only assumption in the above was that the noise in eq. (\ref{noise}) only 
contains error operators in the correctable set $\cal S$. In practice, the 
noise includes both members and non-members of $\cal S$, and the important 
quantity is the probability that the state collapses onto a correctable one 
when the syndrome is extracted. It is here that the theory of 
error-correcting codes enters in: our task is to find encoding and extraction 
operations $E, A$ such that the set $\cal S$ of correctable errors includes 
all the errors most likely to occur. This is a very difficult problem. 

It is a general truth that to permit efficient
stabilization against noise, we have to know something about the noise we 
wish to suppress. The most obvious quasi-realistic assumption is that of 
uncorrelated stochastic noise. That is, at a given time or place the noise 
might have any effect, but the effects on different qubits, or on the same 
qubit at different times, are uncorrelated. This is the quantum equivalent 
of the binary symetric channel, section \ref{s:bin}. 
By assuming uncorrelated stochastic noise we can place all possible error
operators $M$ in a heirarchy of probability: those affecting few qubits
(i.e. only a few terms in the tensor product are different from $I$)
are most likely, while those affecting many qubits at once are
unlikely. Our aim will be to find quantum error correcting codes
(QECCs) such that all errors affecting up to $t$ qubits will be 
correctable. Such a QECC is termed a `$t$-error correcting code'.

The simplest code construction (that discovered by Calderbank and Shor and 
Steane) goes as follows. First we notice that a classical error correcting 
code, such as the Hamming code shown in table 1, can be used to correct 
$X$ errors. The proof relies on eq. (\ref{syn}) which permits the
syndrome extraction $A$ to produce an ancilla state $\ket{s}$ which
depends only on the error $M_s$ and not on the computer's
state $\ket{\phi}$.
This suggests that we store $k$ quantum bits by means of the 
$2^k$ mutually orthogonal $n$-qubit states $\ket{i}$, where the binary 
number $i$ is a member of a classical error correcting code $C$, see 
section \ref{s:ecc}. This will not allow correction of $Z$ errors, however.
Observe that since $Z = HXH$, the correction 
of $Z$ errors is equivalent to rotating the state of each qubit by $H$, 
correcting $X$ errors, and rotating back again. This rotation is called a 
Hadamard transform; it is just a change in basis. The next ingredient is
to notice the following special property (Steane 1996a):
  \beq
\tilde{H} \sum_{i \in C} \ket{i} = 
\frac{1}{\sqrt{2^k}} \sum_{j \in C^{\perp} } \ket{j}  \lab{CC}
  \eeq
where $\tilde{H} \equiv H_1 H_2 H_3 \cdots H_n$.
In words, this says that if we make a quantum state by
superposing all the members of a classical error correcting code $C$, then
the Hadamard-transformed state is just a superposition of all the members
of the dual code $C^{\perp}$. From this it follows, after some 
further steps, that it is possible to correct
both $X$ and $Z$ errors (and therefore also $Y$ errors) if we use
quantum states of the form given in eq. (\ref{CC}), as long as both
$C$ and $C^{\perp}$ are good classical error correcting codes,
i.e. both have good correction abilities.

The simplest QECC constructed by the above recipe requires $n=7$
qubits to store a single ($k=1$) qubit of useful quantum information.
The two orthogonal states required to store the information are built
from the Hamming code shown in table 1:
  \begin{eqnarray}
\ket{0_E} \!\!\!\! &\equiv& \!\!\!\! 
  \ket{0000000} + \ket{1010101} + \ket{0110011} + \ket{1100110}
\nonumber \\
          && \rule{-9ex}{0em}
+ \ket{0001111} + \ket{1011010} + \ket{0111100} + \ket{1101001} 
\label{0E} \\
\ket{1_E} \!\!\!\! &\equiv& \!\!\!\!
  \ket{1111111} + \ket{0101010} + \ket{1001100} + \ket{0011001}
\nonumber \\
          && \rule{-9ex}{0em}
+ \ket{1110000} + \ket{0100101} + \ket{1000011} + \ket{0010110}
\label{1E}
  \end{eqnarray}
Such a QECC has the following remarkable property. Imagine I store a general 
(unknown) state of a single qubit into a spin state $a\ket{0_E} + b 
\ket{1_E}$ of 7 spin-half particles. I then allow you to do anything at all 
to any one of the 7 spins. I could nevertheless extract my original qubit 
state {\em exactly}. Therefore the large perturbation you introduced
did nothing at all to the stored quantum information!

More powerful QECCs can be obtained from more powerful classical codes, and 
there exist quantum code constructions more efficient than the one 
just outlined. Suppose we store $k$ qubits into $n$. There are $3n$ ways for a 
single qubit to be in error, since the error might be one of $X$, $Y$ or $Z$. 
The number of syndrome bits is $n-k$, so if every single-qubit error,
and the error-free case, is to have 
a different syndrome, we require $2^{n-k} \ge 3n + 1$. For $k=1$ this lower 
limit is filled exactly by $n=5$ and indeed such a 5-qubit
single-error correcting code exists (Laflamme {\em et. al.} 1996,
Bennett {\em et. al.} 1996b).

More generally,
the remarkable fact is that for fixed $k/n$, codes exist for which $t/n$ is 
bounded from below as $n \rightarrow \infty$ (Calderbank and Shor 1995, 
Steane 1996b, Calderbank {\em et. al.} 1997). This leads to a quantum 
version of Shannon's theorem (section \ref{s:ecc}), though an exact 
definition of the capacity of a quantum channel remains unclear (Schumacher 
and Nielsen 1996, Barnum {\em et. al.} 1996,
Lloyd 1997, Bennett {\em et. al.} 1996b, Knill and 
Laflamme 1997a). For finite $n$, the probability that the noise produces 
uncorrectable errors scales roughly as $(n \epsilon)^{t+1}$, where 
$\epsilon \ll 1$ is the probability of an arbitrary error on each qubit. 
This represents an extremely powerful noise suppression. We need to be able 
to reduce $\epsilon$ to a sufficiently small value by passive means, and 
then QEC does the rest. For example, consider the case $\epsilon \simeq 
0.001$. With $n=23$ there exisits a code correcting all $t=3$-qubit errors 
(Golay 1949, Steane 1996c). The probability that uncorrectable noise occurs 
is $\sim 0.023^4 \simeq 3 \times 10^{-7}$, thus the noise is suppressed by 
more than three orders of magnitude. 

So far I have described QEC as if the ancilla and the many quantum gates 
and measurements involved were themselves noise-free. Obviously we must 
drop this assumption if we want to form a realistic impression of what 
might be possible in quantum computing. Shor (1996) and Kitaev (1996) 
discovered ways in which all the required operations can be arranged so 
that the correction suppresses more noise than it introduces. 
The essential ideas are to verify states wherever possible, to restrict the 
propagation of errors by careful network design, and to repeat the syndrome 
extraction: for each group of qubits $qc$, the syndrome is extracted several 
times and $qc$ is only corrected once $t+1$ mutually consistent syndromes are 
obtained. Fig. 14 illustrates a fault-tolerant syndrome 
extraction network, i.e. one which restricts the propagation of errors. Note 
that $a$ is verified before it is used, and each qubit in $qc$ only interacts 
with one qubit in $a$. 

In fault-tolerant computing, we cannot apply arbitrary rotations
of a logical qubit, eq. (\ref{V}), in a single step. However, particular
rotations through irrational angles can be carried out, and thus
general rotations are generated to an arbitrary degree of precision
through repetition. Note that the set of computational
gates is now discrete rather than continuous.

Recently the requirements for reliable quantum computing using fault-tolerant 
QEC have been estimated (Preskill 1997, Steane 1997c). They are formidable. 
For example, a computation beyond the capabilities of the best classical 
computers might require $1000$ qubits and $10^{10}$ quantum gates. Without 
QEC, this would require a noise level of order $10^{-13}$ per qubit per gate, 
which we can rule out as impossible. With QEC, the computer would have to be 
made ten or perhaps one hundred times larger, and many thousands of gates 
would be involved in the correctors for each elementary step in the 
computation. However, much more noise could be tolerated: up to about 
$10^{-5}$ per qubit per gate (i.e. in any of the gates, including those in 
the correctors) (Steane 1997c). This is daunting but possible. 

The error correction methods briefly described here are not the only type 
possible. If we know more about the noise, then humbler methods requiring 
just a few qubits can be quite powerful. Such a method was proposed by 
Cirac {\em et. al.} (1996) to deal with the principle noise source in an 
ion trap, which is changes of the motional state during gate operations. 
Also, some joint states of several qubits can have reduced noise if the 
environment affects all qubits together. For example the two states 
$\ket{01} \pm \ket{10}$ are unchanged by environmental coupling of the form 
$\ket{e_0} I_1 I_2 + \ket{e_1} X_1 X_2$. (Palma {\em et. al.} 1996, Chuang 
and Yamamoto 1997). Such states offer a calm eye within the storm of 
decoherence, in which quantum information can be manipulated with relative 
impunity. A practical computer would probably use a combination of methods. 

\section{Discussion}

The idea of `Quantum Computing' has fired many imaginations simply because 
the words themselves suggest something strange but powerful, as if the 
physicists have come up with a second revolution in information processing 
to herald the next millenium. This is a false impression. Quantum 
computing will not replace classical computing for similar reasons that 
quantum physics does not replace classical physics: no one ever consulted 
Heisenberg in order to design a house, and no one takes their car to be 
mended by a quantum mechanic. If large quantum computers are ever made, 
they will be used to address just those special tasks 
which benefit from quantum information processing. 

A more lasting reason to be excited about quantum computing is that it is 
a new and insightful way to think about the fundamental laws of physics. 
The quantum computing community remains fairly small at present, yet the 
pace of progress has been fast and accelerating in the last few years. The 
ideas of classical information theory seem to fit into quantum mechanics 
like a hand into a glove, giving us the feeling that we are uncovering 
something profound about Nature. Shannon's noiseless coding theorem
leads to Schumacher and Josza's quantum coding theorem and
the significance of the qubit as a useful measure of information. This 
enables us to keep track of quantum information, and to be confident
that it is independent of the details of the system in which it is
stored. This is necessary to underpin other concepts such as error correction
and computing. The classical theory of error correction leads to the 
discovery of quantum error correction. This allows a physical process 
previously thought to be impossible, namely the almost perfect recovery of 
a general quantum state, undoing even irreversible processes such as 
relaxation by spontaneous emission. For example, during a long 
error-corrected quantum computation, using fault-tolerant methods,
every qubit in the computer
might decay a million times and yet the coherence of the quantum
information be preserved.

Hilbert's questions regarding the logical structure of 
mathematics encourage us to ask a new type of question about the laws of 
physics. In looking at Schr\"odinger's equation, we can 
neglect whether it is describing an electron or a planet, and just ask 
about the state manipulations it permits. The language of information and 
computer science enables us to frame such questions. Even such a simple 
idea as the quantum gate, the cousin of the classical binary logic gate, 
turns out to be very useful, because it enables us to think clearly about 
quantum state manipulations which would otherwise seem extremely 
complicated or impractical. Such ideas open the way to the design of
quantum algorithms such as those of Shor, Grover and Kitaev. These
show that quantum mechanics allows information processing of a 
kind ruled out in classical physics. It relies on the propagation of a 
quantum state through a huge (exponentially large) number of dimensions of 
Hilbert space. The computation result arises from a controlled 
interference among many computational paths, which even after we have 
examined the mathematical description, still seems wonderful and 
surprising.

The intrinsic difficulty of quantum computation lies in the sensitivity of 
large-scale interference to noise and imprecision. A point often raised 
against the quantum computer is that it is essentially an analogue rather 
than a digital device, and has many limitations as a result. This is 
a misconception. It is true that any quantum system has a continuous state 
space, but so has any classical system, including the circuits of a digital 
computer. The fault-tolerant methods used to permit error correction in a 
quantum computer restrict the set of quantum gates to a discrete set, 
therefore the `legal' states of the quantum computer are discrete, just as in 
a classical digital computer. The really important difference between 
analogue and digital computing is that to increase the precision of a result 
arrived at by analogue means, one must re-engineer the whole computer, 
whereas with digital methods one need merely increase the number of bits and 
operations. The fault-tolerant quantum computer has more in common with a 
digital than an analogue device. 

Shor's algorithm for the factorisation problem stimulated a lot of 
interest in part because of the connection with data encryption. However, 
I feel that the significance of Shor's algorithm is not primarily in its 
possible use for factoring large integers in the distant future. Rather, 
it has acted as a stimulus to the field, proving the existence of a 
powerful new type of computing made possible by controlled quantum 
evolution, and exhibiting some of the new methods. At present, the most 
practically significant achievement in the general area of quantum 
information physics is not in computing at all, but in quantum key 
distribution.

The title `quantum computer' will remain a misnomer for any experimental 
device realised in the next twenty years. It is an abuse of language to call 
even a pocket calculator a `computer', because the word has come to be 
reserved for general-purpose machines which more or less realise Turing's 
concept of the Universal Machine. The same ought to be true for quantum 
computers if we do not want to mislead people. However, small quantum 
information processors may serve useful roles. For example, concepts learned 
from quantum information theory may permit the discovery of useful new 
spectroscopic methods in nuclear magnetic resonance. Quantum key 
distribution could be made more secure, and made possible over larger 
distances, if small `relay stations' could be built which applied 
purification or error correction methods. The relay station could be an ion 
trap combined with a high-$Q$ cavity, which is realisable with current 
technology. It will surely not be long before a quantum state is 
teleported from one laboratory to another, a very exciting prospect. 

The great intrinsic value of a large quantum computer is offset by the 
difficulty of making one. However, few would argue that this prize
does not at least merit a lot of effort to find out just
how unattainable, or hopefully attainable, it is. One of the 
chief uses of a processor which could manipulate a few quantum bits may be to 
help us better understand decoherence in quantum mechanics. This will 
be amenable to experimental investigation during the next few years: rather
than waiting in hope, there is useful work to be done now. 

On the theoretical side, there are two major open questions: the nature of 
quantum algorithms, and the limits on reliability of quantum computing. It 
is not yet clear what is the essential nature of quantum computing, and 
what general class of computational problem is amenable to efficient 
solution by quantum methods. Is there a whole mine of useful quantum 
algorithms waiting to be delved, or will the supply dry up with the few 
nuggets we have so far discovered? Can significant computational
power be achieved with less than 100 qubits? This is by no means
ruled out, since it is hard to simulate even 20 qubits by classical
means. Concerning reliability, great progress 
has been made, so that we can now be cautiously optimistic that quantum 
computing is not an impossible dream. We can identify requirements {\em 
sufficient} to guarantee reliable computing, involving for example 
uncorrelated stochastic noise of order $10^{-5}$ per gate, and a quantum 
computer a hundred times larger than the logical machine embedded within 
it. However, can quantum decoherence be relied upon to have the properties 
assumed in such an estimate, and if not then can error correction methods 
still be found? Conversely, once we know more about the noise,
it may be possible to identify considerably less taxing requirements
for reliable computing.

To conclude with, I would like to propose a more wide-ranging 
theoretical task: to arrive at a set of principles like energy and momentum 
conservation, but which apply to information, and from which much
of quantum mechanics could be derived. Two tests of such
ideas would be whether the EPR-Bell correlations thus became transparent,
and whether they rendered obvious the proper use of terms
such as `measurement' and `knowledge'.

I hope that quantum information physics will be recognised as a valuable 
part of fundamental physics. 
The quest to bring together Turing machines, information, number theory and 
quantum physics is for me, and I hope will be for readers of this review, one 
of the most fascinating cultural endeavours one could have the good fortune 
to encounter. 

I thank the Royal Society and St Edmund Hall, Oxford, for their
support.

Abrams D S and Lloyd S 1997
Simulation of many-body Fermi systems on a universal quantum
computer (preprint quant-ph/9703054)

Aharonov D and Ben-Or M 1996
Fault-tolerant quantum computation with constant error
(preprint quant-ph/9611025)

Aspect A, Dalibard J and Roger G 1982
Experimental test of Bell's inequalities using time-varying analysers,
Phys. Rev. Lett. {\bf 49}, 1804-1807

Aspect A 1991
Testing Bell's inequalities,
Europhys. News. {\bf 22}, 73-75

Barenco A 1995
A universal two-bit gate for quantum computation,
Proc. R. Soc. Lond. A {\bf 449} 679-683

Barenco A and Ekert A K 1995
Dense coding based on quantum entanglement,
J. Mod. Opt. {\bf 42} 1253-1259

Barenco A, Deutsch D, Ekert E and Jozsa R 1995a
Conditional quantum dynamics and quantum gates,
Phys. Rev. Lett. {\bf 74} 4083-4086

Barenco A, Bennett C H, Cleve R, DiVincenzo D P, Margolus N,
Shor P, Sleator T, Smolin J A and Weinfurter H 1995b
Elementary gates for quantum computation,
Phys. Rev. A {\bf 52}, 3457-3467

Barenco A 1996
Quantum physics and computers,
Contemp. Phys. {\bf 37} 375-389

Barenco A, Ekert A, Suominen K A and Torma P 1996
Approximate quantum Fourier transform and decoherence,
Phys. Rev. A {\bf 54}, 139-146

Barenco A, Brun T A, Schak R and Spiller T P 1997
Effects of noise on quantum error correction algorithms,
Phys. Rev. A {\bf 56} 1177-1188

Barnum H, Fuchs C A, Jozsa R and Schumacher B 1996
A general fidelity limit for quantum channels,
Phys. Rev. A {\bf 54} 4707-4711

Beckman D, Chari A, Devabhaktuni S and Preskill J 1996
Efficient networks for quantum factoring,
Phys. Rev. A {\bf 54}, 1034-1063

Bell J S 1964
On the Einstein-Podolsky-Rosen paradox,
Physics {\bf 1} 195-200

Bell J S 1966
On the problem of hidden variables in quantum theory,
Rev. Mod. Phys. {\bf 38} 447-52

Bell J S 1987
{\em Speakable and unspeakable in quantum mechanics}
(Cambridge University Press)

Benioff P, 1980
J. Stat. Phys. {\bf 22} 563

Benioff P 1982a
Quantum mechanical hamiltonian models of Turing machines,
J. Stat. Phys. {\bf 29} 515-546

Benioff P 1982b
Quantum mechanical models of Turing machines that
dissipate no energy,
Phys. Rev. Lett. {\bf 48} 1581-1585

Bennett C H 1973 
Logical reversibility of computation,
IBM J. Res. Develop. {\bf 17} 525-532

Bennett C H 1982 
Int. J. Theor. Phys. {\bf 21} 905

Bennett C H, Brassard G, Briedbart S and Wiesner S 1982
Quantum cryptography, or unforgeable subway tokens,
in {\em Advances in Cryptology: Proceedings of Crypto '82}
(Plenum, New York) 267-275

Bennett C H and Brassard G 1984
Quantum cryptography: public key distribution and coin tossing,
in {\em Proc. IEEE Conf. on Computers, Syst. and Signal Process.}
175-179

Bennett C H and Landauer R 1985
The fundamental physical limits of computation,
Scientific American, July 38-46

Bennett C H 1987
Demons, engines and the second law,
Scientific American vol {\bf 257} no. 5 (November) 88-96

Bennett C H and Brassard G 1989,
SIGACT News {\bf 20}, 78-82

Bennett C H and Wiesner S J 1992
Communication via one- and two-particle operations on
Einstein-Podolsky-Rosen states,
Phys. Rev. Lett. {\bf 69}, 2881-2884

Bennett C H, Bessette F, Brassard G, Savail L and Smolin J 1992
Experimental quantum cryptography,
J. Cryptology {\bf 5}, 3-28

Bennett C H, Brassard G, Cr\'epeau C, Jozsa R, Peres A and
Wootters W K 1993
Teleporting an unknown quantum state via dual classical and
Einstein-Podolsky-Rosen channels,
Phys. Rev. Lett. {\bf 70} 1895-1898

Bennett C H 1995
Quantum information and computation,
Phys. Today {\bf 48 10} 24-30

Bennett C H, Brassard G, Popescu S,
Schumacher B, Smolin J A and Wootters W K 1996a
Purification of noisy entanglement and faithful
teleportation via noisy channels,
Phys. Rev. Lett. {\bf 76} 722-725

Bennett C H, DiVincenzo D P, Smolin J A
and Wootters W K 1996b
Mixed state entanglement and quantum error correction,
Phys. Rev. A {\bf 54} 3825

Bennett C H, Bernstein E, Brassard G and Vazirani U 1997
Strengths and weaknesses of quantum computing,
(preprint quant-ph/9701001)

Berman G P, Doolen G D, Holm D D, Tsifrinovich V I 1994
Quantum computer on a class of one-dimensional Ising
systems,
Phys. Lett. {\bf 193} 444-450

Bernstein E and Vazirani U 1993
Quantum complexity theory,
in {\em Proc. of the 25th Annual ACM Symposium on Theory of Computing}
(ACM, New York) 11-20

Berthiaume A, Deutsch D and Jozsa R 1994
The stabilisation of quantum computation, in
{\em Proceedings of the Workshop on Physics and
Computation, PhysComp 94} 60-62 Los Alamitos: IEEE
Computer Society Press

Berthiaume A and Brassard G 1992a
The quantum challenge to structural complexity theory,
in {\em Proc. of the Seventh Annual Structure in Complexity Theory
Conference}
(IEEE Computer Society Press, Los Alamitos, CA) 132-137 

Berthiaume A and Brassard G 1992b
Oracle quantum computing,
in {\em Proc. of the Workshop on Physics of Computation: PhysComp '92}
(IEEE Computer Society Press, Los Alamitos, CA) 60-62

Boghosian B M and Taylor W 1997
Simulating quantum mechanics on a quantum computer
(preprint quant-ph/9701019)

Bohm D 1951
{\em Quantum Theory}
(Englewood Cliffs, N. J.)

Bohm D and Aharonov Y 1957
Phys. Rev. {\bf 108} 1070

Boyer M, Brassard G, Hoyer P and Tapp A
Tight bounds on quantum searching
(preprint quant-ph/9605034)

Brassard G 1997
Searching a quantum phone book, 
Science {\bf 275} 627-628

Brassard G and Crepeau C 1996
SIGACT News {\bf 27} 13-24

Braunstein S L, Mann A and Revzen M 1992
Maximal violation of Bell inequalities for mixed states,
Phys. Rev. Lett. {\bf 68}, 3259-3261

Braunstein S L and Mann A 1995
Measurement of the Bell operator and quantum teleportation,
Phys. Rev. A {\bf 51}, R1727-R1730

Brillouin L 1956,
{\em Science and information theory}
(Academic Press, New York)

Brune M, Nussenzveig P, Schmidt-Kaler F, Bernardot F, Maali A,
Raimond J M and Haroche S 1994
From Lamb shift to light shifts: vacuum and subphoton cavity fields
measured by atomic phase sensitive detection,
Phys. Rev. Lett. {\bf 72}, 3339-3342

Calderbank A R and Shor P W 1996
Good quantum error-correcting codes exist,
Phys. Rev. A {\bf 54} 1098-1105

Calderbank A R, Rains E M, Shor P W and Sloane N J A 1996
Quantum error correction via codes over $GF(4)$
(preprint quant-ph/9608006)

Calderbank A R, Rains E M, Shor P W and Sloane N J A 1997
Quantum error correction and orthogonal geometry,
Phys. Rev. Lett. {\bf 78} 405-408

Caves C M 1990
Quantitative limits on the ability of a Maxwell Demon
to extract work from heat,
Phys. Rev. Lett. {\bf 64} 2111-2114

Caves C M, Unruh W G and Zurek W H 1990
comment,
Phys. Rev. Lett. {\bf 65} 1387

Chuang I L, Laflamme R, Shor P W and Zurek W H 1995
Quantum computers, factoring, and decoherence,
Science {\bf 270} 1633-1635

Chuang I L and Yamamoto 1997
Creation of a persistent qubit using error correction
Phys. Rev. A {\bf 55}, 114-127

Church A 1936
An unsolvable problem of elementary number theory,
Amer. J. Math. {\bf 58} 345-363

Cirac J I and Zoller P 1995
Quantum computations with cold trapped ions,
{Phys. Rev. Lett.} {\bf 74} 4091-4094

Cirac J I, Pellizari T and Zoller P 1996
Enforcing coherent evolution in dissipative quantum dynamics,
Science {\bf 273}, 1207

Cirac J I, Zoller P, Kimble H J and Mabuchi H 1997
Quantum state transfer and entanglement distribution among distant
nodes of a quantum network,
Phys. Rev. Lett. {\bf 78}, 3221

Clauser J F, Holt R A, Horne M A and Shimony A 1969
Proposed experiment to test local hidden-variable theories,
Phys. Rev. Lett. {\bf 23} 880-884

Clauser J F and Shimony A 1978
Bell's theorem: experimental tests and implications,
Rep. Prog. Phys. {\bf 41} 1881-1927

Cleve R and DiVincenzo D P 1996
Schumacher's quantum data compression as a quantum computation,
Phys. Rev. A {\bf 54} 2636

Coppersmith D 1994
An approximate Fourier transform useful in quantum factoring,
IBM Research Report RC 19642

Cory D G, Fahmy A F and Havel T F 1996
Nuclear magnetic resonance spectroscopy: an experimentally
accessible paradigm for quantum computing, in {\em Proc. of the 4th
Workshop on Physics and Computation} (Complex Systems Institute,
Boston, New England)

Crandall R E 1997
The challenge of large numbers,
Scientific American February 59-62

Deutsch D 1985
Quantum theory, the Church-Turing principle and the
universal quantum computer,
Proc. Roy. Soc. Lond. A {\bf 400} 97-117

Deutsch D 1989
Quantum computational networks,
Proc. Roy. Soc. Lond. A {\bf 425} 73-90

Deutsch D and Jozsa R 1992
Rapid solution of problems by quantum computation,
Proc. Roy. Soc. Lond A {\bf 439} 553-558

Deutsch D, Barenco A \& Ekert A 1995
Universality in quantum computation,
{ Proc. R. Soc. Lond.} A {\bf 449} 669-677

Deutsch D, Ekert A, Jozsa R, Macchiavello C,
Popescu S, and Sanpera A 1996
Quantum privacy amplification and the security of
quantum cryptography over noisy channels,
Phys. Rev. Lett. {\bf 77} 2818

Diedrich F, Bergquist J C, Itano W M and.
Wineland D J 1989
Laser cooling to the zero-point energy of motion,
Phys. Rev. Lett. {\bf 62} 403

Dieks D 1982
Communication by electron-paramagnetic-resonance devices,
Phys. Lett. A {\bf 92} 271

DiVincenzo D P 1995a
Two-bit gates are universal for quantum computation,
Phys. Rev. A {\bf 51} 1015-1022

DiVincenzo D P 1995b
Quantum computation,
Science {\bf 270} 255-261

DiVincenzo D P and Shor P W 1996
Fault-tolerant error correction with efficient quantum codes,
Phys. Rev. Lett. {\bf 77} 3260-3263

Einstein A, Rosen N and Podolsky B 1935
Phys. Rev. {\bf 47}, 777

Ekert A 1991
Quantum cryptography based on Bell's theorem
Phys. Rev. Lett. {\bf 67}, 661-663

Ekert A and Jozsa R 1996
Quantum computation and Shor's factoring algorithm,
Rev. Mod. Phys. {\bf 68} 733

Ekert A and Macchiavello C 1996
Quantum error correction for communication,
Phys. Rev. Lett. {\bf 77} 2585-2588

Ekert A 1997
From quantum code-making to quantum code-breaking,
(preprint quant-ph/9703035)

van Enk S J, Cirac J I and Zoller P 1997
Ideal communication over noisy channels: a quantum optical implementation,
Phys. Rev. Lett. {\bf 78}, 4293-4296

Feynman R P 1982
Simulating physics with computers,
Int. J. Theor. Phys. {\bf 21} 467-488

Feynman R P 1986
Quantum mechanical computers,
Found. Phys. {\bf 16} 507-531;
see also Optics News February 1985, 11-20.

Fredkin E and Toffoli T 1982
Conservative logic,
Int. J. Theor. Phys. {\bf 21} 219-253

Gershenfeld N A and Chuang I L 1997
Bulk spin-resonance quantum computation,
Science {\bf 275} 350-356

Glauber R J 1986, 
in {\em Frontiers in Quantum Optics}, Pike E R and Sarker S, eds
(Adam Hilger, Bristol)

Golay M J E 1949 
Notes on digital coding,
Proc. IEEE {\bf 37} 657

Gottesman D 1996
Class of quantum error-correcting codes saturating the quantum Hamming
bound,
Phys. Rev. A {\bf 54}, 1862-1868

Gottesman D 1997
A theory of fault-tolerant quantum computation
(preprint quant-ph 9702029)

Gottesman D, Evslin J, Kakade S and Preskill J 1996
(to be published)

Greenberger D M, Horne M A and Zeilinger A 1989
Going beyond Bell's theorem,
in {\em Bell's theorem, quantum theory and conceptions of the universe},
Kafatos M, ed, (Kluwer Academic, Dordrecht) 73-76

Greenberger D M, Horne M A, Shimony A and Zeilinger A 1990
Bell's theorem without inequalities,
Am. J. Phys. {\bf 58}, 1131-1143

Grover L K 1997
Quantum mechanics helps in searching for a needle in a haystack,
Phys. Rev. Lett. {\bf 79}, 325-328

Hamming R W 1950 
Error detecting and error correcting codes,
Bell Syst. Tech. J. {\bf 29} 147

Hamming R W 1986
{\em Coding and information theory}, 2nd ed,
(Prentice-Hall, Englewood Cliffs)

Hardy G H and Wright E M 1979
{\em An introduction to the theory of numbers}
(Clarendon Press, Oxford)

Haroche S and Raimond J-M 1996
Quantum computing: dream or nightmare?
Phys. Today August 51-52

Hellman M E 1979
The mathematics of public-key cryptography,
Scientific American {\bf 241} August 130-139

Hill R 1986
{\em A first course in coding theory}
(Clarendon Press, Oxford)

Hodges A 1983
{\em Alan Turing: the enigma}
(Vintage, London)

Hughes R J, Alde D M, Dyer P, Luther G G, Morgan G L ans Schauer M 1995
Quantum cryptography,
Contemp. Phys. {\bf 36} 149-163

J. Mod. Opt. {\bf 41}, no 12 1994
Special issue: quantum communication 

Jones D S 1979
{\em Elementary information theory}
(Clarendon Press, Oxford)

Jozsa R and Schumacher B 1994
A new proof of the quantum noiseless coding theorem,
J. Mod. Optics {\bf 41} 2343

Jozsa R 1997a
Entanglement and quantum computation,
appearing in {\em Geometric issues in the foundations of science},
Huggett S {\em et. al.}, eds, (Oxford University Press)

Jozsa R 1997b
Quantum algorithms and the Fourier transform,
submitted to {\em Proc. Santa Barbara conference on quantum coherence
and decoherence} (preprint quant-ph/9707033)

Keyes R W and Landauer R 1970
IBM J. Res. Develop. {\bf 14}, 152

Keyes R W 1970
Science {\bf 168}, 796

Kholevo A S 1973
Probl. Peredachi Inf {\bf 9}, 3; Probl. Inf. Transm. (USSR) {\bf 9}, 177

Kitaev A Yu 1995
Quantum measurements and the Abelian stablizer problem,
(preprint quant-ph/9511026)

Kitaev A Yu 1996
Quantum error correction with imperfect gates
(preprint)

Kitaev A Yu 1997
Fault-tolerant quantum computation by anyons
(preprint quant-ph/9707021)

Knill E and Laflamme R 1996
Concatenated quantum codes
(preprint quant-ph/9608012)

Knill E, Laflamme R and Zurek W H 1996
Accuracy threshold for quantum computation,
(preprint quant-ph/9610011)

Knill E and Laflamme R 1997
A theory of quantum error-correcting codes,
Phys. Rev. A {\bf 55} 900-911

Knill E, Laflamme R and Zurek W H 1997
Resilient quantum computation: error models and thresholds
(preprint quant-ph/9702058)

Knuth D E 1981
{\em The Art of Computer Programming, Vol. 2: Seminumerical Algorithms},
2nd ed (Addison-Wesley).

Kwiat P G, Mattle K, Weinfurter H, Zeilinger A, Sergienko A and Shih Y 1995
New high-intensity source of polarisation-entangled photon pairs
Phys. Rev. Lett. {\bf 75}, 4337-4341

Laflamme R, Miquel C, Paz J P and Zurek W H 1996
Perfect quantum error correcting code,
Phys. Rev. Lett. {\bf 77}, 198-201

Landauer R 1961 IBM J. Res. Dev. {\bf 5} 183

Landauer R 1991
Information is physical,
Phys. Today May 1991 23-29

Landauer R 1995
Is quantum mechanics useful?
{Philos. Trans. R. Soc. London Ser.} A. {\bf 353} 367-376

Landauer R 1996
The physical nature of information,
Phys. Lett. A {\bf 217} 188

Lecerf Y 1963
Machines de Turing r\'eversibles . R\'ecursive insolubilit\'e
en $n \in N$ de l'equation $u=\theta^n u$, o\`u $\theta$ est un
isomorphisme de codes,
C. R. Acad. Francaise Sci. {\bf 257}, 2597-2600

Levitin L B 1987
in {\em Information Complexity and Control in Quantum Physics},
Blaquieve A, Diner S, Lochak G, eds (Springer, New York) 15-47

Lidar D A and Biham O 1996
Simulating Ising spin glasses on a quantum computer
(preprint quant-ph/9611038)

Lloyd S 1993
A potentially realisable quantum computer,
Science {\bf 261} 1569; see also Science {\bf 263} 695 (1994).

Lloyd S 1995
Almost any quantum logic gate is universal,
Phys. Rev. Lett. {\bf 75}, 346-349

Lloyd S 1996
Universal quantum simulators,
Science {\bf 273} 1073-1078

Lloyd S 1997
The capacity of a noisy quantum channel,
Phys. Rev. A {\bf 55} 1613-1622

Lo H-K and Chau H F 1997
Is quantum bit commitment really possible?,
Phys. Rev. Lett. {\bf 78} 3410-3413

Loss D and DiVincenzo D P 1997
Quantum Computation with Quantum Dots,
submitted to Phys. Rev. A (preprint quant-ph/9701055)

MacWilliams F J and Sloane N J A 1977
{\em The theory of error correcting codes},
(Elsevier Science, Amsterdam)

Mattle K, Weinfurter H, Kwiat P G and Zeilinger A 1996
Dense coding in experimental quantum communication,
Phys. Rev. Lett. {\bf 76}, 4656-4659.

Margolus N 1986
Quantum computation,
Ann. New York Acad. Sci. {\bf 480} 487-497

Margolus N 1990 
Parallel Quantum Computation,
in {\em Complexity, Entropy and the Physics of
Information, Santa Fe Institute Studies in the Sciences
of Complexity,} vol VIII p. 273
ed Zurek W H (Addison-Wesley)

Maxwell J C 1871
{\em Theory of heat} (Longmans, Green and Co, London)

Mayers D 1997
Unconditionally secure quantum bit commitment is impossible,
Phys. Rev. Lett. {\bf 78} 3414-3417

Menezes A J, van Oorschot P C and Vanstone S A 1997
{\em Handbook of applied cryptography}
(CRC Press, Boca Raton)

Mermin N D 1990
What's wrong with these elements of reality?
Phys. Today (June) 9-11

Meyer D A 1996
Quantum mechanics of lattice gas automata I: one particle plane
waves and potentials,
(preprint quant-ph/9611005)

Minsky M L 1967 
{\em Computation: Finite and Infinite Machines}
(Prentice-Hall, Inc., Englewood Cliffs, N. J.; also London 1972)

Miquel C, Paz J P and Perazzo 1996
Factoring in a dissipative quantum computer
Phys. Rev. A {\bf 54} 2605-2613

Miquel C, Paz J P and Zurek W H 1997
Quantum computation with phase drift errors,
Phys. Rev. Lett. {\bf 78} 3971-3974

Monroe C, Meekhof D M, King B E, Jefferts S R,
Itano W M, Wineland D J and Gould P 1995a
Resolved-sideband Raman cooling of a bound atom to the
3D zero-pointenergy,
Phys. Rev. Lett. {\bf 75} 4011-4014

Monroe C, Meekhof D M, King B E, Itano W M
and Wineland D J 1995b
Demonstration of a universal quantum logic gate,
Phys. Rev. Lett. {\bf 75} 4714-4717

Myers J M 1997
Can a universal quantum computer be fully quantum?
Phys. Rev. Lett. {\bf 78}, 1823-1824

Nielsen M A and Chuang I L 1997
Programmable quantum gate arrays,
Phys. Rev. Lett. {\bf 79}, 321-324

Palma G M, Suominen K-A \& Ekert A K 1996
Quantum computers and dissipation,
Proc. Roy. Soc. Lond. A {\bf 452} 567-584

Pellizzari T, Gardiner S A, Cirac J I and Zoller P 1995
Decoherence, continuous observation, and quantum
computing: A cavity QED model,
Phys. Rev. Lett. {\bf 75} 3788-3791

Peres A 1993
{\em Quantum theory: concepts and methods}
(Kluwer Academic Press, Dordrecht)

Phoenix S J D and Townsend P D 1995
Quantum cryptography: how to beat the code breakers using quantum mechanics,
Contemp. Phys. {\bf 36}, 165-195

Plenio M B and Knight P L 1996
Realisitic lower bounds for the factorisation time of large numbers
on a quantum computer,
Phys. Rev. A {\bf 53}, 2986-2990.

Polkinghorne J 1994
{\em Quarks, chaos and christianity}
(Triangle, London)

Preskill J 1997
Reliable quantum computers,
(preprint quant-ph/9705031)

Privman V, Vagner I D and Kventsel G 1997
Quantum computation in quantum-Hall systems,
(preprint, quant-ph/9707017)

Rivest R, Shamir A and Adleman L 1979
On digital signatures and public-key cryptosystems,
MIT Laboratory for Computer Science, Technical Report, MIT/LCS/TR-212

Schroeder M R 1984
{\em Number theory in science and communication}
(Springer-Verlag, Berlin Heidelberg)

Schumacher B 1995
Quantum coding,
Phys. Rev. A {\bf 51} 2738-2747

Schumacher B W and Nielsen M A 1996
Quantum data processing and error correction
Phys Rev A {\bf 54}, 2629

Shankar R 1980
{\em Principles of quantum mechanics}
(Plenum Press, New York)

Shannon C E 1948 
A mathematical theory of communication
Bell Syst. Tech. J. {\bf 27} 379; also p. 623

Shor P W 1994
Polynomial-time algorithms for prime factorization and
discrete logarithms on a quantum computer,
in {\em Proc. 35th Annual Symp. on Foundations of
Computer Science}, Santa Fe, IEEE Computer Society Press;
revised version 1995a preprint quant-ph/9508027

Shor P W 1995b
Scheme for reducing decoherence in quantum computer memory,
Phys. Rev. A {\bf 52} R2493-R2496

Shor P W 1996
Fault tolerant quantum computation,
in {\em Proc. 37th Symp. on Foundations of Computer
Science}, to be published. (Preprint quant-ph/9605011).

Shor P W and Laflamme R 1997
Quantum analog of the MacWilliams identities for classical coding theory,
Phys. Rev. Lett. {\bf 78} 1600-1602

Simon D 1994
On the power of quantum computation,
in {\em Proc. 35th Annual Symposium on Foundations of Computer Science}
(IEEE Computer Society Press, Los Alamitos) 124-134

Slepian D 1974 ed,
{\em Key papers in the development of information theory}
(IEEE Press, New York)

Spiller T P 1996
Quantum information processing: cryptography, 
computation and teleportation,
Proc. IEEE {\bf 84}, 1719-1746

Steane A M 1996a
Error correcting codes in quantum theory,
Phys. Rev. Lett. {\bf 77} 793-797

Steane A M 1996b
Multiple particle interference and quantum error correction,
Proc. Roy. Soc. Lond. A {\bf 452} 2551-2577

Steane A M 1996c
Simple quantum error-correcting codes,
Phys. Rev. A {\bf 54}, 4741-4751

Steane A M 1996d
Quantum Reed-Muller codes,
submitted to IEEE Trans. Inf. Theory (preprint quant-ph/9608026)

Steane A M 1997a
Active stabilisation, quantum computation, and quantum state sythesis,
Phys. Rev. Lett. {\bf 78}, 2252-2255

Steane A M 1997b
The ion trap quantum information processor,
Appl. Phys. B {\bf 64} 623-642

Steane A M 1997c
Space, time, parallelism and noise requirements for reliable
quantum computing
(preprint quant-ph/9708021)

Szilard L 1929 Z. Phys. {\bf 53} 840;
translated in Wheeler and Zurek (1983).

Teich W G, Obermayer K and Mahler G 1988
Structural basis of multistationary quantum systems II. Effective
few-particle dynamics,
Phys. Rev. B {\bf 37} 8111-8121

Toffoli T 1980
Reversible computing,
in {\em Automata, Languages and Programming}, Seventh Colloquium,
Lecture Notes in Computer Science, Vol. 84, de Bakker J W and
van Leeuwen J, eds, (Springer) 632-644

Turchette Q A, Hood C J, Lange W, Mabushi H and Kimble H J 1995
Measurement of conditional phase shifts for quantum logic,
Phys. Rev. Lett. {\bf 75} 4710-4713

Turing A M 1936
On computable numbers, with an application to the
Entschneidungsproblem,
Proc. Lond. Math. Soc. Ser. 2 {\bf 42}, 230 ); see also
Proc. Lond. Math. Soc. Ser. 2 {\bf 43}, 544 )

Unruh W G 1995
Maintaining coherence in quantum computers,
{Phys. Rev.} A {\bf 51} 992-997

Vedral V, Barenco A and Ekert A 1996 
Quantum networks for elementary arithmetic operations,
Phys. Rev. A {\bf 54} 147-153

Weinfurter H 1994
Experimental Bell-state analysis,
Europhys. Lett. {\bf 25} 559-564

Wheeler J A and Zurek W H, eds, 1983
{\em Quantum theory and measurement}
(Princeton Univ. Press, Princeton, NJ)

Wiesner S 1983
Conjugate coding,
SIGACT News {\bf 15} 78-88

Wiesner S 1996
Simulations of many-body quantum systems by a quantum computer
(preprint quant-ph/9603028)

Wineland D J, Monroe C, Itano W M, Leibfried D, King B, and
Meekhof D M 1997
Experimental issues in coherent quantum-state manipulation of trapped
atomic ions,
preprint, submitted to Rev. Mod. Phys.

Wooters W K and Zurek W H 1982
A single quantum cannot be cloned,
Nature {\bf 299}, 802

Zalka C 1996
Efficient simulation of quantum systems by quantum computers,
(preprint quant-ph/9603026)

Zbinden H, Gautier J D, Gisin N, Huttner B, Muller A, Tittle W 1997
Interferometry with Faraday mirrors for quantum cryptography,
Elect. Lett. {\bf 33}, 586-588

Zurek W H 1989 
Thermodynamic cost of computation, algorithmic complexity and the
information metric,
Nature {\bf 341} 119-124

\onecolumn

\centerline{\epsfbox{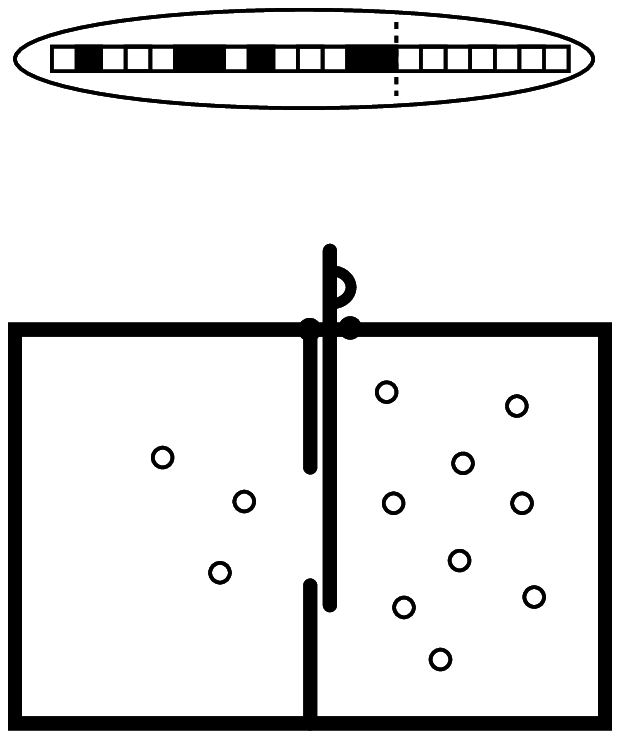}}

Fig. 1. Maxwell's demon. In this illustration the demon sets up a pressure
difference by only raising the partition when more gas molecules approach
it from the left than from the right. This can be done in a completely
reversible manner, as long as the demon's memory stores the random results
of its observations of the molecules. The demon's memory thus gets hotter. 
The irreversible step is not the acquisition of information, but the
loss of information if the demon later clears its memory.

\newpage\centerline{\epsfbox{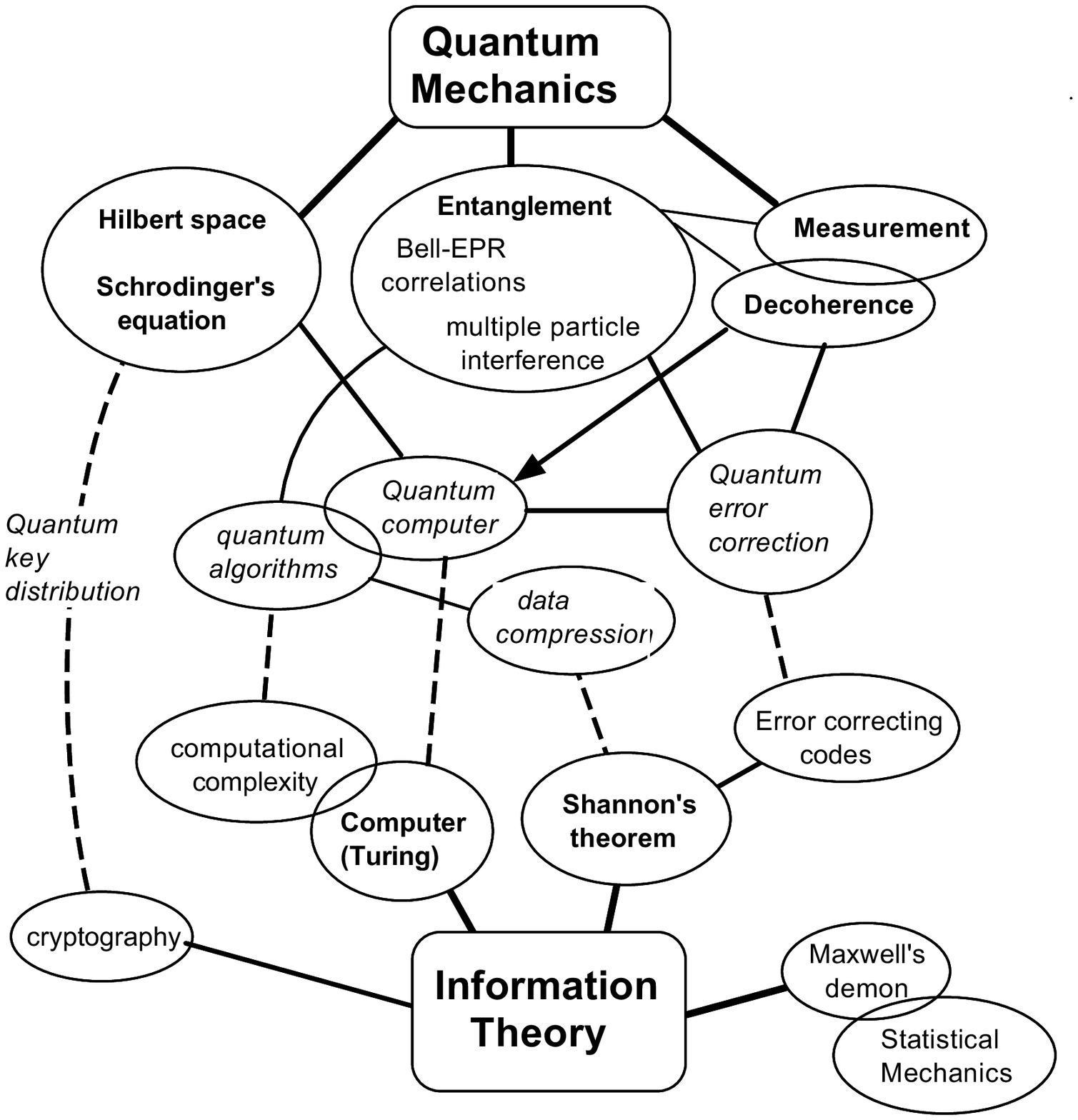}}
Fig. 2. Relationship between quantum mechanics and information theory.
This diagram is not intended to be a definitive statement, 
the placing of entries being to some extent subjective, but it
indicates many of the connections discussed in the article.

\newpage\centerline{\epsfbox{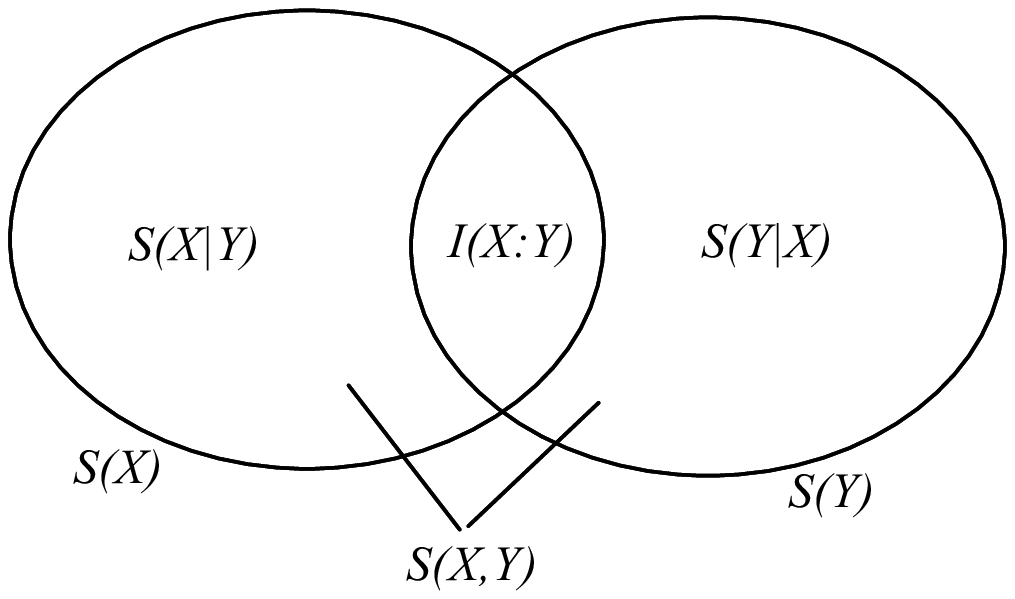}}
Fig. 3.  Relationship between various measures of classical information.

\newpage\centerline{\epsfbox{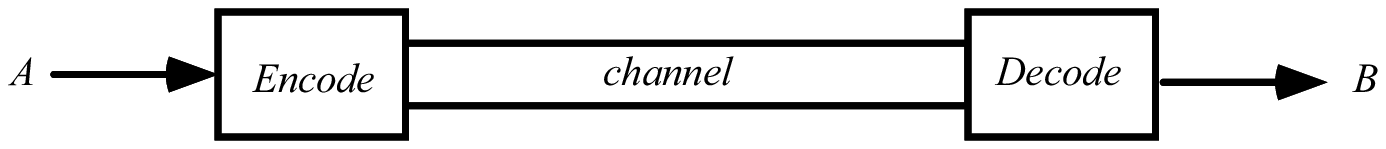}}
Fig. 4. The standard communication channel (``the information theorist's
coat of arms''). The source (Alice) produces information which is
manipulated (`encoded') and then sent over the channel. At the receiver
(Bob) the received values are 'decoded' and the information thus extracted.

\newpage\centerline{\epsfbox{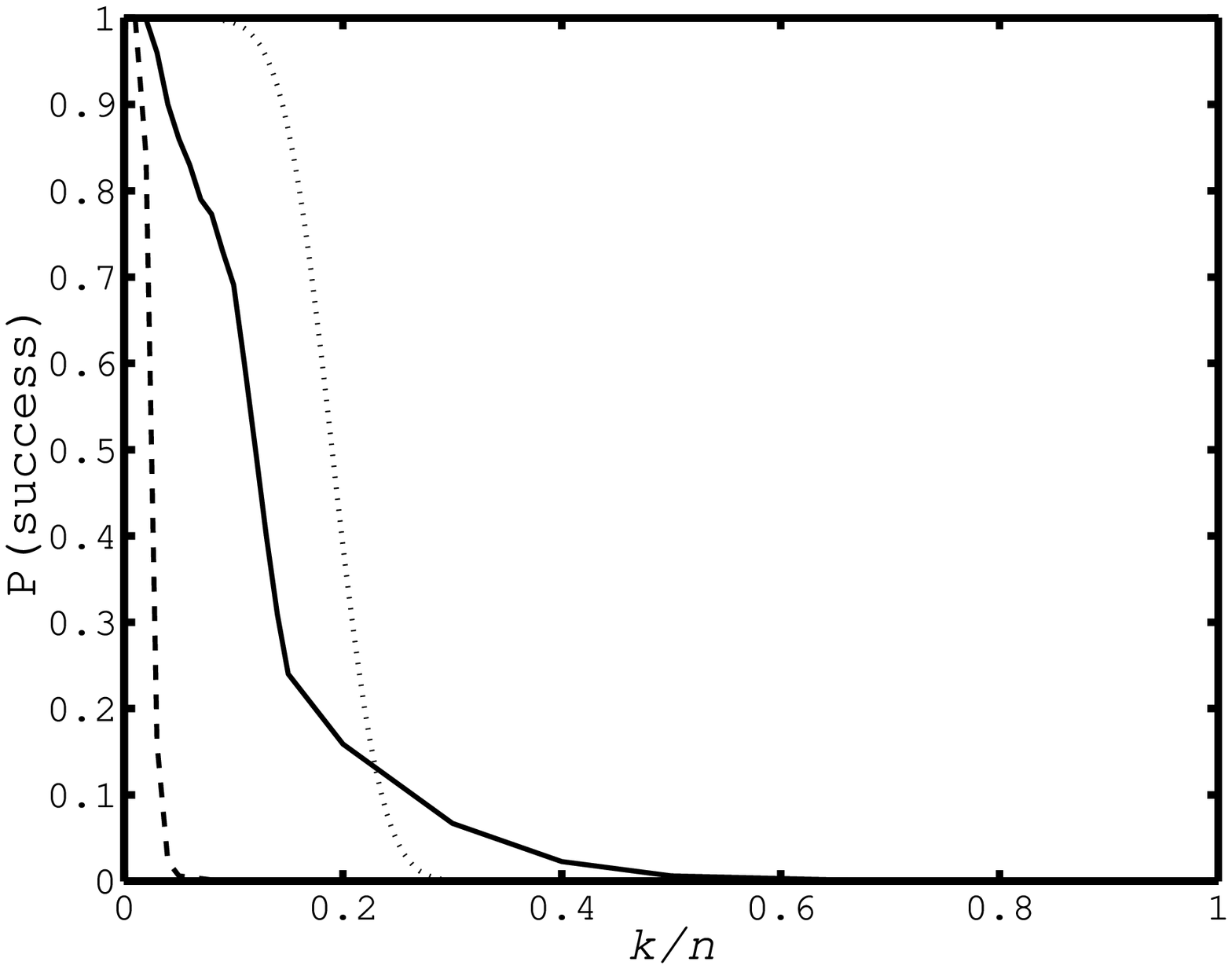}}
Fig. 5. Illustration of Shannon's theorem. Alice sends $n=100$ bits over
a noisy channel, in order to communicate $k$ bits of information to Bob.
The figure shows the probability that Bob interprets the received
data correctly, as a function of $k/n$, when the error probability
per bit is $p=0.25$. 
The channel capacity is $C = 1-H(0.25) \simeq 0.19$.
Dashed line: Alice sends each bit repeated
$n/k$ times. Full line: Alice uses the best linear error-correcting code
of rate $k/n$.
The dotted line gives the performance of error-correcting codes with
larger $n$, to illustrate Shannon's theorem.

\newpage\centerline{\epsfbox{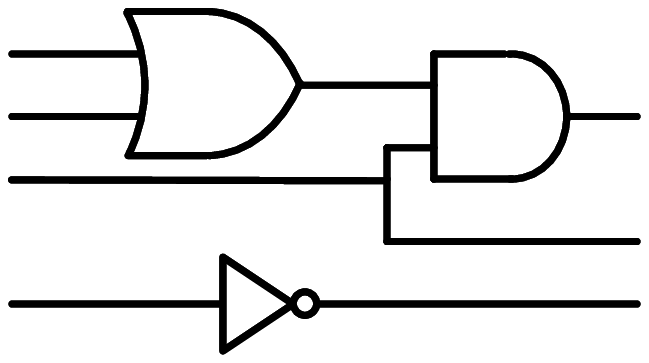}}
Fig. 6. A classical computer can be built from a 
network of logic gates.

\newpage\centerline{\epsfbox{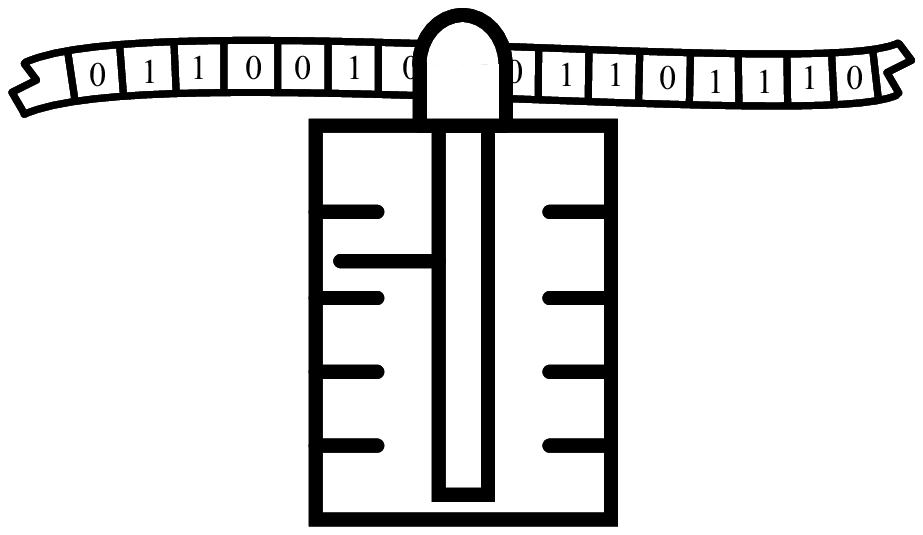}}
Fig. 7. The Turing Machine. This is a conceptual mechanical
device which can be shown to be capable of efficiently simulating all
classical computational methods. The machine has a finite set of
internal states, and a fixed design. It reads one binary symbol at
a time, supplied on a tape. The machine's action on reading a
given symbol $s$ depends only on that symbol and the internal
state $G$. The action consists in overwriting a new symbol $s'$
on the current tape location, changing state to $G'$, and moving
the tape one place in direction $d$ (left or right). The internal
construction of the machine can therefore be specified by a
finite fixed list of rules of the form $(s,G \rightarrow s', G', d)$.
One special internal state is the `halt' state: once in this state
the machine ceases further activity.
An input `programme' on the tape is transformed by the machine
into an output result printed on the tape.

\newpage\centerline{\epsfbox{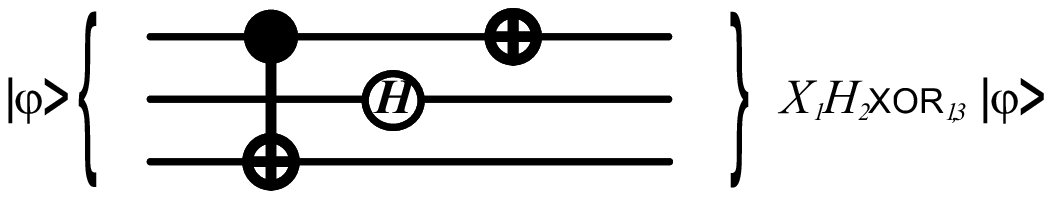}}
Fig. 8. Example `quantum network.' Each horizontal line represents
one qubit evolving in time from left to right. A symbol on
one line represents a single-qubit gate. Symbols on two qubits
connected by a vertical line represent a two-qubit gate operating
on those two qubits. The network shown carries out the operation
$X_1 H_2 {\sc xor}_{1,3} \ket{\phi}$. The $\oplus$ symbol
represents $X$ ({\sc not}), the encircled $H$ is the $H$ gate,
the filled circle linked to $\oplus$ is controlled-{\sc not}. 

\newpage\centerline{\epsfbox{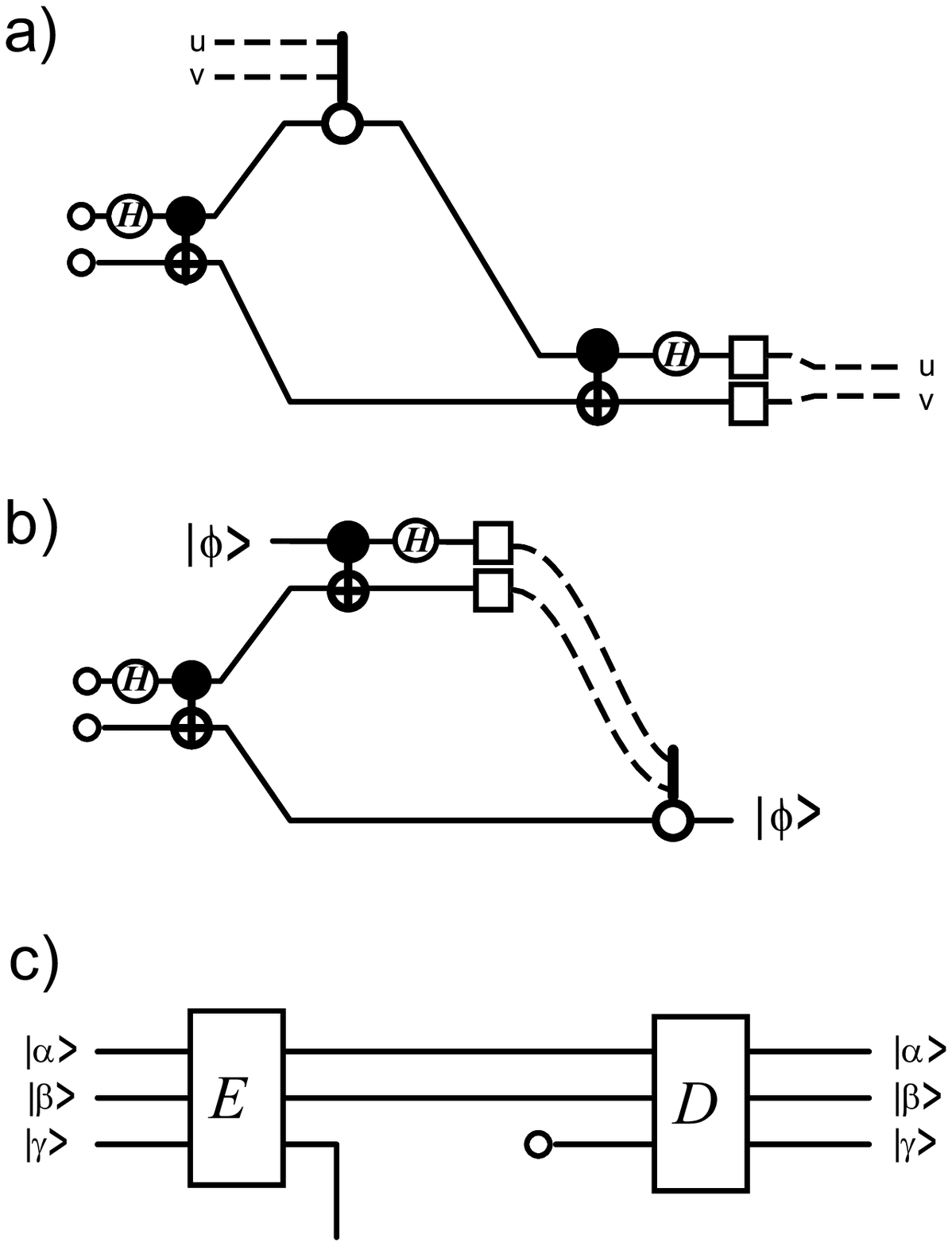}}
Fig. 9. Basic quantum communication concepts. The figure
gives quantum networks for (a) dense coding, (b) teleportation
and (c) data compression. The spatial separation of Alice and Bob
is in the vertical direction; time evolves from left to right
in these diagrams. The boxes represent measurements, the
dashed lines represent classical information.

\newpage\centerline{\epsfbox{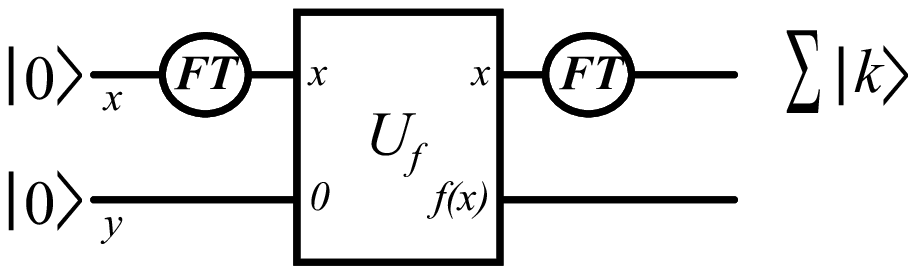}}
Fig. 10. Quantum network for Shor's period-finding algorithm. Here
each horizontal line is a quantum register rather than a single qubit.
The circles at the left represent the preparation of the input
state $\ket{0}$.
The encircled {\cal ft} represents the Fourier transform (see text),
and the box linking the two registers represents a network to
perform $U_f$. The algorithm finishes with a measurement of the $x$
regisiter.

\newpage\centerline{\epsfbox{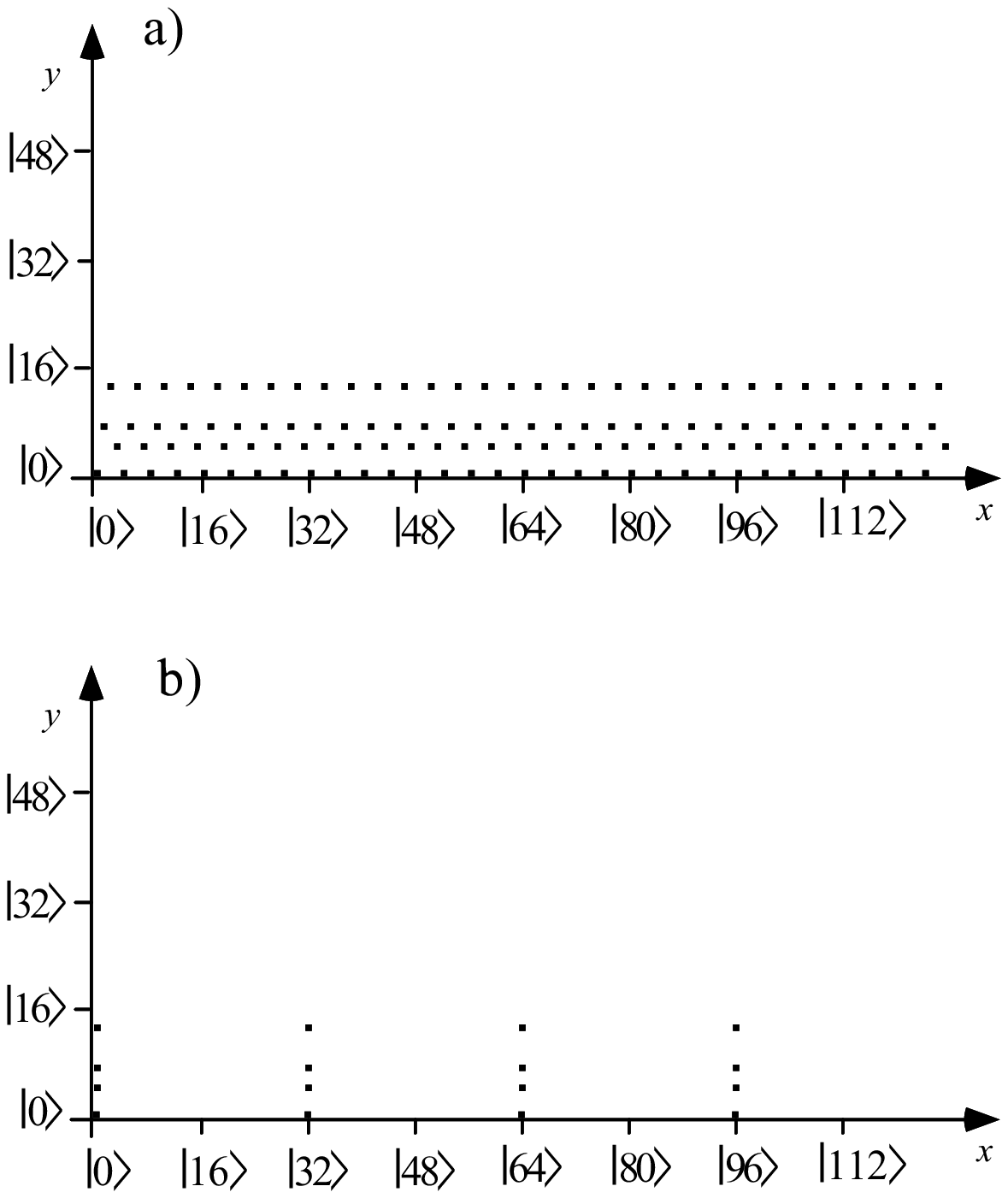}}
Fig. 11. Evolution of the quantum state in Shor's algorithm. The
quantum state is indicated schematically by identifying the
non-zero contributions to the superposition. Thus a general
state $\sum c_{x,y} \ket{x} \ket{y}$ is indicated by placing
a filled square at all those coordinates $(x,y)$ on the
diagram for which $c_{x,y} \neq 0$. (a) eq. (\protect\ref{step2}).
(b) eq. (\protect\ref{step4}). 

\newpage\centerline{\epsfbox{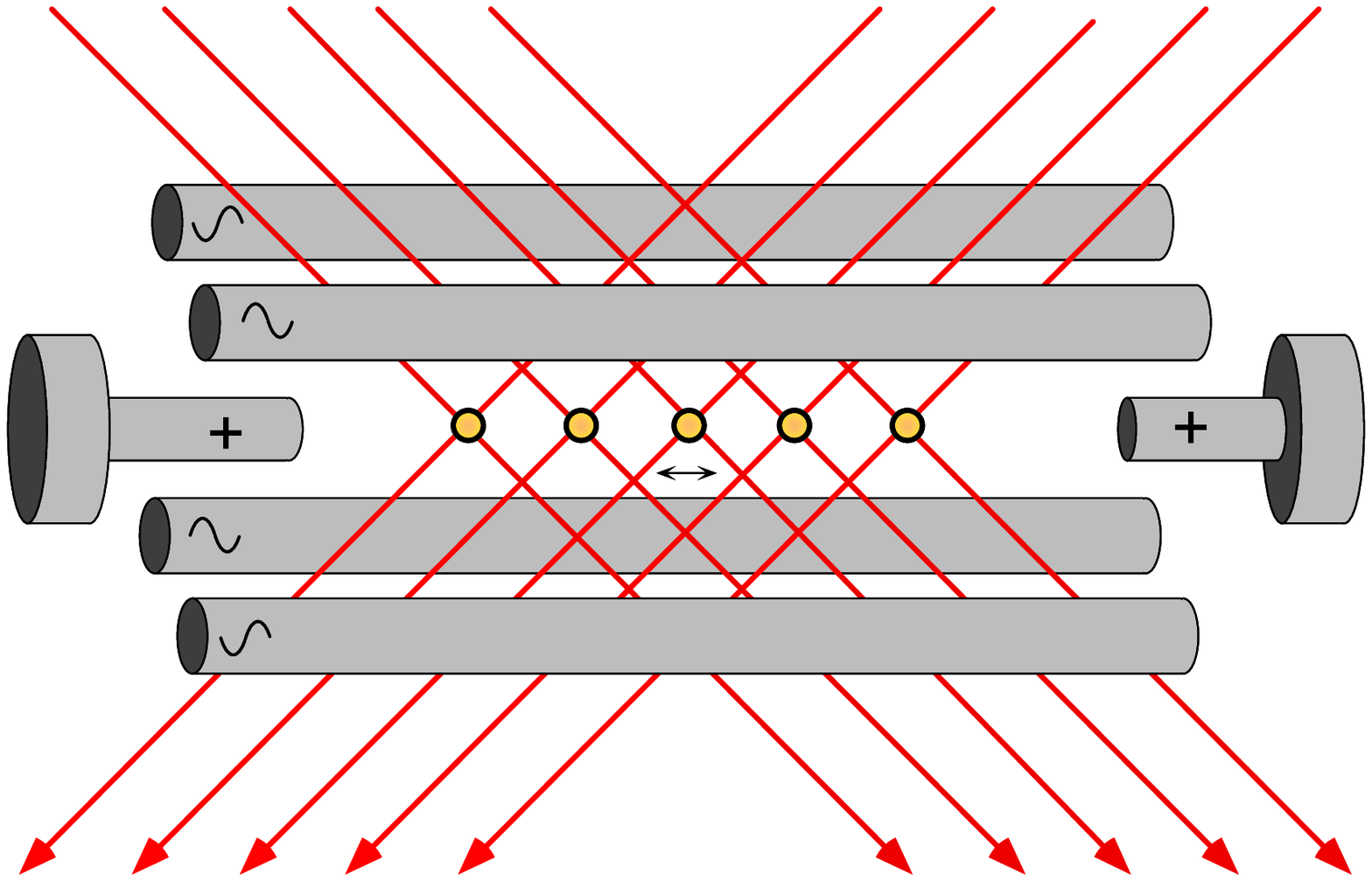}}
Fig. 12. Ion trap quantum information processor. A string of singly-charged 
atoms is stored in a linear ion trap. The ions are separated by $\sim 
20\;\mu$m by their mutual repulsion. Each ion is addressed by a pair of 
laser beams which coherently drive both Raman transitions in the ions, and
also transitions in the state of motion of the string. The motional
degree of freedom serves as a single-qubit `bus' to transport quantum
information among the ions. State preparation is by optical pumping and
laser cooling; readout is by electron shelving and resonance fluorescence,
which enables the state of each ion to be measured with high signal to
noise ratio.

\newpage\centerline{\epsfbox{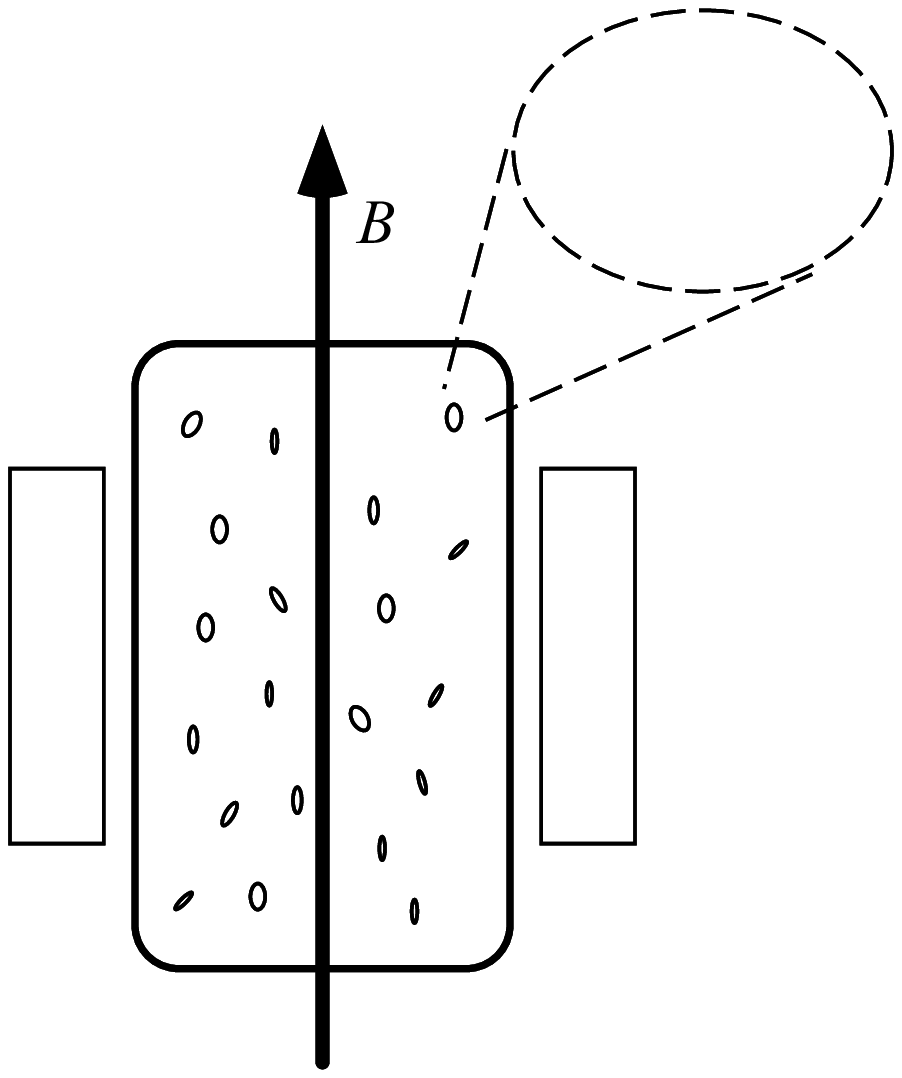}}
Fig. 13. Bulk nuclear spin resonance quantum information processor.
A liquid of $\sim 10^{20}$ `designer' molecules is placed in a sensitive
magnetometer, which can both generate oscillating magnetic fields and
also detect the precession of the mean magnetic moment of the liquid. 
The situation is somewhat like having $10^{20}$ independent
processors, but the initial state is one of thermal equilibrium,
and only the average final state can be detected. 
The quantum information is stored and manipulated in the nuclear
spin states. The spin state energy levels of a given nucleus are
influenced by neighbouring nuclei in the molecule, which enables
{\sc xor} gates to be applied. They are little influenced
by anything else, owing to the small size of a nuclear magnetic moment, 
which means the inevitable dephasing of the processors with respect to each 
other is relatively slow. This dephasing can be undone by `spin echo' 
methods. 

\newpage\centerline{\epsfbox{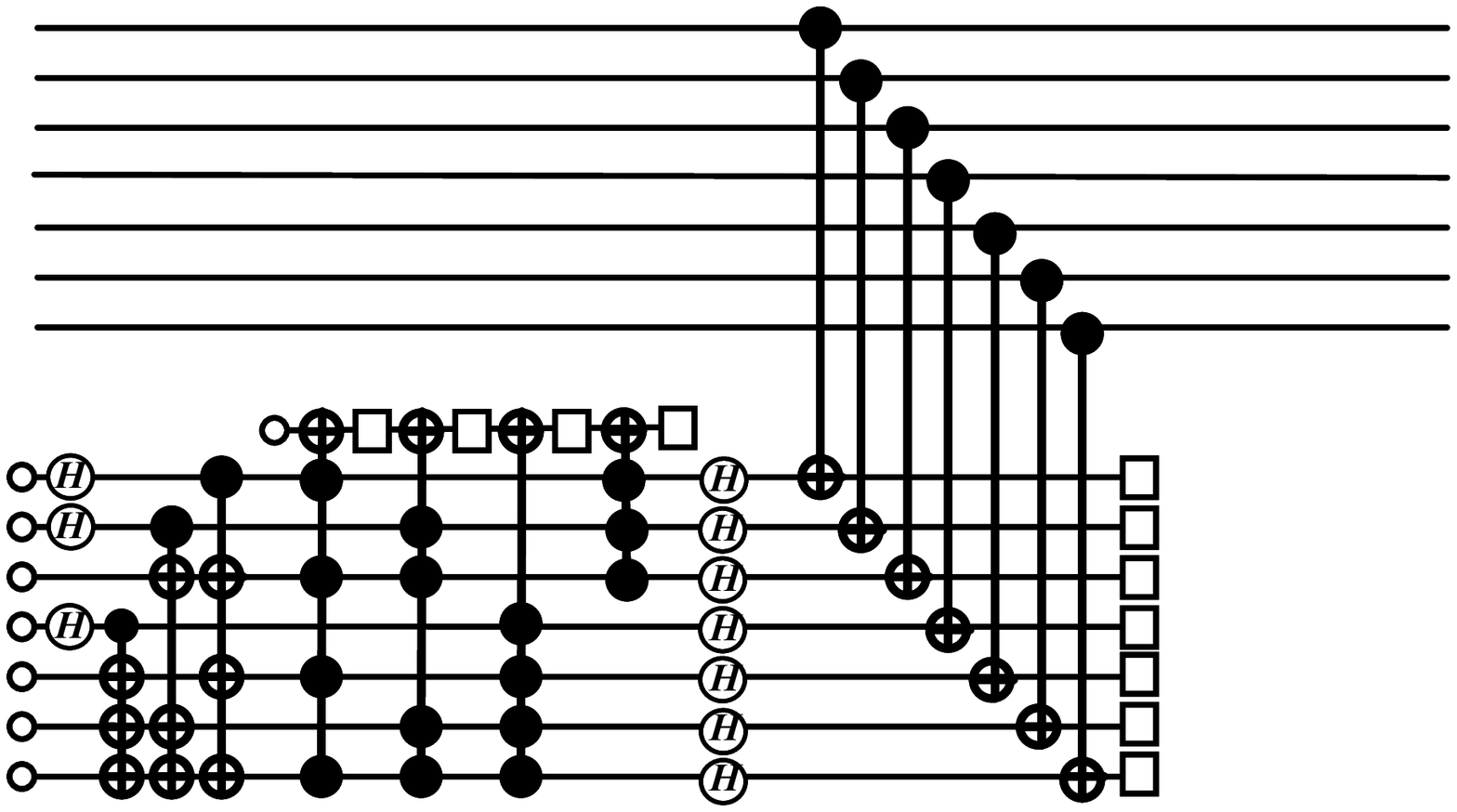}}
Fig. 14. Fault tolerant syndrome extraction, for the QECC given
in equations (\protect\ref{0E}),(\protect\ref{1E}).
The upper 7 qubits are $qc$, the lower are the ancilla $a$.
All gates, measurements and free evolution are assumed to be noisy. 
Only $H$ and 2-qubit {\sc xor} gates are used; when several
{\sc xor}s have the same control or target bit they are shown superimposed,
NB this is a non-standard notation.
The first part of the network, up until the 7 $H$ gates, prepares $a$ in 
$\ket{0_E}$, and also verifies $a$: a small box represents
a single-qubit measurement. If any measurement gives $1$, the
preparation is restarted. The $H$ gates transform the state of $a$
to $\ket{0_E} + \ket{1_E}$. Finally, the 7 {\sc xor} gates between $qc$
and $a$ carry out a single {\sc xor} in the encoded basis $\{\ket{0_E},
\ket{1_E}\}$. This operation carries $X$ errors from $qc$ into $a$,
and $Z$ errors from $a$ into $qc$. The $X$ errors in $qc$
can be deduced from the result of measuring $a$. A further network
is needed to identify $Z$ errors. Such correction never makes $qc$ 
completely noise-free, but when applied between computational steps it 
reduces the accumulation of errors to an acceptable level.

\begin{table}
\begin{center}
\begin{tabular}{lll}
Message & Huffman & Hamming \\
\\
0000 & 10       & 0000000 \\
0001 & 000      & 1010101 \\
0010 & 001      & 0110011 \\
0011 & 11000    & 1100110 \\
0100 & 010      & 0001111 \\
0101 & 11001    & 1011010 \\
0110 & 11010    & 0111100 \\
0111 & 1111000  & 1101001 \\
1000 & 011      & 1111111 \\
1001 & 11011    & 0101010 \\
1010 & 11100    & 1001100 \\
1011 & 111111   & 0011001 \\
1100 & 11101    & 1110000 \\
1101 & 111110   & 0100101 \\
1110 & 111101   & 1000011 \\
1111 & 1111001  & 0010110
\end{tabular}
\end{center}
\caption{Huffman and Hamming codes. The left column shows the sixteen
possible 4-bit messages, the other columns show the encoded version
of each message. The Huffman code is for data compression: the
most likely messages have the shortest encoded forms; the code
is given for the case that each message bit is three times more
likely to be zero than one. The Hamming code is an error correcting
code: every codeword differs from all the others in at least 3 places,
therefore any single error can be corrected. The Hamming code
is also linear: all the words are given by linear combinations of
$1010101,\;0110011,\;0001111,\;1111111$. They satisfy the
parity checks $1010101,\;0110011,\;0001111$.}
\end{table}

\end{document}